\documentclass[aps,prl,twocolumn]{revtex4}
\usepackage{bm,color}
\usepackage{graphicx}
\usepackage{amsmath}
\usepackage{color}

\begin{document}
\title {Fundamental Theory of Current-Induced Motion of Magnetic Skyrmions}
\author{Yuto Ohki}
\affiliation{Department of Applied Physics, Waseda University, Okubo, Shinjuku-ku, Tokyo 169-8555, Japan}
\affiliation{Department of Physics and Mathematics, Aoyama Gakuin University, Sagamihara, Kanagawa 229-8558, Japan}
\author{Masahito Mochizuki}
\affiliation{Department of Applied Physics, Waseda University, Okubo, Shinjuku-ku, Tokyo 169-8555, Japan}
\begin{abstract}
Magnetic skyrmions are topological spin textures that appear in magnets with broken spatial inversion symmetry as a consequence of competition between the (anti)ferromagnetic exchange interactions and the Dzyaloshinskii-Moriya interactions in a magnetic field. In the research of spintronics, the current-driven dynamics of skyrmions has been extensively studied aiming at their applications to next-generation spintronic devices. However, current-induced skyrmion motion exhibits diverse behaviors depending on various factors and conditions such as the type of skyrmion, driving mechanism, system geometry, direction of applied current, and type of the magnet. While this variety attracts enormous research interest of fundamental science and enriches their possibilities of technical applications, it is, at the same time, a source of difficulty and complexity that hinders their comprehensive understandings. In this article, we discuss fundamental and systematic theoretical descriptions of current-induced motion of skyrmions driven by the spin-transfer torque and the spin-orbit torque. Specifically, we theoretically describe the behaviors of current-driven skyrmions depending on the factors and conditions mentioned above by means of analyses using the Thiele equation. Furthermore, the results of the analytical theory are visually demonstrated and quantitatively confirmed by micromagnetic simulations using the Landau-Lifshitz-Gilbert-Slonczewski equation. In particular, we discuss dependence of the direction and velocity of motion on the type of skyrmion (Bloch type and N\'{e}el type) and its helicity, the system geometry (thin plate and nanotrack), the direction of applied current (length and width direction of the nanotrack) and its spin-polarization orientation, and the type of magnet (ferromagnet and antiferromagnet). The comprehensive theory provided by this article is expected to contribute significantly to research on the manipulation and control of magnetic skyrmions by electric currents for future spintronics applications.
\end{abstract}
\maketitle

\section{1. Introduction}
There is a growing demand for high-performance memory devices that realize robustness, high information density and power savings simultaneously~\cite{Compagnoni2017,SunG2017,AChen2016,Meena2014,Qureshi2009}. In particular, magnetic memory devices, which use magnetizations in magnets as information carriers, are considered to be superior to semiconductor memory devices in terms of infinite endurance, resistance to disturbances such as radiation and thermal fluctuations, nonvolatile nature and so on~\cite{Barla2021,Joshi2016}. For this reason, magnetic memory devices are expected to be used as ubiquitous devices that can withstand long-term stable operation with a small energy supply in various harsh environments. Because of such social demands, research on the basic technologies and fundamental sciences of next-generation magnetic memory devices such as racetrack memory~\cite{Blasing2020,Parkin2015,Parkin2008} and magnetoresistive memory~\cite{Ikegawa2020,Bhatti2017} is now being pursued intensively in the fields of spintronics, and some of them have even reached practical and commercial uses.

In these magnetic memory devices, magnetic domains in ferromagnets are mainly used as information carriers. In recent years, however, magnetic textures called skyrmions~\cite{Muhlbauer2009,YuXZ2010,SekiBook2016}, which are characterized by their topological properties~\cite{Nagaosa2013,SeidelBook2016}, have begun to show up as a candidate for new information carriers~\cite{Fert2013}. In particular, attempts to realize high-performance magnetic memory by establishing methods to create, drive and control skyrmions with electric currents instead of magnetic fields have attracted much attention~\cite{Koshibae2015,Finocchio2016,KangW2016,Fert2017,XichaoZ2020}, and the current-induced motion of skyrmions has become one of the most important subjects in the spintronics research~\cite{Jonietz2010,YuXZ2012,Iwasaki2013a,Iwasaki2013b,Sampaio2013,Iwasaki2014a,WooS2016,ZhangX2017,Litzius2020,Reichhardt2022,ZhangX2023a,Pham2024}, aiming at the application of skyrmions in racetrack memory~\cite{Tomasello2014,ZhangX2015,YuG2017,Maccariello2018,ZhuD2018,HeB2023}. In the early studies, it was focused on that skyrmions can be driven with a ultralow electric current density, which is five or six orders of magnitude smaller than that required to drive ferromagnetic domains~\cite{Jonietz2010,YuXZ2012,Iwasaki2013a}. However, as the research progressed, some critical problems that stand in the way of memory-device applications became apparent. The skyrmion Hall effect is one of the most critical problems as discussed later~\cite{Iwasaki2013a,Iwasaki2013b,ZangJ2011,Everschor-Sitte2014,JiangW2017,Litzius2017}. At the same time, various proposals have been made to suppress the skyrmion Hall effect. The attempts involve the usages of magnetic stripe domains~\cite{XingX2020,SongM2020,Muller2017a,Knapman2021,HeZ2024}, guiding lanes with designed potential wells~\cite{Purnama2015,Muller2017b,LaiP2017,CaiN2021,Toscano2020,Kern2022}, patterned structures~\cite{ZhangX2021,ZhaoL2024}, bipartite antiferromagnetic skyrmions~\cite{Barker2016,ZhangX2016a,Gobel2017,Akosa2018}, synthetic antiferromagnetic skyrmions~\cite{Pham2024,ZhangX2016b,XiaJ2019,Dohi2019,MaM2022,ChenR2022,Panigrahy2022}, and ferrimagnetic skyrmions~\cite{WooS2018,Hirata2019,LiuY2023,XuT2023,BoL2024a,Silva2024}, as well as engineering the spin-orbit coupling~\cite{Gobel2019,Akosa2019}.

It is known that the current-induced dynamics of skyrmions exhibits a variety of behaviors depending on conditions and situations, e.g., (1) the type of skyrmion (Bloch type or N\'{e}el type) and its helicity, (2) the type of magnetic torque (spin-transfer torque or spin-orbit torque), (3) the system geometry (thin plate or nanotrack), (4) the direction of applied electric current  (longitudinal or width direction of the nanotrack) and its spin-polarization orientation, and (5) the type of magnet (ferromagnet or antiferromagnet). Depending on these conditions and factors, the properties of current-induced motion of skyrmions vary, that is, the skyrmion may or may not move, the direction of motion may be parallel or perpendicular to the current, the velocity may be fast or slow and so on. The current-induced motion of skyrmions has such complicated aspects that even experts working in this research field often have misunderstanding. Now that experimental research on the current-induced motion of skyrmions is booming, it would be useful to provide a comprehensive theory to summarize them here.

In this article, we provide systematic theoretical description on the current-induced motion of skyrmions based on a general theoretical framework using the Thiele equation~\cite{Thiele1973,Everschor2012,Schulz2012}. The Thiele equation is an equation of motion describing the center-of-mass motion of magnetic texture, under the assumption that its shape does not change during the motion. Although the equation leads to simple simultaneous linear equations that are easy to solve, it describes well the motion of magnetic textures dependent on the characteristics of the system. The analysis based on the Thiele equation is highly versatile and can be applied directly to the current-induced dynamics of various magnetic textures such as magnetic domain walls, magnetic helices and magnetic vortices, in addition to skyrmions.

The rest of this paper is structured as follows. In Sec.~2, we explain fundamental properties of magnetic skyrmions. In Sec.~3, we introduce the Landau-Lifshitz-Gilbert-Slonczewski equation, which describes dynamics of current-driven magnetizations. Here we explain typical three types of torques which are exerted from electric currents to magnetizations, i.e., the spin-transfer torque, the nonadiabatic torque, and the spin-orbit torque. In Sec.~4, we derive the Thiele equation from the LLGS equation, which describes center-of-mass motion of the magnetic texture. In Secs.~5, 6, and 7, we analyze the current-induced motion of ferromagnetic and antiferromagnetic skyrmions driven by the spin-transfer torque and the spin-orbit torque using the Thiele equation. In Sec.~8, we perform micromagnetic simulations based on the LLGS equation to confirm and visualize the results of the Thiele analyses. Sec.~9 is devoted to summary and discussion. An appendix is also added for qualitative calculations of some physical quantities in the Thiele equation.

\section{2. Magnetic Skyrmions}
\begin{figure}
\includegraphics[scale=1.0]{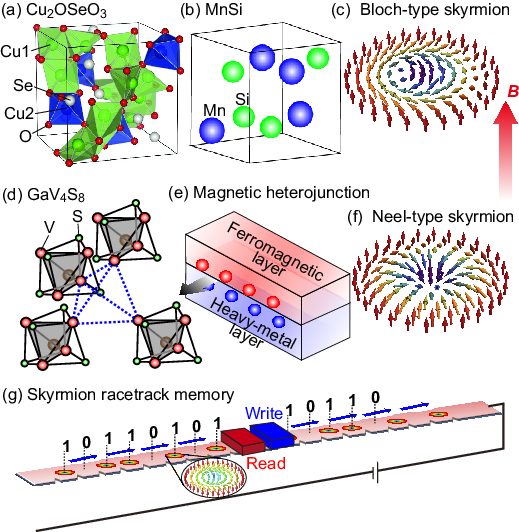}
\caption{(a),~(b) Examples of magnets with chiral cubic crystal structures which host Boch-type skyrmions, that is, (a) insulating Cu$_2$OSeO$_3$ and (b) metallic MnSi. (c) Magnetization configuration of a Bloch-type skyrmion. (d),~(e) Examples of magnetic systems which host N\'{e}el-type skyrmions, that is, (d)  lacunar spinel compounds GaV$_4$S$_8$ and GaV$_4$Se$_8$ with a polar crystal structure and (e) magnetic bilayer heterojunction with ferromagnetic and heavy-metal layers. (f) Magnetization configuration of a N\'{e}el-type skyrmion. (g) Skyrmion racetrack memory as one of the promising forms of magnetic memory devices using magnetic skyrmions.}
\label{Fig01}
\end{figure}
In ferromagnets that have a crystal structure with broken spatial inversion symmetry, the Dzyaloshinskii-Moriya (DM) interaction, which favors a helical alignment of magnetizations with a pitch angle of 90$^\circ$, becomes active~\cite{Dzyaloshinskii1957,Moriya1960a,Moriya1960b,Fert1980}. This interaction strongly competes with the ferromagnetic exchange interaction, which favors a parallel alignment of magnetizations. As a consequence of this competition, a helimagnetic state with a moderate pitch angle becomes stabilized in the absence of external magnetic field. When a steady magnetic field is applied to this helimagnetic state, skyrmions with circular magnetization configurations emerge in the plane perpendicular to the magnetic field.

In the magnetic texture of a skyrmion, the magnetizations at the periphery are aligned parallel to the external magnetic field to maximize the energy gain associated with the Zeeman interaction, while the magnetizations at the core are aligned antiparallel to the external magnetic field. In the region between the periphery and the core, the magnetizations are gradually rotating. Depending on the way of the rotation, skyrmions can be classified into three types mainly~\cite{Nagaosa2013}. The type is determined by the structure of Dzyaloshinskii-Moriya vectors, which depends on the way of the breaking of spatial inversion symmetry of the system~\cite{Bogdanov1989,Bogdanov1994}. In magnets with a chiral cubic crystal structure such as MnSi~\cite{Muhlbauer2009}, Fe$_{1-x}$Co$_x$Si~\cite{YuXZ2010,Munzer2010}, FeGe~\cite{YuXZ2011}, Cu$_2$OSeO$_3$~\cite{Seki2012a,Seki2012b}, and $\beta$-Mn type Co-Zn-Mn alloy~\cite{Tokunaga2015,Karube2017}, Bloch-type skyrmions characterized by a vortex-like magnetization configuration emerge [Figs.~\ref{Fig01}(a)-(c)]. On the contrary, polar magnets such as  GaV$_4$S$_8$~\cite{Kezsmarki2015}, GaV$_4$Se$_8$~\cite{Fujima2017} and VOSe$_2$O$_5$~\cite{Kurumaji2017,Kurumaji2021}, magnetic heterostructures~\cite{ChenG2015,Moreau-Luchaire2016,Soumyanarayanan2017}, and atomic layers~\cite{Romming2013,Heinze2011,Wiesendanger2016} host N\'{e}el-type skyrmions characterized by a fountain-like magnetization configuration [Figs.~\ref{Fig01}(d)-(f)]. In addition, antiskyrmions with antivortex magnetization configurations have been found in an inverse Heusler alloy Mn$_{1.4}$Pt$_{0.9}$Pd$_{0.1}$Sn~\cite{Nayak2017} and schreibersite compounds (Fe,Ni)$_3$P~\cite{Karube2021,Karube2022}.

\begin{figure}
\includegraphics[scale=1.0]{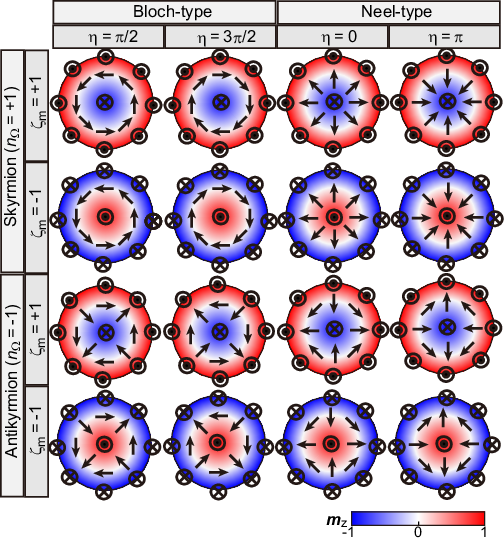}
\caption{Classification of (anti)skyrmions in terms of three quantities that characterize the magnetization configuration, i.e., the vorticity ($n_\Omega=\pm 1$), the sign of core magnetization ($\zeta_{\rm m}=\pm 1$), and the helicity $\eta$. (Anti)skyrmions with $\eta$=$\pi/2$ and $3\pi/2$ are called Bloch type, whereas those with $\eta$=$0$ and $\pi$ are called N\'{e}el type.}
\label{Fig02}
\end{figure}
The types of (anti)skyrmions with various spatial magnetization configurations are characterized and classified by differences in three quantities, i.e., the vorticity ($n_\Omega=\pm 1$), the sign of core magnetization ($\zeta_{\rm m}=\pm 1$), and the helicity $\eta$ (see Fig.~\ref{Fig02}). The vorticity $n_\Omega$ is $+1$ for skyrmions, whereas $-1$ for antiskyrmions. When the perpendicular magnetic field $\bm B_{\rm ex}=(0, 0, B_z)$ is applied, the sign of core magnetization $\zeta_{\rm m}$ is $-1$ for $B_z>0$, whereas $+1$ for $B_z<0$. The helicity $\eta$ is a continuous variable, and the (anti)skyrmions with $\eta$=$\pi/2$ or $3\pi/2$ are called Bloch type, whereas those with $\eta$=$0$ or $\pi$ are called N\'{e}el type. Normalized magnetization vectors $\bm m(r, \phi)$ constituting a(n) (anti)skyrmion is given in the polar representation form as,
\begin{align}
\begin{pmatrix}
m_x \\
m_y \\
m_z 
\end{pmatrix}
=
\begin{pmatrix}
\sin{\Theta(r)}\cos{\Phi(\phi)} \\
\sin{\Theta(r)}\sin{\Phi(\phi)} \\
\cos{\Theta(r)}
\end{pmatrix}.
\end{align}
Here the origin of the polar coordinate $(r, \phi)$ is the center of the skyrmion. The spatial profiles of $\Theta(r)$ and $\Phi(\phi)$ are approximately given by,
\begin{align}
&\Theta(r)=\pi\zeta_{\rm m} \left( \frac{r}{r_{\rm sk}} + \dfrac{\zeta_{\rm m} - 1}{2} \right), \\
&\Phi(\phi)=n_\Omega \phi + \eta,
\end{align}
where $r_{\rm sk}$ is the radius of the (anti)skyrmion.

All of these skyrmions have a quantized topological invariant $Q$ in common. The invariant $Q$ for a two-dimensional magnetic structure is given by the following equation in the continuum model of magnets,
\begin{eqnarray}
Q=\frac{1}{4\pi}\int dxdy \left( 
\frac{\partial\bm m}{\partial x} \times 
\frac{\partial\bm m}{\partial y} \right)
\cdot \bm m  \;=\zeta_{\rm m}n_\Omega \; (=\pm 1).
\end{eqnarray}
Here $\bm m(\bm r)$ is the unit magnetization vector. The sign of $Q$ is determined by the type of skyrmion and the direction of core magnetization.  When a magnetic field is applied in the positive (negative) perpendicular direction, the core magnetization is pointed in the negative (positive) perpendicular direction, i.e., $\zeta_{\rm m}=-1$ ($\zeta_{\rm m}=+1$), irrespective of the skyrmion type. In this situation, the sign of $Q$ is negative (positive) for skyrmions with $n_\Omega=+1$, whereas it is positive (negative) for antiskyrmions with $n_\Omega=-1$.

For long-period magnetic textures such as skyrmions and helices realized in chiral magnets or non-centrosymmetric magnets, spatial modulation of the magnetizations is decoupled from underlying crystal structures, and a continuum spin model, which does not take into account real crystal structures, provides a good description for the magnetic system. Furthermore, the continuum model can be reduced to a lattice spin model by dividing the spatial volume of the bulk magnet into cubic or cuboid cells. In the case of quasi two-dimensional systems such as thin-plate and nanotrack systems, we use a two-dimensional lattice spin model with square or rectangular meshes. In the square-lattice spin model, the above topological invariant is given by,
\begin{eqnarray}
Q=\frac{1}{4\pi}\sum_{(i,j,k)} \bm m_i \cdot (\bm m_j \times \bm m_k).
\end{eqnarray}
The summation is taken over combinations of three adjacent sites $( i, j, k)$ in counterclockwise order, covering the entire lattice. As a result, the topological invariant $Q$ corresponds to the sum of solid angles spanned by three adjacent magnetization vectors constituting the magnetic structure. Because the magnetization vectors constituting a skyrmion are pointing in all directions wrapping a sphere, the sum of the solid angles takes $\pm 4\pi$, i.e., the surface area of a unit sphere. Because the topological invariant $Q$ corresponds to this sum divided by $4\pi$, a single skyrmion has a quantized topological invariant of $Q=\pm 1$.

The study aiming at the use of skyrmions as information carriers in future memory devices has been a main stream of the skyrmion research since their discovery in chiral magnets in 2009~\cite{Muhlbauer2009,YuXZ2010} and the prediction of their emergence in magnetic heterojunction system in 2013~\cite{Fert2013}. In particular, skyrmions in B20 alloys (e.g., MnSi, FeGe, Fe$_{1-x}$Co$_x$Si) with a chiral cubic crystal structure have several advantageous properties for application to information carriers in magnetic memory devices, that is, (1) nanometric small size, (2) topologically protected stability, and (3) manipulability with ultralow power consumption. The realization of high-density and low-power-consuming memory devices exploiting these properties has been expected. On the other hand, skyrmions in magnetic heterojunctions are suited for systematic studies because they are rather large in size (typically micrometric size) and thus are easy to be observed, which triggered the entry of a large number of researchers into this field. However, the current density required to drive the micrometrically large skyrmions is almost the same as that of ferromagnetic domain walls, so it does not currently have advantages over other magnetic textures for memory applications. However, it was recently reported experimentally that nanometircally small skyrmions can be realized even in the magnetic heterojunction system with fabricated stacking layers~\cite{Moreau-Luchaire2016,Soumyanarayanan2017}, which are comparable to the nanometric skyrmions in chiral magnets in size.

The most feasible form of the skyrmion-based memory is the skyrmion racetrack memory in which ferromagnetic domains of racetrack memory are replaced by skyrmions [Fig.~\ref{Fig01}(g)]~\cite{Tomasello2014,ZhangX2015,YuG2017,Maccariello2018,ZhuD2018,HeB2023}. Immediately after the discovery of skyrmions in chiral magnets, it was demonstrated that the threshold current density for driving skyrmions is five or six orders of magnitude smaller than that for ferromagnetic domain walls~\cite{Jonietz2010,YuXZ2012,Iwasaki2013a}. Current drive of ferromagnetic domains in racetrack memory requires a huge current density of 10$^{11}$-10$^{12}$ A/m$^2$~\cite{Blasing2020,Parkin2015,Parkin2008}, which causes large power consumption and instability of magnetic information due to Joule heating. In contrast, skyrmions can be driven with a small current density of 10$^5$-10$^6$ A/m$^2$ in thin-plate geometry, which can suppress the energy consumption and instability due to Joule heating. Because of this ultralow threshold current density together with their topologically protected stability, skyrmions have attracted a great deal of research interest as promising building blocks for high-performance magnetic memory devices.

Motivated by this aspect, the current-induced dynamics of skyrmions has been intensively studied~\cite{Jonietz2010,YuXZ2012,Iwasaki2013a,Iwasaki2013b,Sampaio2013,Iwasaki2014a,WooS2016,ZhangX2017,Litzius2020,Reichhardt2022,ZhangX2023a,Pham2024}, but as the research progressed, it became clear that things were not as simple as had been thought in the early stage. There are mainly two reasons for the experimentally observed high mobility of skyrmions with a ultralow current density~\cite{Iwasaki2013a}. One is its magnetization configuration closed within a nanometric area, which enables current-driven skyrmions to efficiently avoid impurities and defects that could cause pinning during. The other is a finite topological invariant, whose effect can be understood from the Thiele equation as discussed later. However, theoretical analyses have revealed that current-driven skyrmions attain the high mobility only in limited situations, that is, when the skyrmions move in a thin-plate geometry away from its edges without feeling any repulsive edge potentials~\cite{Iwasaki2013a}.

On the contrary, skyrmions in nanotrack geometry inevitably feel repulsive potentials from the edges. When the repulsive potential acts strongly, the threshold current density for driving skyrmions becomes almost the same as that for driving ferromagnetic domain walls~\cite{Iwasaki2013a}. Furthermore, it has turned out that the current-driven skyrmions in the nanotrack geometry exhibits subsequent motion perpendicular to the current, which is called the skyrmion Hall effect~\cite{Iwasaki2013a,Iwasaki2013b,ZangJ2011,JiangW2017,Litzius2017}, in addition to the motion along the current. Eventually, if the current density is large, the skyrmion is pushed strongly to the upper (or lower) edge of the nanotrack, resulting in its annihilation through being absorbed by the edge. These properties of current-driven skyrmions are major problems that hinder the memory application of skyrmions. However, through intensive research to date, several methods have been proposed to partially solve these problems~\cite{XingX2020,SongM2020,Muller2017a,Knapman2021,HeZ2024,Purnama2015,Muller2017b,LaiP2017,CaiN2021,Toscano2020,Kern2022,ZhangX2021,ZhaoL2024,Barker2016,ZhangX2016a,Gobel2017,Akosa2018,Pham2024,ZhangX2016b,XiaJ2019,Dohi2019,MaM2022,ChenR2022,Panigrahy2022,WooS2018,Hirata2019,LiuY2023,XuT2023,BoL2024a,Silva2024,Gobel2019,Akosa2019}. The purpose of this paper is to construct a universal theoretical framework describing the current-induced dynamics of skyrmions under various conditions comprehensively.

In this article, we provide systematic and comprehensive theoretical descriptions of fundamental properties of motion of skyrmions driven by electric currents. The current-induced motion of skyrmions has turned out to sensitively depend on the type of skyrmion (Bloch type or N\'{e}el type) and its helicity, driving torque (spin-transfer torque or spin-orbit torque), system shape (thin plate or nanotrack), direction of applied electric current (length or width direction of the nanotrack) and its spin-polarization orientation, type of magnet (ferromagnet or antiferromagnet) and so on. We systematically discuss dependence of the behaviors of current-induced skyrmion motion on these factors and conditions by means of theoretical analyses using the Thiele equation, which is an equation of center-of-mass motion for driven magnetic textures. Subsequently, the results of the analytical theory are visually demonstrated and quantitatively confirmed by micromagnetic simulations based on the Landau-Lifshitz-Gilbert-Slonczewski equation, which is a time-evolution equation of current-driven magnetizations. We expect that the systematic and comprehensive understandings of the fundamental properties of current-induced motion of skyrmions established by this article will be a firm ground to future research on the manipulation and control of skyrmions with electric currents for spintronics applications.

\section{3. LLGS Equation}
\begin{figure}
\includegraphics[scale=1.0]{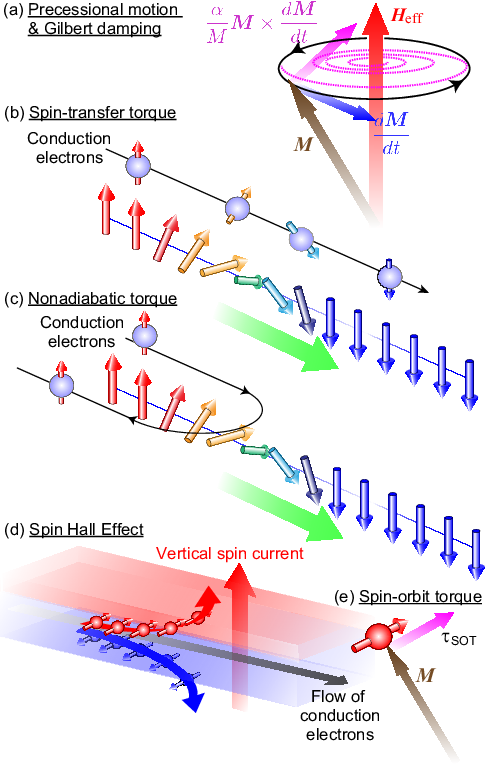}
\caption{Schematics of the effect of each term in the LLGS equation. (a) Precession of the magnetization $\bm M$ around the effective local magnetic field $\bm B_{\rm eff}$ described by the first term and the Gilbert damping due to the energy dissipation described by the second term. (b) Spin-transfer torque described by the third term. (c) Nonadiabatic torque described by the fourth term. (d) Vertical spin current generated via the spin Hall effect. (e) Spin-orbit torque described by the fifth term. For details of the effect and role of each term, see text.}
\label{Fig03}
\end{figure}
When an electric current is injected into a metallic ferromagnet, conduction electrons of the current become spin-polarized through exchange interactions with the ferromagnetic magnetization. On the other hand, an electric current injected into a heavy-metal layer of ferromagnet/heavy-metal bilayer heterojunction is converted to a pure spin current flowing perpendicular to the interface via the spin Hall effect due to the spin-orbit coupling. The generated spin-polarized current or spin current exerts several kinds of torques on magnetization vectors constituting the noncollinear magnetic textures such as ferromagnetic domain walls, magnetic helices, magnetic vortices and skyrmions in magnets and induces their motion. There are mainly three kinds of spin torques for current-driven dynamics of magnetizations~\cite{Brataas2012}. 

The first is the spin-transfer torque~\cite{Slonczewski1996,ZhangLi2004,Tatara2004}. A concept of the spin-transfer torque was proposed by Slonczewski in 1996, which is referred to as the Slonczewski model~\cite{Slonczewski1996}. He considered a system composed of two parallel ferromagnetic films connected by a nonmagnetic metallic spacer and a situation in which an electric current is injected perpendicular to the films. The injected current becomes spin-polarized when it flows through the first hard ferromagnetic film due to the exchange coupling between the conduction electron spins and the ferromagnetic magnetization. This spin-polarized current exerts torque to the magnetization in the second soft ferromagnetic film via the transfer of spin angular momenta, which results in mesoscopic precession and switching of the magnetization. On the other hand, another concept of the spin-transfer torque was proposed by Zhang and Li in 2004, which is referred to as the Zhang-Li model~\cite{ZhangLi2004}. They proposed that the spin-transfer torque also occurs in magnetic systems with spatial gradients of magnetization or for noncollinear magnetic structures. The spin angular momenta of conduction electrons of the spin-polarized current are transferred to the spatially varying magnetization, which induces motion of the magnetic textures.

The transfers of spin angular momenta occur in both adiabatic and nonadiabatic ways in the Zhang-Li model. The adiabatic contribution is nothing but the spin-transfer torque argued above. On the other hand, the nonadiabatic contribution gives rise to the second torque, which is referred to as the nonadiabatic torque~\cite{ZhangLi2004,Tatara2004}. This torque drives noncollinear magnetization textures via nonadiabatic transfer of angular momenta from conduction-electron spins to the magnetizations. The third is the spin-orbit torque, which is a torque exerted by a vertical spin current generated by the spin Hall effect due to the interface spin-orbit coupling in the magnetic heterojunction system~\cite{Manchon2019,ShaoQ2021,KimKW2024}. Among these three torques, the nonadiabatic torque always works when the spin-transfer torque works. Hereafter, we refer to the driving mechanism with these two kinds of spin torques as the spin-transfer torque mechanism. On the contrary, the driving mechanism with the spin-orbit torque, which works in the magnetic heterojunction system, is referred to as the spin-orbit torque mechanism.

The time evolution of magnetization vectors in ferromagnets in the presence of electric current is described by the Landau-Lifshitz-Gilbert-Slonczewski (LLGS) equation~\cite{Slonczewski1996}:
\begin{eqnarray}
\frac{d \bm M}{dt}
&=&-\gamma \bm M \times \bm B^{\rm eff}
+\frac{\alpha}{M} \bm M \times \frac{d\bm M}{dt}
\nonumber \\
& &+\frac{\gamma\hbar a^3p}{2eM}(\bm j_{\rm e} \cdot \bm \nabla)\bm M
\nonumber \\
& &-\beta\frac{\gamma\hbar a^3p}{2eM}\cdot\frac{1}{M}\left[ \bm M \times (\bm j_{\rm e} \cdot \bm \nabla)\bm M \right]
\nonumber \\
& &+\frac{\gamma\hbar a^3 |\theta_{\rm SH}|j_{\rm e}}{2eM}\cdot\frac{1}{Md}
\left[ \bm M \times (\bm \sigma \times \bm M) \right].
\label{eq:LLG1}
\end{eqnarray}
Here the ferromagnetic system is divided into cubic cells, each of which has a volume of $a^3$, and the magnetic moments $\bm M=\bm M_i$ in the $i$th cell (or $\bm M=\bm M(\bm r)$ in the cell at the position $\bm r$) is considered. The unit of the magnetic moment $\bm M$ is A$\cdot$m$^2$, and its magnitude is
\begin{math} 
|\bm M|=M=a^3M_{\rm s}.
\end{math}
Here $M_{\rm s}$ is the saturation magnetization in the unit of A$\cdot$m$^{-1}$. 

The first terms is the gyrotropic term, which describes the precession of magnetization vector $\bm M$ around the effective local magnetic field $\bm B_{\rm eff}$, while the second term is the Gilbert-damping term, which describes gradual decrease in radius of precession due to energy dissipation phenomenologically [Fig.~\ref{Fig03}(a)]. The effective local magnetic field $\bm B^{\rm eff}$ includes, in addition to the real externally applied magnetic field, contributions from coupling with surrounding magnetizations through exchange and DM interactions and those from magnetic anisotropies, which is given by,
\begin{eqnarray}
\bm B_i^{\rm eff}
=-\frac{\partial \mathcal{H}}{\partial \bm M_i},
\label{eq:Beff1a}
\end{eqnarray}
or
\begin{eqnarray}
\bm B^{\rm eff}(\bm r)=-a^3\frac{\delta \mathcal{E}(\bm r)}{\delta \bm M(\bm r)},
\label{eq:Beff1b}
\end{eqnarray}
where $\mathcal{H}$ is the Hamiltonian of the lattice spin model with cells of volume $a^3$ as lattice sites, and $\mathcal{E}(\bm r)$ is the local energy density. Here $\gamma=g\mu_{\rm B}/\hbar(>0)$ is the gyromagnetic ratio, $e(>0)$ is the elementary charge of electron, and $\alpha$ is the dimensionless Gilbert-damping coefficient.

The third to fifth terms describe contributions of the electric current $\bm j_{\rm e}$ to the magnetization dynamics. Specifically, they describe the spin-transfer torque [Fig.~\ref{Fig03}(b)], the nonadiabatic torque [Fig.~\ref{Fig03}(c)], and the spin-orbit torque [Fig.~\ref{Fig03}(d)], respectively. For the former two torques, we consider those of the Zhang-Li model~\cite{ZhangLi2004} instead of the Slonczewski model~\cite{Slonczewski1996}. The variable $p$  in the third and fourth terms is a quantity called the spin polarization that represents the degree to which the electric current is spin polarized. The coefficient $\beta$ in the fourth term is the strength of the nonadiabatic torque. The variable $\theta_{\rm SH}$  in the fifth term is called the spin Hall angle. Its absolute value represents the conversion efficiency $j_{\rm s}/j_{\rm e}$ from the in-plane electric current density $j_{\rm e}$ to the vertical spin-current density $j_{\rm s}$, while its sign represents the sign of the spin polarization of the vertical spin current. The vector $\bm \sigma$ is a unit directional vector representing the direction of the spin polarization of the spin current, which is usually parallel to the heterojunction interface and orthogonal to the injected in-plane electric current. When the electric current is applied in the longitudinal direction ($x$-direction) of the nanotrack system, it should be $\bm \sigma=(0,\pm1,0)\parallel \bm y$. On the contrary, when the electric current is applied in the transverse direction ($y$-direction), it should be $\bm \sigma=(\pm1,0,0)\parallel \bm x$. The sign depends on the sign of the spin Hall angle $\theta_{SH}$ governed by the spin-orbit coupling.

Dividing both sides of Eq.~(\ref{eq:LLG1}) by $M=|\bm M|$ and considering that $M=a^3 M_{\rm s}$, the LLGS equation can be rewritten by using unit magnetization vectors $\bm m_i$ or $\bm m(\bm r)$ as,
\begin{eqnarray}
\frac{d \bm m}{dt}
&=&-\gamma \bm m \times \bm B^{\rm eff}
+\alpha \bm m \times \frac{d\bm m}{dt}
\nonumber \\
& &+\frac{\gamma\hbar p}{2eM_{\rm s}}(\bm j_{\rm e} \cdot \bm \nabla)\bm m
-\beta\frac{\gamma\hbar p}{2eM_{\rm s}}
\left[ \bm m \times (\bm j_{\rm e} \cdot \bm \nabla)\bm m \right]
\nonumber \\
& &+\frac{\gamma\hbar |\theta_{\rm SH}|j_{\rm e}}{2eM_{\rm s}}\cdot\frac{1}{d}
\left[ \bm m \times (\bm \sigma \times \bm m) \right],
\label{eq:LLG2}
\end{eqnarray}
with 
\begin{eqnarray}
\bm B_i^{\rm eff}=-\frac{1}{a^3 M_{\rm s}}\frac{\partial \mathcal{H}}{\partial \bm m_i},
\label{eq:Beff2a}
\end{eqnarray}
or
\begin{eqnarray}
\bm B^{\rm eff}(\bm r)=-\frac{1}{M_{\rm s}}\frac{\delta \mathcal{E}(\bm r)}{\delta \bm m(\bm r)}.
\label{eq:Beff2b}
\end{eqnarray}

Here a typical form of the Hamiltonian $\mathcal{H}$ for the skyrmion-hosting magnetic materials is given by,
\begin{align}
\mathcal{H}=&
-J \sum_{\langle i,j\rangle} \bm m_i \cdot \bm m_j
-\bm B_{\rm ext} \cdot \sum_i \bm m_i
\nonumber \\
&+D\sum_{i}\sum_{\gamma=x,y,z}
\hat{\bm \gamma} \cdot (\bm m_i \times \bm m_{i+ \hat{\bm \gamma}}).
\label{eq:model1}
\end{align}
The first three terms describe the ferromagnetic exchange interactions, the Zeeman interactions, and the DM interactions, respectively. On the other hand, the local energy density $\mathcal{E}(\bm r)$  is given by,
\begin{align}
\mathcal{E}
&=\frac{J}{2a}(\nabla \bm m)^2 - M_{\rm s}(\bm B_{\rm ext} \cdot \bm m)
-\frac{D}{a^2} \bm m \cdot (\bm \nabla \times \bm m).
\label{eq:model2}
\end{align}
Details of the Hamiltonian are explained later in Sec.8.

\section{4. Thiele Equation}
\subsection{4.1 Derivation}
The Thiele equation is an equation of motion that describes the center-of-mass motion of the magnetization texture in magnets when it is driven by an external field such as current, magnetic field, and electric field~\cite{Thiele1973}. This equation is derived by a mathematical transformation of the LLG and LLGS equations under an assumption that the magnetization texture does not change its shape during the motion. In addition, gradual variation in space is assumed for the magnetization configuration. The procedure of deriving the Thiele equation is briefly presented in the following.

By taking the outer product with $\bm m$ from left for both sides of Eq.~(\ref{eq:LLG2}), we obtain,
\begin{eqnarray}
\bm m \times \dot{\bm m}&=&\bm m \times \left[\bm m \times \frac{\gamma}{M_{\rm s}}\left(\frac{\delta \mathcal{E}}{\delta \bm m}\right)\right]+\alpha \bm m \times (\bm m \times \dot{\bm m})
\nonumber \\
&-&\bm m \times (\bm v_{\rm s} \cdot \bm \nabla)\bm m
+\beta \bm m \times \left[\bm m \times (\bm v_{\rm s} \cdot \bm \nabla)\bm m \right]
\nonumber \\
&+&\frac{v_{\rm s}^\perp}{d} \bm m \times
\left[ \bm m \times (\bm \sigma \times \bm m) \right].
\label{eq:THeq01}
\end{eqnarray}
Here $\bm v_{\rm s}$ and $v_{\rm s}^\perp$ are the normalized in-plane spin-polarized current and the normalized vertical spin current, respectively, both of which have a velocity dimension,
\begin{align}
&\bm v_{\rm s} \equiv -\frac{\gamma\hbar p}{2eM_{\rm s}} \bm j_{\rm e},
\label{eq:vs1}
\\
&v_{\rm s}^\perp \equiv \frac{\gamma\hbar |\theta_{\rm SH}|}{2eM_{\rm s}} j_{\rm e}
=\frac{\gamma\hbar}{2eM_{\rm s}} j_{\rm s}.
\label{eq:vs2}
\end{align}
Here $j_{\rm s}(=|\theta_{\rm SH}|j_{\rm e})$ is the spin-current density. The spin polarization $\bm \sigma$ of the vertical spin current depends on the direction of the electric current $\bm j_{\rm e}$ and the sign of the spin Hall angle $\theta_{\rm SH}$ as,
\begin{eqnarray}
\bm \sigma={\rm sgn}(\theta_{\rm SH})\frac{\bm j_{\rm e} \times \hat{\bm z}}{|\bm j_{\rm e} \times \hat{\bm z}|}.
\label{eq:sigma}
\end{eqnarray}
The normalized vertical spin current $v_{\rm s}^\perp$ defined by Eq.~(\ref{eq:vs2}) is always positive by definition. When we define this quantity with $\theta_{\rm SH}$ instead of $|\theta_{\rm SH}|$, it should have a sign, which represents the spin-polarization direction. However, to avoid confusion, $v_{\rm s}^\perp$ is always considered in its absolute value in Eq.~(\ref{eq:THeq01}), and the spin-polarization direction is considered through the sign of $\bm \sigma$.

Each term in Eq.~(\ref{eq:THeq01}) is transformed using the following vector formula,
\begin{eqnarray}
\bm A \times (\bm B \times \bm C)=(\bm A \cdot \bm C)\bm B - (\bm A \cdot \bm B)\bm C.
\label{eq:vecform1}
\end{eqnarray}
Since the length of $\bm m$ is constant to be unity, the relation $\bm m \cdot \bm m$=1 holds. Furthermore, since $\dot{\bm m}$ has only an azimuthal component, the relation $\bm m \cdot \dot{\bm m}=0$ holds. As a result, we obtain the following equation,
\begin{eqnarray}
\bm m \times \dot{\bm m}&=&\left[\bm m \cdot \frac{\gamma}{M_{\rm s}}\left(\frac{\delta \mathcal{E}}{\delta \bm m}\right)\right]\bm m -\frac{\gamma}{M_{\rm s}} \frac{\delta \mathcal{E}}{\delta \bm m}-\alpha \dot{\bm m}
\nonumber \\
&-&\bm m \times (\bm v_{\rm s} \cdot \bm \nabla)\bm m
+\beta \left[\bm m \cdot (\bm v_{\rm s} \cdot \bm \nabla)\bm m \right]\bm m\nonumber \\
&-&\beta (\bm v_{\rm s} \cdot \bm \nabla)\bm m
+\frac{v_{\rm s}^\perp}{d} \bm m \times \bm \sigma.
\label{eq:THeq02}
\end{eqnarray}

If the magnetization texture is rigid and not deformed during the motion, the unit magnetization vector $\bm m(\bm r, t)$ at time $t$ and position $\bm r$ is determined only by the relative coordinate $\bm \xi(t) \equiv \bm r-\bm R(t)$ measured from the center-of-mass coordinate $\bm R(t)$. This aspect is expressed by the following equation,
\begin{eqnarray}
\bm m(\bm r, t)=\bm m(\bm \xi(t))=\bm m(\bm r-\bm R(t)).
\label{eq:Relat01}
\end{eqnarray}
In this case, the following equation holds,
\begin{eqnarray}
\dot{\bm m}
=\sum_\nu \dot{\xi}_\nu \frac{\partial \bm m}{\partial \xi_\nu}
=-\sum_\nu \dot{R}_\nu \frac{\partial \bm m}{\partial \xi_\nu}
=-(\bm v_{\rm d} \cdot \bm \nabla) \bm m,
\label{eq:Relat02}
\end{eqnarray}
where $\bm R=(X, Y, Z)$ is the coordinate of the center-of-mass of the magnetization texture, $\bm \xi=(x, y, z)$ is the coordinate relative to the center-of-mass position $\bm R$, and $\dot{\bm R}=(\dot{X}, \dot{Y}, \dot{Z}) \equiv \bm v_{\rm d}$ is the drift velocity of the center-of-mass motion. Hereafter, the symbol $\sum_\nu$ for the sum over $\nu$ ($=x,y,z$) is omitted. Substituting Eq.~(\ref{eq:Relat02}) into Eq.~(\ref{eq:THeq02}) and taking the inner product with $\displaystyle{-\frac{\partial \bm m}{\partial \xi_\mu}}$ for both sides, the following equation is obtained after some algebra,
\begin{align}
&\bm m \cdot \left(\frac{\partial \bm m}{\partial \xi_\mu} \times \frac{\partial \bm m}{\partial \xi_\nu} \right) \left(v_{{\rm s}\nu} - v_{{\rm d}\nu}\right)
\nonumber \\
&\hspace{0.5cm}
-\frac{\partial \bm m}{\partial \xi_\mu} \cdot \frac{\partial \bm m}{\partial \xi_\nu} \left(\beta v_{{\rm s}\nu} - \alpha v_{{\rm d}\nu} \right)
\nonumber \\
&\hspace{0.5cm}
-\frac{\gamma}{M_{\rm s}} \frac{\partial \bm m}{\partial \xi_\mu} \cdot \frac{\delta \mathcal{E}}{\delta \bm m}
-\frac{v_{\rm s}^\perp}{d} \bm \sigma \cdot \left(
\frac{\partial \bm m}{\partial \xi_\mu} \times \bm m \right)
=0.
\end{align}
In the process, the vector formula
\begin{eqnarray}
\bm A \cdot (\bm B \times \bm C)=\bm B \cdot (\bm C \times \bm A),
\label{eq:vecform1}
\end{eqnarray}
was used to transform the first term
\begin{eqnarray}
-\frac{\partial \bm m}{\partial \xi_\mu} \cdot \left[\bm m \times  \left(v_{{\rm s}\nu} - v_{{\rm d}\nu}\right) \frac{\partial \bm m}{\partial \xi_\nu} \right]
\end{eqnarray}
into
\begin{eqnarray}
\bm m \cdot \left(\frac{\partial \bm m}{\partial \xi_\mu} \times \frac{\partial \bm m}{\partial \xi_\nu} \right) \left(v_{{\rm s}\nu} - v_{{\rm d}\nu}\right).
\end{eqnarray}
In this transformation, we also used the fact that $\displaystyle{\frac{\partial \bm m}{\partial \xi_\mu} \cdot \bm m=0}$, since the spatial variation $\displaystyle{\frac{\partial \bm m}{\partial \xi_\mu}}$ of the unit magnetization vector $\bm m$ always occurs in the direction orthogonal to $\bm m$.

Both sides of this equation are volume-integrated, and the integral with respect to $z$ is replaced by a multiplication by the number of layers $d/a$, where $a$ is the lattice constant. Furthermore, dividing the entire equation by $d/a$ yields the following equation,
\begin{eqnarray}
G_{\mu\nu} \left(v_{{\rm s}\nu} - v_{{\rm d}\nu}\right) - D_{\mu\nu} \left(\beta v_{{\rm s}\nu} - \alpha v_{{\rm d}\nu} \right) - J_\mu^{\rm spin} - F_\mu=0.
\nonumber \\
\label{eq:Thiele0}
\end{eqnarray}
with
\begin{eqnarray}
\label{eq:G}
G_{\mu\nu}
&=&\int_{\rm UC} \bm m \cdot \left(\frac{\partial \bm m}{\partial \xi_\mu} \times \frac{\partial \bm m}{\partial \xi_\nu}\right) \; dxdy \\
\nonumber \\
\label{eq:D}
D_{\mu\nu}
&=&\int_{\rm UC} \left(
\frac{\partial \bm m}{\partial \xi_\mu} \cdot \frac{\partial \bm m}{\partial \xi_\nu} 
\right) \; dxdy \\
\nonumber \\
\label{eq:J}
J_\mu^{\rm spin}
&=&-\frac{v_{\rm s}^\perp}{d} \bm \sigma \cdot
\int_{\rm UC} \left(
\frac{\partial \bm m}{\partial \xi_\mu} \times \bm m \right) \; dxdy \\
\nonumber \\
\label{eq:F}
F_\mu
&=&\frac{\gamma}{M_{\rm s}}
\int_{\rm UC}  \left(
\frac{\partial \bm m}{\partial \xi_\mu} \cdot \frac{\delta \mathcal{E}}{\delta \bm m} 
\right) \; dxdy
\nonumber \\
&=&-\frac{\gamma}{M_{\rm s}}
\int_{\rm UC} \frac{\partial \mathcal{E}}{\partial R_\mu} \;dxdy
\equiv -\nabla_{\bm R}V.
\end{eqnarray}
Here we used the fact that $\bm \xi=\bm r - \bm R(t)$ for the second half of the transformation of $F_\mu$.

In the case of magnetic textures such as skyrmions and magnetic vortices, the following holds for $G_{ij}$,
\begin{eqnarray}
G_{\mu\nu}
&=&\int_{\rm UC} \bm m \cdot \left(\frac{\partial \bm m}{\partial \xi_\mu} \times \frac{\partial \bm m}{\partial \xi_\nu}\right) \; dxdy
\nonumber \\
&=&\left\{
\begin{array}{ll}
\mathcal{G}  & \;\;\; {\rm for}\; (\xi_\mu, \xi_\nu)=(x, y)\\
-\mathcal{G} & \;\;\; {\rm for}\; (\xi_\mu, \xi_\nu)=(y, x)\\
0            & \;\;\; {\rm otherwise} \\
\end{array} \right.
\end{eqnarray}
The first term of Eq.~(\ref{eq:Thiele0}) can be written as,
\begin{eqnarray}
G_{\mu\nu} \left(v_{{\rm s}\nu} - v_{{\rm d}\nu}\right)
&=&
\left\{
\begin{array}{ll}
 \mathcal{G}\left(v_{{\rm s}y} - v_{{\rm d}y}\right) & \;\;\; {\rm for}\; \mu=x \\
-\mathcal{G}\left(v_{{\rm s}x} - v_{{\rm d}x}\right) & \;\;\; {\rm for}\; \mu=y \\
0                   & \;\;\; {\rm for}\; \mu=z \\
\end{array} \right.
\nonumber \\
&=&-\bm G \times \left(\bm v_{\rm s} - \bm v_{\rm d}\right).
\end{eqnarray}
Here the vector $\bm G$ is defined as,
\begin{eqnarray}
\bm G =(0, 0, \mathcal{G}).
\end{eqnarray}
For $D_{ij}$, the following holds,
\begin{eqnarray}
D_{\mu\nu}
&=&\int_{\rm UC} \frac{\partial \bm m}{\partial \xi_\mu} \cdot \frac{\partial \bm m}{\partial \xi_\nu} \; dxdy
\nonumber \\
&=&\left\{
\begin{array}{ll}
\mathcal{D} & \;\;\; {\rm for}\; (\xi_\mu, \xi_\nu)=(x, x),\;(y, y) \\
0           & \;\;\; {\rm otherwise} \\
\end{array} \right.
\end{eqnarray}
Thereby, the second term in Eq.~(\ref{eq:Thiele0}) can be written as,
\begin{eqnarray}
-D_{\mu\nu} \left(\beta v_{{\rm s}\nu} - \alpha v_{{\rm d}\nu}\right)
=-\mathcal{D} \left(\beta \bm v_{\rm s} - \alpha \bm v_{\rm d}\right).
\end{eqnarray}
Putting the above together, Eq.~(\ref{eq:Thiele0}) is rewritten in the form,
\begin{eqnarray}
\bm G \times \left(\bm v_{\rm s} - \bm v_{\rm d}\right) +\mathcal{D}\left(\beta \bm v_{\rm s} - \alpha \bm v_{\rm d} \right) + \bm J^{\rm spin } + \bm F =0.
\label{eq:Thiele1}
\end{eqnarray}
Rewritting 
\begin{math}
\displaystyle
\bm \nabla_{\rm R}=\frac{\partial}{\partial R_\mu}=(\frac{\partial}{\partial X}, \frac{\partial}{\partial Y}, \frac{\partial}{\partial Z})
\end{math}
as $\bm \nabla$, we obtain the following Thiele equation,
\begin{eqnarray}
\bm G \times \left(\bm v_{\rm s} - \bm v_{\rm d}\right) +\mathcal{D}\left(\beta \bm v_{\rm s} - \alpha \bm v_{\rm d} \right)  + \bm J^{\rm spin } - \bm \nabla V=0.
\label{eq:Thiele2}
\end{eqnarray}
There are two main origins of the potential force $\bm F$$(=-\bm \nabla V)$ acting on the magnetization texture, i.e., the contribution from edges, structures and/or obstacles in the system and the contribution from pinning by impurities and/or defects. We describe these two contributions explicitly as,
\begin{eqnarray}
\bm F=\bm F_{\rm sys}+\bm F_{\rm imp} \hspace{0.5cm}
(\bm \nabla V=\bm \nabla V_{\rm sys}+\bm \nabla V_{\rm imp}).
\end{eqnarray}
The value of the component $\mathcal{G}$ depends on the magnetization texture. It takes $\mathcal{G}=4\pi Q$ for skyrmions in ferromagnets, where $Q(=\pm 1)$ is a quantity called topological charge. The sign of $Q$ is determined by the sign of core magnetization and the vorticity. Specifically, $Q=-1$ ($Q=+1$) when the core magnetization is pointing downward (upward) for the skyrmions. On the contrary, for the antiskyrmions, $Q=+1$ ($Q=-1$) when the core magnetization is pointing downward (upward). Besides, $\mathcal{G}=2\pi Q$ for magnetic vortices (called merons), and $\mathcal{G}=0$ for non-topological magnetic textures such as helimagnetic states, ferromagnetic domain walls, stripe domains, and ferromagnetic states.

The derived Thiele equation does not contain the acceleration term with an inertia mass, which is absent as far as the magnetic texture is not deformed as assumed in its derivation process. When the magnetic texture is dynamically or oscillationally deformed, an effective mass appears in its dynamics. Subsequently, the Thiele equation is generalized by phenomenologically introducing the effective mass as~\cite{Barker2016,Makhfudz2012,Tveten2013,Schutte2014b},
\begin{align}
&-\mathcal{M}\dot{\bm v}_{\rm d}+\bm G \times \left(\bm v_{\rm s} - \bm v_{\rm d}\right) 
\notag \\
&\hspace{1.0cm}
+\mathcal{D}\left(\beta \bm v_{\rm s} - \alpha \bm v_{\rm d} \right) + \bm J^{\rm spin } + \bm F =0.
\label{eq:Thiele1b}
\end{align}
However, the acceleration term does not affect the terminal velocity of the magnetic textures at steady motion with $\dot{\bm v}_{\rm d}=0$. Hence, we will discuss the current-induced motion of skyrmions using the Thiele equation without the acceleration term in the following.

\subsection{4.2 Ferromagnetic systems}
The Thiele equation can describe the motion of magnetic textures in various ferromagnetic systems. Here we explain typical two examples, i.e., the non-heterojunction system (i.e., bulk materials and single-layer samples) and the multilayer heterojunction system here by focusing on their differences.\\
\\
\noindent
\underline{\bf Ferromagnetic non-heterojunction system}\\
In the normal magnetic system, which is not a magnetic heterojunction, there is no interfacial spin-orbit coupling caused by the broken spatial inversion symmetry at the interface, and thus the vertical spin current is absent. Therefore, there is no need to consider the spin-orbit torque in this case. Instead, only the spin-transfer torque and nonadiabatic torque due to the in-plane spin-polarized electric current should be considered. Assuming $\bm J^{\rm spin}=0$ in Eq.~(\ref{eq:Thiele2}), the Thiele equation for the non-heterojunction system is given by,
\begin{eqnarray}
\bm G \times \left(\bm v_{\rm s} - \bm v_{\rm d}\right) +\mathcal{D}\left(\beta \bm v_{\rm s} - \alpha \bm v_{\rm d} \right) - \bm \nabla V=0.
\label{eq:ThieleA}
\end{eqnarray}
\\
\noindent
\underline{\bf Ferromagnetic heterojunction system}\\
In the magnetic/heavy-metal heterojunction system, the spin-orbit coupling originating from the broken spatial inversion symmetry at the interface generates a spin current flowing vertical to the interface, which exerts spin-orbit torque to the magnetization. On the contrary, effects of the spin-transfer torque and the nonadiabatic torque due to the in-plane electric current are small because the noncollinear magnetic texture is localized at a few atomic layers near the interface. In this case, setting $\bm v_{\rm s}=0$ in Eq.~(\ref{eq:Thiele2}), the Thiele equation for the magnetic heterojunction system, in which the vertical spin current governs the magnetization dynamics, is given by,
\begin{eqnarray}
\bm G \times \bm v_{\rm d} + \alpha\mathcal{D}\bm v_{\rm d} - \bm J^{\rm spin} + \bm \nabla V=0.
\label{eq:ThieleB}
\end{eqnarray}

In the following, we focus on differences between thin-plate and nanotrack systems in the current-induced motion of skyrmions in them. The thin-plate system is a (quasi) two-dimensional system whose area is sufficiently large as compared to the skyrmion size. In such a system, skyrmions can be sufficiently away from system edges, and they can move without being affected by repulsive potentials from the edges. On the other hand, the nanotrack system is a system which is long in the $x$-direction (longitudinal direction), while narrow in the $y$-direction (width direction). In such a system, skyrmions move under the influence of repulsive potentials from upper and/or lower edges along the $y$ axis. 

We also focus on differences in motion between the Bloch-type and N\'{e}el-type skyrmions and those between ferromagnets and antiferromagnets. It should also be mentioned that the current-driven magnetic texture with a finite topological charge often has a velocity component perpendicular to the current in addition to the parallel component, which results in the Hall motion (the skyrmion Hall effect). The Hall angle $\theta_{\rm H}$ which represents the magnitude of this effect is defined by,
\begin{eqnarray}
\theta_{\rm H}=\tan^{-1}\left(\frac{v_{{\rm d}y}}{v_{{\rm d}x}}\right)
\approx \frac{v_{{\rm d}y}}{v_{{\rm d}x}}.
\end{eqnarray}

\subsection{4.3 Antiferromagnetic systems}
The spatial derivatives of the magnetizations in the Thiele equation indicate that gradual variations of magnetization profile in space are assumed for this equation. It seems that the antiferromagnetic states with staggered alignment of bipartite magnetizations contradict to this assumption. However, as long as the spin-orbit torque mechanism is considered, theoretical descriptions based on the Thiele equation can be applied even to the bipartite antiferromagnetic system as argued in the following. 

Both the bulk N\'{e}el-type antiferromagnets and the synthetic antiferromagnets with antiferromagnetically coupled two ferromagnetic layers have two sublattice magnetizations $\bm m_{\rm A}$ and $\bm m_{\rm B}$, which are oriented in opposite directions to each other where the relation $\bm m_{\rm A}=-\bm m_{\rm B}$ holds. When the spin-orbit torque is exerted to the magnetizations, the current-driven dynamics of each sublattice magnetization can be described by a set of the LLGS equations,
\begin{align}
\frac{d \bm m_{\rm A}}{dt}
&=-\gamma \bm m_{\rm A} \times \bm B_{\rm A}^{\rm eff}
+\alpha \bm m_{\rm A} \times \frac{d\bm m_{\rm A}}{dt}
\nonumber \\
&+\frac{\gamma\hbar \theta_{\rm SH}j_{\rm e}}{2eM_{\rm As}}\cdot\frac{1}{d}
\left[ \bm m_{\rm A} \times (\bm \sigma \times \bm m_{\rm A}) \right],
\\
\frac{d \bm m_{\rm B}}{dt}
&=-\gamma \bm m_{\rm B} \times \bm B_{\rm B}^{\rm eff}
+\alpha \bm m_{\rm B} \times \frac{d\bm m_{\rm B}}{dt}
\nonumber \\
&+\frac{\gamma\hbar \theta_{\rm SH}j_{\rm e}}{2eM_{\rm Bs}}\cdot\frac{1}{d}
\left[ \bm m_{\rm B} \times (\bm \sigma \times \bm m_{\rm B}) \right].
\label{eq:LLGSAB}
\end{align}
This decomposition of the equation is possible because the spin-orbit torques work only locally. Similar to the ferromagnetic case, we assume rigidity of the magnetic textures during the motion. We also note that configurations of the two sublattice magnetizations move at the same drift velocity $\bm v_{\rm d}$, because they are inseparably coupled in antiferromagnets. In the spirit of Eq.~(\ref{eq:Relat01}), these aspects can be expressed by the following equations,
\begin{align}
\bm m_{\rm A}(\bm r, t)=\bm m_{\rm A}(\bm r-\bm R(t)),
\quad 
\bm m_{\rm B}(\bm r, t)=\bm m_{\rm B}(\bm r-\bm R(t)),
\label{eq:Relat01AB}
\end{align}
and
\begin{align}
\dot{\bm m}_{\rm A}=-(\bm v_{\rm d} \cdot \bm \nabla) \bm m_{\rm A},
\quad
\dot{\bm m}_{\rm B}=-(\bm v_{\rm d} \cdot \bm \nabla) \bm m_{\rm B}.
\label{eq:Relat02AB}
\end{align}
Following the procedure in the ferromagnetic case, we can derive the Thiele equation for magnetic textures in antiferromagnets as,
\begin{align}
&(\bm G_{\rm A}+\bm G_{\rm B}) \times \bm v_{\rm d} 
+\alpha(\mathcal{D}_{\rm A}+\mathcal{D}_{\rm B})\bm v_{\rm d}
-(\bm J_{\rm A}^{\rm spin}+\bm J_{\rm B}^{\rm spin}) 
\notag \\
&\hspace{0.5cm}+\bm \nabla (V_{\rm A} + V_{\rm B})=0.
\label{eq:ThieleAF1}
\end{align}
Here $\bm G_{\rm A,B}$, $\mathcal{D}_{\rm A,B}$, $\bm J_{\rm A,B}^{\rm spin}$ and $V_{\rm A,B}$ are given by replacing $\bm m$ with $\bm m_{\rm A,B}$ in Eqs.~(\ref{eq:G})-(\ref{eq:F}). Because of the relation $\bm m_{\rm A}=-\bm m_{\rm B}$, these equations indicate the following relations,
\begin{align}
&\bm G_{\rm A}=-\bm G_{\rm B},
\\
&\mathcal{D}_{\rm A}=\mathcal{D}_{\rm B}(=\mathcal{D}),
\\
&\bm J_{\rm A}^{\rm spin}=\bm J_{\rm B}^{\rm spin}(=\bm J^{\rm spin}).
\end{align}
Also the following relation holds for the sublattice saturation magnetizations in antiferromagnets,
\begin{align}
M_{\rm As}=M_{\rm Bs}.
\end{align}
As a results, we obtain the following Thiele equation for the antiferromagnetic system driven by the spin-orbit torque,
\begin{align}
\alpha\mathcal{D}\bm v_{\rm d} - \bm J^{\rm spin} + \bm \nabla V=0,
\label{eq:ThieleAF2}
\end{align}
where
\begin{align}
\bm \nabla V \equiv \bm \nabla (V_{\rm A} + V_{\rm B})/2.
\label{eq:VAB}
\end{align}
We discuss the current-induced motion of antiferromagnetic skyrmions using this equation later.

\begin{table*}[tbp]
\caption{Expressions of the drift velocity of the current-induced skyrmion motion depending on the types of skyrmion (Bloch-type and N\'{e}el-type), the types of magnets (ferromagnets and antiferromagnets), the driving mechanism (the spin-transfer torque mechanism and the spin-orbit torque mechanism), the system geometry (thin plate and nanotrack), and the direction of the electric current (length and width directions of the nanotrack). The spin-polarization orientations $\bm \sigma$ of the vertical spin current in the magnetic bilayer heterojunction system are indicated in parentheses in the first column. Here STT and SOT denote the spin-transfer torque mechanism and the spin-orbit torque mechanism, respectively, while FM and AFM denote the ferromagnetic and antiferromagnetic, respectively. The $x$ and $y$ axes are defined to be the length and width directions of the nanotrack system. In the nanotrack system, the velocity component in the width direction is always zero, i.e., $v_{{\rm d}y}=0$ in steady motion after a sufficient duration.}
\tabcolsep = 0.4cm
\renewcommand\arraystretch{3.0}
\begin{tabular}{c|c|ccccc}
\hline
Torque & & STT & SOT & SOT & SOT & SOT \\
Sk. type & & FM-Bloch/N\'{e}el & FM-Bloch & FM-N\'{e}el & AFM-Bloch & AFM-N\'{e}el \\
Helicity $\eta$ & & -- & $\pi/2$, $3\pi/2$ & 0, $\pi$ & $\pi/2$, $3\pi/2$ & 0, $\pi$\\
\hline
\hline
Thin plate & $v_{{\rm d}x}$
& $\displaystyle v_{{\rm s}x}$ 
& $\displaystyle \sigma_y \sin\eta \frac{\mathcal{J}_{\rm FM}}{\mathcal{G}}$ 
& $\displaystyle -\sigma_y \cos\eta\frac{\alpha\mathcal{D}\mathcal{J}_{\rm FM}}{\mathcal{G}^2}$ 
& 0
& $\displaystyle -\sigma_y \cos\eta\frac{\mathcal{J}_{\rm AFM}}{\alpha\mathcal{D}}$ \\
$\bm j_{\rm e}$$\parallel$$\bm x$ ($\bm \sigma$$\parallel$$\bm y$) &  $v_{{\rm d}y}$
& $\displaystyle (\alpha-\beta)\frac{\mathcal{D}}{\mathcal{G}}v_{{\rm s}x}$ 
& $\displaystyle \sigma_y \sin\eta \frac{\alpha\mathcal{D}\mathcal{J}_{\rm FM}}{\mathcal{G}^2}$ 
& $\displaystyle \sigma_y \cos\eta \frac{\mathcal{J}_{\rm FM}}{\mathcal{G}}$ 
& $\displaystyle \sigma_y \sin\eta \frac{\mathcal{J}_{\rm AFM}}{\alpha\mathcal{D}}$ 
& 0 \\
\hline
Nanotrack  & $v_{{\rm d}x}$
& $\displaystyle \frac{\beta}{\alpha}v_{{\rm s}x}$ 
& 0 
& $-\displaystyle \sigma_y \cos\eta\frac{\mathcal{J}_{\rm FM}}{\alpha\mathcal{D}}$ 
& 0 
& $-\displaystyle \sigma_y \cos\eta \frac{\mathcal{J}_{\rm AFM}}{\alpha\mathcal{D}}$  \\
$\bm j_{\rm e}$$\parallel$$\bm x$ ($\bm \sigma$$\parallel$$\bm y$) & $v_{{\rm d}y}$
& 0 & 0 & 0 & 0 & 0 \\
\hline
Nanotrack & $v_{{\rm d}x}$
& $-\displaystyle \frac{\mathcal{G}}{\alpha\mathcal{D}}v_{{\rm s}y}$ 
& $\displaystyle \sigma_x \sin\eta \frac{\mathcal{J}_{\rm FM}}{\alpha\mathcal{D}}$ 
& 0
& $\displaystyle \sigma_x \sin\eta \frac{\mathcal{J}_{\rm AFM}}{\alpha\mathcal{D}}$
& 0  \\
$\bm j_{\rm e}$$\parallel$$\bm y$ ($\bm \sigma$$\parallel$$\bm x$) & $v_{{\rm d}y}$
& 0 & 0 & 0 & 0 & 0\\
\hline
\end{tabular}
\label{tabl:MDLPRMS}
\end{table*}

\section{5. Dynamics of Ferromagnetic Skyrmions Driven by the Spin-Transfer Torque}
\subsection{5.1 General Aspects of STT-Driven Motion of FM Skyrmions}
\begin{figure*}
\includegraphics[scale=1.0]{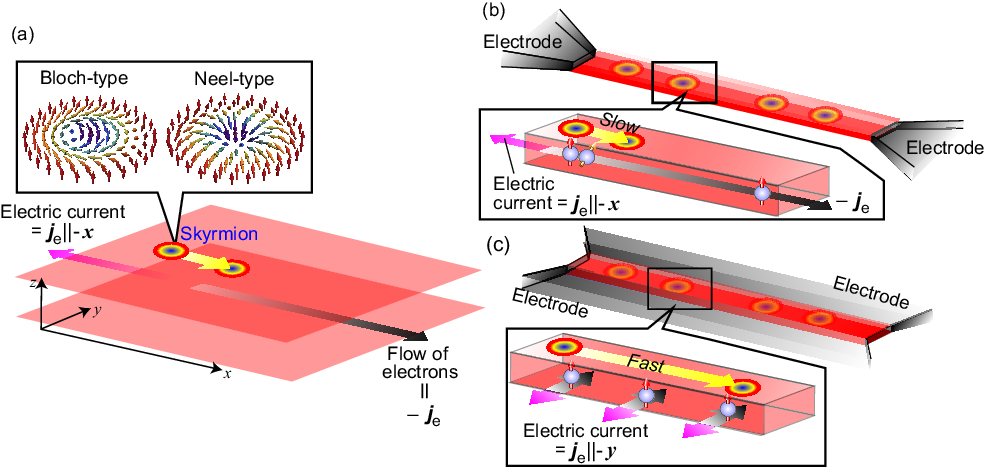}
\caption{Schematics of the skyrmion motion induced by the spin-transfer torque and the nonadiabatic torque exerted by in-plane spin-polarized electric currents in ferromagnetic thin-plate and nanotrack systems. In this case, there is no difference in direction and velocity of the motion between the Bloch-type and N\'{e}el-type skyrmions. (a) In the thin-plate system without influence of edge potentials, the skyrmion moves nearly along the electric current but also exhibits the Hall motion with a subsequent perpendicular velocity component. (b),~(c) The electric current is applied (b) in the length direction ($\bm j_{\rm e}$$\parallel$$-\bm x$) and (c) in the width direction ($\bm j_{\rm e}$$\parallel$$-\bm y$) of the nanotrack. In both cases, the skyrmion moves in the length direction. However, in the latter case, its drift velocity is two to three orders of magnitude faster than the velocity in the former case.}
\label{Fig04}
\end{figure*}
First, we consider the case that the electric current is applied to (non-heterojunction) ferromagnetic thin-plate and nanotrack systems [see Fig.~\ref{Fig04}]. In this case, the magnetic texture is driven mainly by the spin-transfer torque and the nonadiabatic torque. This situation is well described by the Thiele equation in Eq.~(\ref{eq:ThieleA}). We incorporate the following (quasi) two-dimensional conditions into Eq.~(\ref{eq:ThieleA}),
\begin{align}
\label{eq:2Dcnd1}
&\bm v_{\rm s}=(v_{{\rm s}x},v_{{\rm s}y},0),\\
\notag \\
\label{eq:2Dcnd2}
&\bm v_{\rm d}=(v_{{\rm d}x},v_{{\rm d}y},0),\\
\notag \\
\label{eq:2Dcnd3}
&\bm \nabla V=(\partial_xV,\partial_yV,0).
\end{align}
\begin{widetext}
Solving the equation for the drift velocity $\bm v_{\rm d}=(v_{{\rm d}x},v_{{\rm d}y})$, we obtain the following solution,
\begin{align}
&\left(\begin{array}{cc} v_{{\rm d}x} \\ v_{{\rm d}y}
\end{array}
\right)
=\frac{1}{\alpha^2 \mathcal{D}^2+\mathcal{G}^2}
\left(\begin{array}{cc} \left(\alpha \beta \mathcal{D}^2+\mathcal{G}^2 \right) v_{{\rm s}x}- \left(\alpha -\beta \right) \mathcal{D} \mathcal{G} v_{{\rm s}y} - \alpha \mathcal{D} \partial_xV- \mathcal{G} \partial_yV \\
\left(\alpha -\beta \right) \mathcal{D} \mathcal{G} v_{{\rm s}x}+\left(\alpha \beta \mathcal{D}^2+\mathcal{G}^2 \right) v_{{\rm s}y}+\mathcal{G} \partial_xV - \alpha \mathcal{D} \partial_yV
\end{array}
\right).
\label{eq:5-2}
\end{align}
By investigating this solution under various conditions, we discuss the current-induced motion of skyrmions driven by the spin-transfer torque mechanism. In the nanotrack system, we investigate both the cases that the electric current flows in the length direction $\bm j_{\rm e}$$\parallel$$\bm x$ [Fig.~\ref{Fig04}(b)] and the width direction $\bm j_{\rm e}$$\parallel$$\bm y$ [Fig.~\ref{Fig04}(c)]. Some of the formulas for velocities have been partially derived and proposed in literatures~\cite{Iwasaki2013a,Iwasaki2013b,Iwasaki2014a,ZhangX2017b}.
\end{widetext}

\subsection{5.2 STT-Driven Motion of FM Skyrmions in the Thin-Plate System with $\bm j_{\rm e}$$\parallel$$\bm x$}
In the thin-plate system, the skyrmion is not affected by edge potentials, so that we can set $\bm \nabla V=\bm 0$. We consider the case that the spin-polarized electric current flows in the $x$ direction, i.e., $v_{{\rm s}x} \ne 0$ and $v_{{\rm s}y}=0$ [Fig.~\ref{Fig04}(a)]. Substituting these conditions into Eq.~(\ref{eq:5-2}) and neglecting the second-order quantities with respect to $\alpha$ and $\beta$, we obtain,
\begin{eqnarray}
&v_{{\rm d}x} \approx v_{{\rm s}x},
\hspace{0.3cm}
v_{{\rm d}y} \approx (\alpha-\beta)\frac{\mathcal{D}}{\mathcal{G}}v_{{\rm s}x},\\
\label{eq:5-3}
\nonumber \\
&\tan \theta_{\rm H} \approx (\alpha-\beta)\frac{\mathcal{D}}{\mathcal{G}}.
\label{eq:5-4}
\end{eqnarray}
The expression of $v_{{\rm d}x}$ indicates that when the spin-transfer torque is a driving mechanism, skyrmions in the thin-plate system move at a high velocity, almost as fast as the normalized current velocity $\bm v_{\rm s}$. For $\alpha \ne \beta$, the motion is accompanied by a Hall motion with $v_{{\rm d}y}\ne 0$ in addition to the velocity $v_{{\rm d}x}$ parallel to the current. Furthermore, the expressions of $v_{{\rm d}y}$ and the Hall angle $\theta_{\rm H}$ indicate that the Hall motion disappears when $\alpha=\beta$. However, if there is a pinning potential due to impurities and/or defects etc., the skyrmion Hall effect shows up even at $\alpha=\beta$. 

\subsection{5.3 STT-Driven Motion of FM Skyrmions in the Nanotrack System with $\bm j_{\rm e}$$\parallel$$\bm x$}
In the nanotrack system, the skyrmion is subject from a confinement potential from the edges, which is described by the conditions $\partial_xV=0$ and $\partial_yV\ne 0$. Considering the case that the electric current $\bm j_{\rm e}$$\parallel$$-\bm x$ flows in the length direction of the nanotrack, we set $v_{{\rm s}x} \ne 0$ and $v_{{\rm s}y}=0$ [Fig.~\ref{Fig04}(b)]. Furthermore, after a sufficient time, the velocity in the $y$ direction should be zero ($v_{{\rm d}y}=0$) because of the confinement. Substituting these conditions into Eq.~(\ref{eq:5-2}), we obtain,
\begin{equation}
v_{{\rm d}x} \approx \frac{\beta}{\alpha}v_{{\rm s}x},
\hspace{0.5cm}
v_{{\rm d}y}=0.
\label{eq:5-5}
\end{equation}
These expressions indicate that in the nanotrack system with an electric current in the length direction, the velocity of the skyrmion depends strongly on the values of $\alpha$ and $\beta$. When $\alpha=\beta$, the drift velocity is the same as that for the thin-plate system, i.e., $v_{{\rm d}x}=v_{{\rm s}x}$.

\subsection{5.4 STT-Driven Motion of FM Skyrmions in the Nanotrack System with $\bm j_{\rm e}$$\parallel$$\bm y$}
Considering the case that the electric current $\bm j_{\rm e}$$\parallel$$\bm y$ flows in the width direction of the nanotrack, we set $v_{{\rm s}x}=0$ and $v_{{\rm s}y} \ne 0$ [Fig.~\ref{Fig04}(c)]. Again, we consider that the velocity in the $y$ direction should be zero ($v_{{\rm d}y}=0$) after a sufficient time. Substituting these conditions as well as $\partial_xV=0$ and $\partial_yV\ne 0$ into Eq.~(\ref{eq:5-2}), we obtain,
\begin{eqnarray}
v_{{\rm d}x} \approx -\frac{\mathcal{G}}{\alpha \mathcal{D}} v_{{\rm s}y},
\hspace{0.5cm}
v_{{\rm d}y}=0.
\label{eq:5-6}
\end{eqnarray}
In this case, the skyrmion moves in the length direction along the nanotrack edge. Note that $|\mathcal{G}/\mathcal{D}|$ has a magnitude of nearly unity because $\mathcal{G}=-4\pi$, while the Gilbert-damping coefficient $\alpha$(=$10^{-3}$-$10^{-2}$) is sufficiently smaller than unity. Therefore, the drift velocity $\displaystyle{v_{{\rm d}x}(\approx -\frac{\mathcal{G}}{\alpha \mathcal{D}} v_{{\rm s}y}})$ is two to three orders of magnitude faster than the drift velocity $\displaystyle{v_{{\rm d}x}=v_{{\rm s}x}}$ in the thin-plate system and the drift velocity $\displaystyle{v_{{\rm d}x} \approx \frac{\beta}{\alpha}v_{{\rm s}x}}$ in the nanotrack system with $\bm j_{\rm e}$$\parallel$$\bm x$. In this case, the electric current pushes the skyrmion to the system edge to realize this quick motion. If the roughness of the system edge is high or the electric current is too strong, the skyrmion is absorbed by the system edge and eventually disappears.\\

In the meanwhile, the results for the nanotrack system show $v_{{\rm d}y}=0$ irrespective of the current direction (length or width direction), but this does not mean that the skyrmion Hall effect is absent in the nanotrack system. If skyrmions in a nanotrack do not feel repulsive potentials from edges at the beginning of their motion, they should show Hall motion at the same Hall angle as in the case of the thin-plate system. As a result, the skyrmions approach a system edge and, eventually, feel a repulsive edge potential. At this point, the skyrmions can no longer generate any further Hall motion, and the velocity in the width direction becomes zero, i.e., $v_{{\rm d}y}=0$. The above obtained results should be regarded as a drift velocity $\bm v_{\rm d}$ of skyrmions in the steady motion after certain duration.

\section{6. Dynamics of Ferromagnetic Skyrmions Driven by the Spin-Orbit Torque}
\subsection{6.1 General Aspects of SOT-Driven Motion of FM Skyrmions}
\begin{figure*}
\includegraphics[scale=1.0]{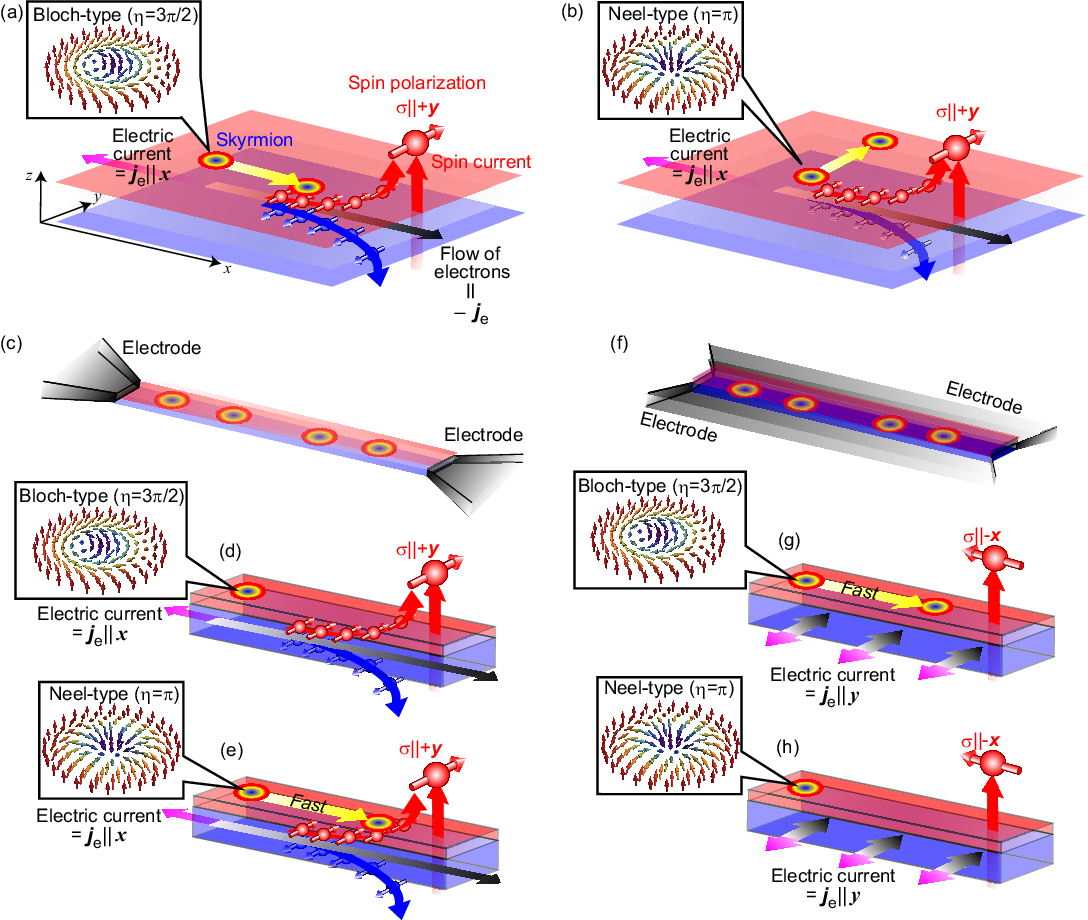}
\caption{Schematics of the skyrmion motion induced by the spin-orbit torque exerted by a vertical spin current in the ferromagnetic thin-plate and nanotrack systems. There are apparent differences in direction and velocity of the motion between the Bloch-type and N\'{e}el-type skyrmions. 
(a),~(b) In the thin-plate system without influence of edge potentials, the Bloch-type skyrmion moves nearly parallel to the current direction (perpendicular to the spin polarization) as shown in (a), whereas the N\'{e}el-type skyrmion moves nearly perpendicular to the current direction (parallel to the spin polarization) as shown in (b). (c)-(e) In the nanotrack system with an electric current in the length direction ($\bm j_{\rm e}$$\parallel$$\bm x$), the vertical spin current is spin-polarized in the $y$ direction ($\bm \sigma$$\parallel$$\bm y$). In this case, the Bloch-type skyrmion cannot move as shown in (d), whereas the N\'{e}el-type skyrmion moves quickly in the length direction as shown in (e). (f)-(h) In the nanotrack system with an electric current in the width direction ($\bm j_{\rm e}$$\parallel$$\bm y$), the vertical spin current is spin-polarized in the $x$ direction ($\bm \sigma$$\parallel$$\bm x$). In this case, the Bloch-type skyrmion moves quickly in the length direction as shown in (g), whereas the N\'{e}el-type skyrmion cannot move as shown in (h).  Note that the relationship between the signs of $\bm j_{\rm e}$ and $\bm \sigma$ should be reversed upon the sign reversal of the spin Hall angle $\theta_{SH}$ or the sign reversal of the interfacial spin-orbit coupling.}
\label{Fig05}
\end{figure*}
Next, we consider the case that the electric current is applied to thin-plate and nanotrack ferromagnet/heavy-metal heterojunction systems [see Fig.~\ref{Fig05}]. In this case, the interfacial spin-orbit coupling generates a vertical spin current via the spin Hall effect. The generated spin current exerts the spin-orbit torque to the magnetization, which drives the magnetic texture. This situation is well described by the Thiele equation in Eq.~(\ref{eq:ThieleB}). We incorporate the (quasi)two-dimensional conditions Eqs.~(\ref{eq:2Dcnd1})-(\ref{eq:2Dcnd3}) into Eq.~(\ref{eq:ThieleB}). Solving the equation for the drift velocity $\bm v_{\rm d}=(v_{{\rm d}x},v_{{\rm d}y})$, we obtain the following solution,
\begin{align}
\left(\begin{array}{cc} v_{{\rm d}x} \\ v_{{\rm d}y}
\end{array}
\right)=
\frac{1}{\alpha^2 \mathcal{D}^2+\mathcal{G}^2}
\left(\begin{array}{cc} -\alpha \mathcal{D} & -\mathcal{G} \\ \mathcal{G} & -\alpha \mathcal{D}
\end{array}
\right)
\left(\begin{array} {cc} \partial_xV-J^{\rm spin}_x \\ \partial_yV-J^{\rm spin}_y
\end{array}
\right).
\nonumber \\
\label{eq:6-3}
\end{align}
By investigating this solution under various conditions, we discuss the current-induced motion of skyrmions driven by the spin-orbit torque mechanism. We investigate the motion particularly focusing on differences between the Bloch-type and N\'{e}el-type skyrmions. In the nanotrack system, we investigate both the cases that the electric current flows in the length direction $\bm j_{\rm e}$$\parallel$$\bm x$ [Figs.~\ref{Fig05}(c)-(e)] and the width direction $\bm j_{\rm e}$$\parallel$$\bm y$ [Figs.~\ref{Fig05}(f)-(h)]. The difference in the current direction gives difference in the spin-polarization of the vertical spin current. Specifically, when the electric current flows in the $\bm x$ direction, i.e., $\bm j_{\rm e}$$\parallel$$\bm x$, the spin current is spin-polarized in the $\bm y$ direction, i.e., $\bm \sigma$$\parallel$$\bm y$. On the other hand, the electric current flows in the $\bm y$ direction, i.e., $\bm j_{\rm e}$$\parallel$$\bm y$, it is spin-polarized in the $\bm x$ direction, i.e., $\bm \sigma$$\parallel$$\bm x$. The relationship between the signs of $\bm j_{\rm e}$ and $\bm \sigma$ should be reversed upon the sign of the spin Hall angle $\theta_{\rm H}$ or the sign of the interfacial spin-orbit coupling is reversed. Some of the formulas of velocities discussed in the following have been in part derived and proposed in literature~\cite{XingX2020,LuoJ2023}.

\vspace{0.5cm}
When the spin polarization is $\bm \sigma$$\parallel$$\bm y$, the vector quantities $\bm J^{\rm spin}$ in Eq.~(\ref{eq:J}) can be written, for respective types of skyrmions, as follows.\\

\noindent
-- Bloch-type ferromagnetic skyrmions ($\eta$=$\pi/2$, $3\pi/2$)
\begin{eqnarray}
\left(\begin{array}{cc} J^{\rm spin}_x \\ J^{\rm spin}_y
\end{array}
\right)=
\left(\begin{array}{cc} 0 \\ \sigma_y \sin\eta \mathcal{J}_{\rm FM}
\end{array}
\right),
\label{eq:6-4}
\end{eqnarray}

\noindent
-- N\'{e}el-type ferromagnetic skyrmions ($\eta$=0, $\pi$)
\begin{eqnarray}
\left(\begin{array}{cc} J^{\rm spin}_x \\ J^{\rm spin}_y
\end{array}
\right)=
\left(\begin{array}{cc} -\sigma_y \cos\eta \mathcal{J}_{\rm FM} \\ 0
\end{array}
\right).
\label{eq:6-5}
\end{eqnarray}

On the other hand, when the spin polarization is $\bm \sigma$$\parallel$$\bm x$, they can be written, for respective types of skyrmions, as follows.\\

\noindent
-- Bloch-type ferromagnetic skyrmions ($\eta$=$\pi/2$, $3\pi/2$)
\begin{eqnarray}
\left(\begin{array}{cc} J^{\rm spin}_x \\ J^{\rm spin}_y
\end{array}
\right)=
\left(\begin{array}{cc} \sigma_x \sin\eta \mathcal{J}_{\rm FM} \\ 0
\end{array}
\right),
\label{eq:6-6}
\end{eqnarray}

\noindent
-- N\'{e}el-type ferromagnetic skyrmions ($\eta$=0, $\pi$)
\begin{eqnarray}
\left(\begin{array}{cc} J^{\rm spin}_x \\ J^{\rm spin}_y
\end{array}
\right)=
\left(\begin{array}{cc} 0 \\ \sigma_x \cos\eta \mathcal{J}_{\rm FM}
\end{array}
\right).
\label{eq:6-7}
\end{eqnarray}

\noindent
Here $\bm J^{\rm spin}_\mu$ is defined by Eq.~(\ref{eq:J}), and the constant $\mathcal{J}_{\rm FM}(>0)$ can be evaluated by performing the spatial integration in this equation (see appendix). 

\subsection{6.2 SOT-Driven Motion of FM Skyrmions in the Thin-Plate System with $\bm j_{\rm e}$$\parallel$$\bm x$ and $\bm \sigma$$\parallel$$\bm y$}
In the thin-plate heterojunction system, the skyrmion is not affected by the edge potentials, so that we can set $\bm \nabla V=\bm 0$. Substituting this condition into Eq.~(\ref{eq:6-3}), we obtain,
\begin{eqnarray}
\left(\begin{array}{cc} v_{{\rm d}x} \\ v_{{\rm d}y}
\end{array}
\right)
&=&
\frac{1}{\alpha^2 \mathcal{D}^2+\mathcal{G}^2}
\left(\begin{array}{cc} \alpha \mathcal{D} J^{\rm spin}_x+\mathcal{G} J^{\rm spin}_y \\
-\mathcal{G} J^{\rm spin}_x+\alpha \mathcal{D} J^{\rm spin}_y
\end{array}
\right).
\nonumber \\
\label{eq:6-8}
\end{eqnarray}
We consider the case that the spin polarization of the vertical spin current is $\bm \sigma$$\parallel$$\bm y$ [Figs.~\ref{Fig05}(a) and (b)]. Substituting Eqs.~(\ref{eq:6-4}) and (\ref{eq:6-5}) into Eq.~(\ref{eq:6-8}) and neglecting the second-order quantities with respect to $\alpha$ and $\beta$, we obtain the following equations for the Bloch-type and N\'{e}el-type ferromagnetic skyrmions, respectively,\\

\noindent
-- Bloch-type ferromagnetic skyrmions ($\eta$=$\pi/2$, $3\pi/2$)
\begin{align}
&v_{{\rm d}x} \approx \sigma_y \sin\eta \frac{\mathcal{J}_{\rm FM}}{\mathcal{G}},
\\
&v_{{\rm d}y} \approx \sigma_y \sin\eta \frac{\alpha\mathcal{D}\mathcal{J}_{\rm FM}}{\mathcal{G}^2}.
\label{eq:6-9}
\end{align}

\noindent
-- N\'{e}el-type ferromagnetic skyrmions ($\eta$=0, $\pi$)
\begin{align}
&v_{{\rm d}x} \approx -\sigma_y \cos\eta \frac{\alpha\mathcal{D}\mathcal{J}_{\rm FM}}{\mathcal{G}^2},
\\
&v_{{\rm d}y} \approx \sigma_y \cos\eta \frac{\mathcal{J}_{\rm FM}}{\mathcal{G}}.
\label{eq:6-10}
\end{align}

\noindent
The signs of $v_{{\rm d}x}$ and $v_{{\rm d}y}$ can be known from $\mathcal{J}_{\rm FM}>0$, $\mathcal{D}>0$, $\alpha>0$ and $\mathcal{G}=-4\pi<0$ (when $\bm B$$\parallel$$+\bm z$). If the magnetic field is reversed to be $\bm B$$\parallel$$-\bm z$, the sign of $v_{{\rm d}x}$ is reversed for the Bloch-type skyrmions, while the sign of $v_{{\rm d}y}$ is reversed for the N\'{e}el-type skyrmions because the sign of $\mathcal{G}$ is reversed to be positive, i.e., $\mathcal{G}=+4\pi>0$. Furthermore, if the direction of the current is reversed, the spin polarization $\bm \sigma$ is reversed, so the signs of $v_{{\rm d}x}$ and $v_{{\rm d}y}$ should be reversed.\\

\noindent
These expressions indicate the following aspects when skyrmions are driven by the spin-orbit torque exerted by a vertical spin current in the thin-plate ferromagnet/heavy-metal heterojunction system:
\begin{itemize}
\item Bloch-type skyrmions move nearly perpendicular to the spin polarization $\bm \sigma$ of the vertical spin current.
\item N\'{e}el-type skyrmions move nearly parallel to the spin polarization $\bm \sigma$ of the vertical spin current.
\item The drift velocities of Bloch-type and N\'{e}el-type skyrmions are identical.
\end{itemize}

\subsection{6.3 SOT-Driven Motion of FM Skyrmions in the Nanotrack System with $\bm j_{\rm e}$$\parallel$$\bm x$ and $\bm \sigma$$\parallel$$\bm y$}
In the narrow nanotrack system, the potential in the system satisfies the conditions $\partial_xV=0$ and $\partial_yV\ne 0$. Substituting these conditions into Eq.~(\ref{eq:6-3}), we obtain,
\begin{eqnarray}
\left(\begin{array}{cc} v_{{\rm d}x} \\ v_{{\rm d}y}
\end{array}
\right)=
\frac{1}{\alpha^2 \mathcal{D}^2+\mathcal{G}^2}
\left(\begin{array}{cc} -\alpha \mathcal{D} & -\mathcal{G} \\ \mathcal{G} & -\alpha \mathcal{D}
\end{array}
\right)
\left(\begin{array}{cc} -J^{\rm spin}_x \\ -J^{\rm spin}_y+\partial_yV
\end{array}
\right).
\nonumber \\
\label{eq:6-11}
\end{eqnarray}
When the spin polarization of the vertical spin current is $\bm \sigma$$\parallel$$\bm y$, substituting Eqs.~(\ref{eq:6-4}) and (\ref{eq:6-5}) into Eq.~(\ref{eq:6-11}), we obtain the following equations for the Bloch-type and N\'{e}el-type ferromagnetic skyrmions, respectively,\\

\noindent
-- Bloch-type ferromagnetic skyrmions ($\eta$=$\pi/2$, $3\pi/2$)
\begin{eqnarray}
\left(\begin{array}{cc} v_{{\rm d}x} \\ v_{{\rm d}y}
\end{array}
\right)
&=&
-\frac{1}{\alpha^2 \mathcal{D}^2+\mathcal{G}^2}
\left(\begin{array}{cc} \mathcal{G}\left(-\sigma_y \sin\eta \mathcal{J}_{\rm FM}+\partial_yV \right) \\
\alpha \mathcal{D} \left(-\sigma_y \sin\eta \mathcal{J}_{\rm FM}+\partial_yV \right)
\end{array}
\right).
\nonumber \\
\label{eq:6-12}
\end{eqnarray}
After sufficient time has passed, the velocity in the $y$-direction should become zero, i.e., $v_{{\rm d}y}=0$. Then, the lower equation of Eq.~(\ref{eq:6-12}) indicates that the following condition holds,
\begin{eqnarray}
\partial_yV=\sigma_y \sin\eta \mathcal{J}_{\rm FM}.
\label{eq:6-13}
\end{eqnarray}
Substituting this condition into the upper equation of Eq.~(\ref{eq:6-12}), we obtain,
\begin{eqnarray}
v_{{\rm d}x}=0.
\label{eq:6-14}
\end{eqnarray}
This result indicates that Bloch-type ferromagnetic skyrmions in the nanotrack heterojunction system cannot be driven the spin-orbit torque exerted from a vertical spin current with $\bm \sigma$$\parallel$$\bm y$.\\

\noindent
-- N\'{e}el-type ferromagnetic skyrmions ($\eta$=0, $\pi$)
\begin{eqnarray}
\left(\begin{array}{cc} v_{{\rm d}x} \\ v_{{\rm d}y}
\end{array}
\right)
&=&
\frac{1}{\alpha^2 \mathcal{D}^2+\mathcal{G}^2}
\left(\begin{array}{cc} -\sigma_y \cos\eta \alpha \mathcal{D}\mathcal{J}_{\rm FM} 
-\mathcal{G} \partial_yV \\
\sigma_y \cos\eta \mathcal{G} \mathcal{J}_{\rm FM} -\alpha \mathcal{D} \partial_yV 
\end{array}
\right).
\nonumber \\
\label{eq:6-15}
\end{eqnarray}
After sufficient time has passed, the following relation holds because $v_{{\rm d}y}=0$ in the lower equation of Eq.~(\ref{eq:6-15}),
\begin{eqnarray}
\partial_yV=\sigma_y \cos\eta \frac{\mathcal{G}\mathcal{J}_{\rm FM}}{\alpha \mathcal{D}}.
\label{eq:6-16}
\end{eqnarray}
Substituting this condition into the upper equation of Eq.~(\ref{eq:6-15}), we obtain,
\begin{align}
v_{{\rm d}x}
&=-\frac{\sigma_y \cos\eta \mathcal{J}_{\rm FM}}{\alpha^2 \mathcal{D}^2+\mathcal{G}^2}
\left(\alpha \mathcal{D}+\frac{\mathcal{G}^2}{\alpha \mathcal{D}}
\right)
\notag \\
&=-\sigma_y \cos\eta \frac{\mathcal{J}_{\rm FM}}{\alpha \mathcal{D}}.
\label{eq:6-17}
\end{align}
This expression indicates that N\'{e}el-type ferromagnetic skyrmions in the nanotrack heterojunction system move quickly along the nanotrack edge in the length direction when they are driven by the spin-orbit torque exerted from a vertical spin current with $\bm \sigma$$\parallel$$\bm y$.

\subsection{6.4 SOT-Driven Motion of FM Skyrmions in the Nanotrack System with $\bm j_{\rm e}$$\parallel$$\bm y$ and $\bm \sigma$$\parallel$$\bm x$}
When the spin polarization of the vertical spin current is $\bm \sigma$$\parallel$$\bm x$, substituting Eqs.~(\ref{eq:6-6}) and (\ref{eq:6-7}) into Eq.~(\ref{eq:6-11}), we obtain the following equations for the Bloch-type and N\'{e}el-type ferromagnetic skyrmions, respectively.\\

\noindent
-- Bloch-type ferromagnetic skyrmions ($\eta$=$\pi/2$, $3\pi/2$)
\begin{eqnarray}
\left(\begin{array}{cc} v_{{\rm d}x} \\ v_{{\rm d}y}
\end{array}
\right)
&=&
\frac{1}{\alpha^2 \mathcal{D}^2+\mathcal{G}^2}
\left(\begin{array}{cc} \sigma_x \sin\eta \alpha \mathcal{D} \mathcal{J}_{\rm FM} -\mathcal{G} \partial_yV \\
-\sigma_x \sin\eta \mathcal{G} \mathcal{J}_{\rm FM} -\alpha \mathcal{D} \partial_yV 
\end{array}
\right).
\nonumber \\
\label{eq:6-18}
\end{eqnarray}
After sufficient time has passed, the following relation holds because $v_{{\rm d}y}=0$ in the lower equation of Eq.~(\ref{eq:6-18}),
\begin{eqnarray}
\partial_yV=-\sigma_x \sin\eta \frac{\mathcal{G}\mathcal{J}_{\rm FM}}{\alpha \mathcal{D}}.
\label{eq:6-19}
\end{eqnarray}
Substituting this condition into the upper equation of Eq.~(\ref{eq:6-18}), we obtain,
\begin{align}
v_{{\rm d}x}
&=\frac{\sigma_x \sin\eta \mathcal{J}_{\rm FM}}{\alpha^2 \mathcal{D}^2+\mathcal{G}^2}
\left(\alpha \mathcal{D}+\frac{\mathcal{G}^2}{\alpha \mathcal{D}}
\right)
\notag \\
&=\sigma_x \sin\eta \frac{\mathcal{J}_{\rm FM}}{\alpha \mathcal{D}}.
\label{eq:6-20}
\end{align}
This expression indicates that Bloch-type ferromagnetic skyrmions in the nanotrack system move quickly along the nanotrack edge in the length direction when they are driven by the spin-orbit torque exerted from a vertical spin current with $\bm \sigma$$\parallel$$\bm x$.\\

\noindent
-- N\'{e}el-type ferromagnetic skyrmions ($\eta$=0, $\pi$)
\begin{eqnarray}
\left(\begin{array}{cc} v_{{\rm d}x} \\ v_{{\rm d}y}
\end{array}
\right)
&=&\frac{1}{\alpha^2 \mathcal{D}^2+\mathcal{G}^2}
\left(\begin{array}{cc} \mathcal{G} \left(\sigma_x \cos\eta \mathcal{J}_{\rm FM}-\partial_yV \right) \\
\alpha \mathcal{D} \left(\sigma_x \cos\eta \mathcal{J}_{\rm FM}-\partial_yV \right)
\end{array}
\right).
\nonumber \\
\label{eq:6-21}
\end{eqnarray}
After sufficient time has passed, the following relation holds because $v_{{\rm d}y}=0$ in the lower equation of Eq.~(\ref{eq:6-21}),
\begin{eqnarray}
\partial_yV=\sigma_x \cos\eta \mathcal{J}_{\rm FM}.
\label{eq:6-22}
\end{eqnarray}
Substituting this condition into the upper equation of Eq.~(\ref{eq:6-21}), we obtain,
\begin{eqnarray}
v_{{\rm d}x}=0.
\label{eq:6-23}
\end{eqnarray}
This result indicates that N\'{e}el-type ferromagnetic skyrmions in the nanotrack system cannot be driven by the spin-orbit torque when $\bm \sigma$$\parallel$$\bm x$.

\section{7. Dynamics of Antiferromagnetic Skyrmions Driven by the Spin-Orbit Torque}
\subsection{7.1 General Aspects of SOT-Driven Motion of AFM Skyrmions}

\begin{figure*}
\includegraphics[scale=0.5]{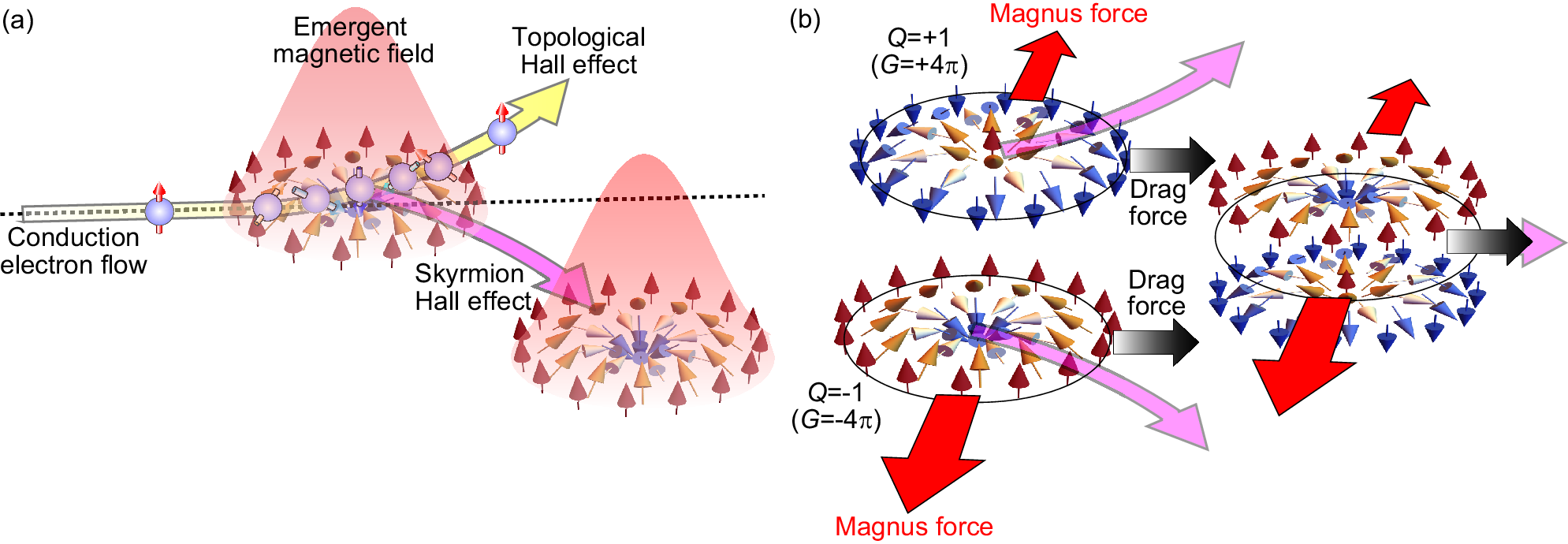}
\caption{(a) Schematics of the current-induced motion of ferromagnetic skyrmion driven by the spin-transfer torque mechanism. Conduction electrons of the electric current injected into the ferromagnet becomes spin-polarized parallel to the background ferromagnetic magnetization through exchange coupling to them. Spin angular momenta of the conduction electrons are transferred to the noncollinear skyrmion magnetizations which drives the skyrmion dominantly towards the current direction. The flow of the conduction electrons is deflected by the emergent magnetic field generated by the noncollinear skyrmion magnetizations characterized by a quantized topological charge, which results in the Hall motion of the conduction electrons (topological Hall effect). The reaction force of this topological Hall effect acts on the skyrmion, which causes subsequent transverse motion of skyrmion perpendicular to the injected electric current (skyrmion Hall effect). (b) Vanishing of the skyrmion Hall effect for antiferromagnetic skyrmions. An antiferromagnetic skyrmion can be regarded as a pair of ferromagnetic skyrmions with opposite topological charges. Consequently, opposite contributions of driving forces to the transverse motion from these two skyrmions cancel each other, resulting in the absence of skyrmion Hall effect.}
\label{Fig06}
\end{figure*}
Finally, we discuss the current-induced motion of Bloch-type and N\'{e}el-type antiferromagnetic skyrmions using the Thiele equation when they are driven by the spin-orbit torque exerted by a vertical spin current. There are mainly two different systems that host antiferromagnetic skyrmions. One is the bipartite antiferromagnet system, in which the skyrmion structures are formed by rotating antiparallel spin pairs. Another system is the synthetic bilayer antiferromagnetic system which is composed of antiferromagnetically coupled two ferromagnetic layers. For both cases, the antiferromagnetic skyrmion can be regarded as a combination of two ferromagnetic skyrmions with opposite topological charges $Q=-1$ and $Q=+1$. 

One of the important properties of antiferromagnetic skyrmions is the absence of skyrmion Hall effect. In the case of ferromagnetic skyrmions, their current-induced motion is inevitably accompanied by subsequent transverse motion perpendicular to the current in addition to dominant longitudinal motion along the current. This is caused by the reaction force of the topological Hall effect, for which the flow of conduction electrons are deflected by emergent magnetic field generated by the topological spin configuration of skyrmions [Fig.~\ref{Fig06}(a)]. On the contrary, the opposite topological charges of two ferromagnetic skyrmions constituting the antiferromagnetic skyrmion cancel each other, and, therefore, the antiferromagnetic skyrmions have zero topological charge. As a result, the antiferromagnetic skyrmions do not exhibit the topological Hall effect, and thus the skyrmion Hall effect is absent [Fig.~\ref{Fig06}(b)].

Because antiferromagnetic skyrmions have a zero topological charge $Q$=0 and, hence, a zero gyromagnetic vector with $\mathcal{G}=0$, Eq.~(\ref{eq:ThieleB}) is reduced to be,
\begin{eqnarray}
\alpha \mathcal{D} \bm v_{\rm d}+\bm \nabla V - \bm J^{\rm spin}=\bm 0.
\label{eq:7-1}
\end{eqnarray}
This equation is the Thiele equation for antiferromagnetic skyrmions driven by the spin-orbit torque. According to the discussion in Subsection 4.3, the quantities $\mathcal{D}$ and $\bm J^{\rm spin}$ are defined using the sublattice magnetizations $\bm m_{\rm A}$ or $\bm m_{\rm B}$ (see also appendix). Solving this equation for $\bm v_{\rm d}$, we obtain,
\begin{eqnarray}
\left(\begin{array}{cc} v_{{\rm d}x} \\ v_{{\rm d}y}
\end{array}
\right)=
\frac{1}{\alpha \mathcal{D}}
\left(\begin{array}{cc}
J^{\rm spin}_x-\partial_xV \\ J^{\rm spin}_y-\partial_yV 
\end{array}
\right).
\label{eq:7-2}
\end{eqnarray}
By investigating this solution under various conditions, we discuss the motion of antiferromagnetic skyrmions driven by the spin-orbit torque. Some of the formulas of velocities discussed in the following have been in part derived and proposed in literature~\cite{ZhangX2016b,ShenL2019a,ShenL2019b}.

\vspace{0.5cm}
When the electric current is applied in the $x$ direction ($\bm j_{\rm e}$$\parallel$$\bm x$), the vertical spin current is spin-polarized in the $y$ direction ($\bm \sigma$$\parallel$$\bm y$). In this case, the vector quantities $\bm J^{\rm spin}$ in Eq.~(\ref{eq:J}) can be written for the Bloch-type and N\'{e}el-type antiferromagnetic skyrmions as follows. \\

\noindent
-- Bloch-type antiferromagnetic skyrmions ($\eta$=$\pi/2$, $3\pi/2$)
\begin{eqnarray}
\left(\begin{array}{cc} J^{\rm spin}_x \\ J^{\rm spin}_y
\end{array}
\right)=
\left(\begin{array}{cc} 0 \\ \sigma_y \sin\eta \mathcal{J}_{\rm AFM}
\end{array}
\right),
\label{eq:7-3}
\end{eqnarray}

\noindent
-- N\'{e}el-type antiferromagnetic skyrmions ($\eta$=0, $\pi$)
\begin{eqnarray}
\left(\begin{array}{cc} J^{\rm spin}_x \\ J^{\rm spin}_y
\end{array}
\right)=
\left(\begin{array}{cc} -\sigma_y \cos\eta \mathcal{J}_{\rm AFM} \\ 0
\end{array}
\right).
\label{eq:7-4}
\end{eqnarray}
Here the helicity $\eta$ for the antiferromagnetic skyrmion is defined as that of its sublattice ferromagnetic skyrmion whose core magnetization points downwards where $\zeta_{\rm m}=-1$.

On the other hand, when the spin polarization is $\bm \sigma$$\parallel$$\bm x$, they can be written, for respective types of skyrmions, as follows.\\

\noindent
-- Bloch-type antiferromagnetic skyrmions ($\eta$=$\pi/2$, $3\pi/2$)
\begin{eqnarray}
\left(\begin{array}{cc} J^{\rm spin}_x \\ J^{\rm spin}_y
\end{array}
\right)=
\left(\begin{array}{cc} \sigma_x \sin\eta \mathcal{J}_{\rm AFM} \\ 0
\end{array}
\right),
\label{eq:7-5}
\end{eqnarray}

\noindent
-- N\'{e}el-type antiferromagnetic skyrmions ($\eta$=0, $\pi$)
\begin{eqnarray}
\left(\begin{array}{cc} J^{\rm spin}_x \\ J^{\rm spin}_y
\end{array}
\right)=
\left(\begin{array}{cc} 0 \\ \sigma_x \cos\eta \mathcal{J}_{\rm AFM}
\end{array}
\right).
\label{eq:7-6}
\end{eqnarray}
Again $\eta$ is the helicity of the sublattice ferromagnetic skyrmion whose core magnetization points downwards where $\zeta_{\rm m}=-1$. The constant $\mathcal{J}_{\rm AFM}(>0)$ is calculated by performing the spatial integration in Eq.~(\ref{eq:J}) (see appendix). 

\subsection{7.2 SOT-Driven Motion of AFM Skyrmions in the Thin-Plate System with $\bm j_{\rm e}$$\parallel$$\bm x$ and $\bm \sigma$$\parallel$$\bm y$}
In the thin-plate system, the skyrmion can move without any influence of edge potentials. Hence, we can set $\bm \nabla V=\bm 0$. Substituting this condition into Eq.~(\ref{eq:7-2}), we obtain,
\begin{eqnarray}
\left(\begin{array}{cc} v_{{\rm d}x} \\ v_{{\rm d}y}
\end{array}
\right)=
\frac{1}{\alpha \mathcal{D}}
\left(\begin{array}{cc}
J^{\rm spin}_x \\ J^{\rm spin}_y 
\end{array}
\right).
\label{eq:7-7}
\end{eqnarray}
We consider the case that the electric current is applied in the $x$ direction as $\bm j_{\rm e}$$\parallel$$\bm x$ and the spin polarization of the vertical spin current is $\bm \sigma$$\parallel$$\bm y$. Substituting Eqs.~(\ref{eq:7-3}) and (\ref{eq:7-4}) into Eq.~(\ref{eq:7-5}), we obtain the following equations for the Bloch-type and N\'{e}el-type antiferromagnetic skyrmions, respectively.\\

\noindent
-- Bloch-type antiferromagnetic skyrmions ($\eta$=$\pi/2$, $3\pi/2$)
\begin{eqnarray}
v_{{\rm d}x}=0, \hspace{0.5cm}
v_{{\rm d}y}=\sigma_y \sin\eta \frac{\mathcal{J}_{\rm AFM}}{\alpha \mathcal{D}}
\label{eq:7-8}
\end{eqnarray}

\noindent
-- N\'{e}el-type antiferromagnetic skyrmions ($\eta$=0, $\pi$)
\begin{eqnarray}
v_{{\rm d}x}=-\sigma_y \cos\eta \frac{\mathcal{J}_{\rm AFM}}{\alpha \mathcal{D}},
\hspace{0.5cm} v_{{\rm d}y}=0
\label{eq:7-9}
\end{eqnarray}

\noindent
These results indicate the following aspects when antiferromagnetic skyrmions are driven by the spin-orbit torque in the thin-plate system:
\begin{itemize}
\item Bloch-type antiferromagnetic skyrmions move parallel to the spin polarization $\bm \sigma$ of the vertical spin current.
\item N\'{e}el-type antiferromagnetic skyrmions move perpendicular to the spin polarization $\bm \sigma$ of the vertical spin current.
\item Both Bloch-type and N\'{e}el-type antiferromagnetic skyrmions do not show the skyrmion Hall effect.
\item The driving velocity of antiferromagnetic skyrmion is inversely proportional to the Gilbert-damping coefficient $\alpha$ irrespective of its type (Bloch-type or N\'{e}el-type).
\end{itemize}

\subsection{7.3 SOT-Driven Motion of AFM Skyrmions in the Nanotrack System with $\bm j_{\rm e}$$\parallel$$\bm x$ and $\bm \sigma$$\parallel$$\bm y$}
\noindent
In the nanotrack system, substituting the conditions $\partial_xV=0$ and $\partial_yV\ne 0$ into Eq.~(\ref{eq:7-2}), we obtain,
\begin{eqnarray}
\left(\begin{array}{cc} v_{{\rm d}x} \\ v_{{\rm d}y}
\end{array}
\right)=\frac{1}{\alpha \mathcal{D}}
\left(\begin{array}{cc} J^{\rm spin}_x \\ J^{\rm spin}_y-\partial_yV
\end{array}
\right).
\label{eq:7-10}
\end{eqnarray}
When the electric current is applied in the $x$ direction ($\bm j_{\rm e}$$\parallel$$\bm x$) and, consequently, the vertical spin current is spin-polarized in the $y$ direction ($\bm \sigma$$\parallel$$\bm y$), substituting Eqs.~(\ref{eq:7-3}) and (\ref{eq:7-4}) into Eq.~(\ref{eq:7-10}), we obtain the following equations for the Bloch-type and N\'{e}el-type antiferromagnetic skyrmions, respectively.\\

\noindent
-- Bloch-type antiferromagnetic skyrmions ($\eta$=$\pi/2$, $3\pi/2$)
\begin{eqnarray}
\left(\begin{array}{cc} v_{{\rm d}x} \\ v_{{\rm d}y}
\end{array}
\right)=\frac{1}{\alpha \mathcal{D}}
\left(\begin{array}{cc} 0 \\ \sigma_y \sin\eta \mathcal{J}_{\rm AFM}-\partial_yV
\end{array}
\right).
\label{eq:7-11}
\end{eqnarray}
Because $v_{{\rm d}y}=0$ after sufficient time has passed, we obtain,
\begin{eqnarray}
v_{{\rm d}x}=0, \hspace{0.5cm} v_{{\rm d}y}=0.
\label{eq:7-12}
\end{eqnarray}
This result indicates that Bloch-type antiferromagnetic skyrmions in the nanotrack system cannot move when they are driven by the spin-orbit torque with electric current injected in the length direction, i.e., $\bm j_{\rm e}$$\parallel$$\bm x$, which generates a vertical spin current with the spin polarization $\bm \sigma$$\parallel$$\bm y$.\\

\noindent
-- N\'{e}el-type antiferromagnetic skyrmions ($\eta$=0, $\pi$)
\begin{eqnarray}
\left(\begin{array}{cc} v_{{\rm d}x} \\ v_{{\rm d}y}
\end{array}
\right)=\frac{1}{\alpha \mathcal{D}}
\left(\begin{array}{cc} -\sigma_y \cos\eta \mathcal{J}_{\rm AFM} \\ -\partial_yV
\end{array}
\right).
\label{eq:7-13}
\end{eqnarray}
Again because $v_{{\rm d}y}=0$ after sufficient time has passed, we obtain,
\begin{eqnarray}
v_{{\rm d}x}=-\sigma_y \cos\eta \frac{\mathcal{J}_{\rm AFM}}{\alpha \mathcal{D}},
\hspace{0.5cm} v_{{\rm d}y}=0.
\label{eq:7-14}
\end{eqnarray}
Considering that the Gilbert-damping coefficient $\alpha$ is much smaller than unity, this result indicates that N\'{e}el-type antiferromagnetic skyrmions in the nanotrack move very quickly along the nanotrack without Hall motion when they are driven by the spin-orbit torque exerted from a vertical spin current with $\bm \sigma$$\parallel$$\bm y$, which is generated by the electric current $\bm j_{\rm e}$$\parallel$$\bm x$.

\subsection{7.4 SOT-Driven Motion of AFM Skyrmions in the Nanotrack System with $\bm j_{\rm e}$$\parallel$$\bm y$ and $\bm \sigma$$\parallel$$\bm x$}
When the spin polarization of the vertical spin current is $\bm \sigma$$\parallel$$\bm x$ as $\bm j_{\rm e}$$\parallel$$\bm y$, substituting Eqs.~(\ref{eq:7-5}) and (\ref{eq:7-6}) into Eq.~(\ref{eq:7-10}), we obtain the following equations for the Bloch-type and N\'{e}el-type antiferromagnetic skyrmions, respectively.\\

\noindent
-- Bloch-type antiferromagnetic skyrmions ($\eta$=$\pi/2$, $3\pi/2$)
\begin{eqnarray}
\left(\begin{array}{cc} v_{{\rm d}x} \\ v_{{\rm d}y}
\end{array}
\right)=\frac{1}{\alpha \mathcal{D}}
\left(\begin{array}{cc} \sigma_x \sin\eta \mathcal{J}_{\rm AFM} \\ -\partial_yV
\end{array}
\right).
\label{eq:7-15}
\end{eqnarray}
Because $v_{{\rm d}y}=0$ after sufficient time has passed, we obtain,
\begin{eqnarray}
v_{{\rm d}x}=\sigma_x \sin\eta \frac{\mathcal{J}_{\rm AFM}}{\alpha \mathcal{D}},
\hspace{0.5cm} v_{{\rm d}y}=0.
\label{eq:7-16}
\end{eqnarray}
This result indicates that Bloch-type antiferromagnetic skyrmions in the nanotrack system move very quickly along the nanotrack without Hall motion when they are driven by the spin-orbit torque exerted from a vertical spin current with $\bm \sigma$$\parallel$$\bm x$, which is generated by the electric current $\bm j_{\rm e}$$\parallel$$\bm y$.\\

\noindent
-- N\'{e}el-type antiferromagnetic skyrmions ($\eta$=0, $\pi$)
\begin{eqnarray}
\left(\begin{array}{cc} v_{{\rm d}x} \\ v_{{\rm d}y}
\end{array}
\right)=\frac{1}{\alpha \mathcal{D}}
\left(\begin{array}{cc} 0 \\ \sigma_x \cos\eta \mathcal{J}_{\rm AFM}-\partial_yV
\end{array}
\right).
\label{eq:7-17}
\end{eqnarray}
Because $v_{{\rm d}y}=0$ after sufficient time has passed, we obtain,
\begin{eqnarray}
v_{{\rm d}x}=0, \hspace{0.5cm} v_{{\rm d}y}=0.
\label{eq:7-18}
\end{eqnarray}
This result indicates that N\'{e}el-type antiferromagnetic skyrmions in the nanotrack cannot move when they are driven by the spin-orbit torque exerted from a vertical spin current with $\bm \sigma$$\parallel$$\bm x$, which is generated by the electric current $\bm j_{\rm e}$$\parallel$$\bm y$.

\section{8. Micromagnetic simulations}
\subsection{8.1 Spin Model for Magnetic Skyrmions}
\begin{figure}
\includegraphics[scale=1.0]{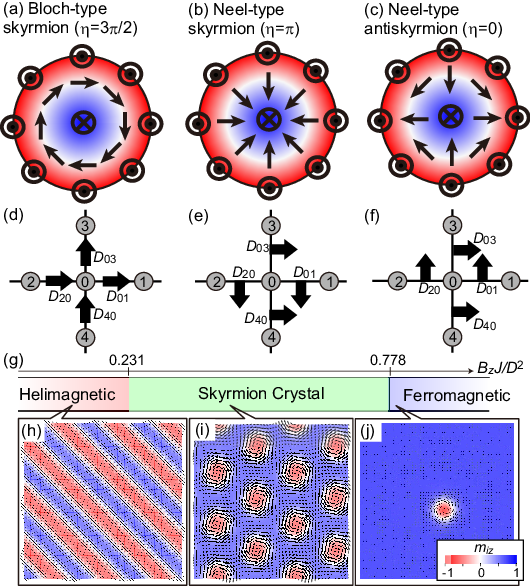}
\caption{(a)-(c) Magnetization configurations of various types of skyrmions, i.e., (a) Bloch-type skyrmion ($\eta$=$3\pi/2$) characterized by a vortex-like magnetization configuration, which appears mainly in chiral magnets, (b) N\'{e}el-type skyrmion ($\eta$=$\pi$) characterized by a fountain-like magnetization configuration, which appears mainly in polar magnets and ferromagnet/heavy-metal bilayer heterojunctions, and (c) Antivortex-type skyrmion, which appears in certain kinds of inverse Heusler compounds and metallic magnets. (d)-(f) Structures of the DM vectors in the lattice spin models for the skyrmions in (a)-(c), respectively. Thick arrows represent the DM vectors $\bm D_{ij}$ for the DM-interaction term $\mathcal{H}_{\rm DM}=-\sum_{<i,j>}\bm D_{ij} \cdot (\bm m_i \times \bm m_j)$ on the bond connecting the sites $i$ and $j$. (g) Theoretical ground-state phase diagram of the spin model in Eq.~(\ref{eq:Lmodel}) as a function of external magnetic field $B_z$. (h)-(j) Magnetization configurations of three magnetic phases, i.e., (h) helimagnetic (cycloidal) phase, (i) skyrmion crystal phase, and (j) ferromagnetic phase with an isolated skyrmion as a defect.}
\label{Fig07}
\end{figure}
A continuum spin model for bulk chiral cubic magnets was proposed by Bak and Jensen in 1980~\cite{Bak1980} as,
\begin{eqnarray}
\mathcal{H}&=&\int d\bm r \left[\frac{J}{2a}(\nabla \bm m)^2 
-\frac{D}{a^2}\bm m \cdot (\nabla \times\ \bm m) \right.
\nonumber \\
&-&\frac{1}{a^3}\bm B \cdot \bm m 
+\frac{A_1}{a^3}(m_x^4 + m_y^4 + m_z^4)
\nonumber \\
&-&\left. \frac{A_2}{2a}[(\nabla_x m_x)^2+(\nabla_y m_y)^2+(\nabla_z m_z)^2]
\right],
\label{eqn:Cmodel}
\end{eqnarray}
where $\bm m(\bm r)$ is the unit magnetization vector at the spatial position $\bm r$, and $a$ is the lattice constant of cubic lattice or the length of cubic cell. The first and the second terms describe the ferromagnetic exchange interactions and the DM interactions, respectively. The third term describes the Zeeman coupling with external magnetic field $\bm B$.  In addition, several magnetic anisotropies allowed by cubic symmetry of the crystal are considered. Note that in addition to the terms considered here, there are various other types of magnetic anisotropies which are compatible with cubic symmetry. Starting from this continuum model, the following classical Heisenberg model on a cubic lattice can be derived by dividing the continuum space into cubic cells~\cite{YiSD2009,Buhrandt2013}.
\begin{align}
&\mathcal{H}=\mathcal{H}_{\rm ex}+\mathcal{H}_{\rm DM}+\mathcal{H}_{\rm Zeeman}
+\mathcal{H}_{\rm aniso.1}+\mathcal{H}_{\rm aniso.2}
\label{eq:Lmodel}
\end{align}
with
\begin{align}
&\mathcal{H}_{\rm ex}=-J \sum_{i} \bm m_i \cdot 
(\bm m_{i+\hat{x}} + \bm m_{i+\hat{y}} + \bm m_{i+\hat{z}}),
\label{eq:Lmodel1}
\\
&\mathcal{H}_{\rm DM}=\sum_{\langle i,j \rangle}\bm D_{ij} \cdot (\bm m_i \times \bm m_j),
\label{eq:Lmodel2}
\\
&\mathcal{H}_{\rm Zeeman}=-\bm B \cdot \sum_i \bm m_i,
\label{eq:Lmodel3}
\\
&\mathcal{H}_{\rm aniso.1}=A_1 \sum_{i} [(m_{i}^x)^4+(m_{i}^y)^4+(m_{i}^z)^4],
\label{eq:Lmodel4}
\\
&\mathcal{H}_{\rm aniso.2}=-A_2 \sum_{i} (m_{i}^xm_{i+\hat{x}}^x+m_{i}^ym_{i+\hat{y}}^y+m_{i}^ym_{i+\hat{z}}^z),
\label{eq:Lmodel5}
\end{align}
where $\bm m_i$ is the unit magnetization vector at the $i$th cell or the $i$th lattice site, which is related with the classical spin vector $\bm S_i$ as $\bm m_i=-\bm S_i/|\bm S_i|$.

The above spin model in three dimensions describes bulk samples of chiral magnets. On the contrary, spin models in two dimensions are also useful because skyrmions get enhanced stability in thin-plate samples and magnetic heterojunction system. The two-dimensional model is important also because the pseudo two-dimensional samples are advantageous and well-studied for device applications. In two-dimensions, it is known that experimental phase diagrams and most of the experimental results can be reproduced quantitatively even without considering the magnetic anisotropy terms. Therefore, the classical Heisenberg models on cubic lattices are often used for theoretical studies of magnetic skyrmions. In this model, various types of skyrmions can be reproduced by varying the spatial structure of the DM vectors [see also Fig.~\ref{Fig07}(a)-(f)]:\\
\\
\underline{-- Bloch-type skyrmions}
\begin{align}
\mathcal{H}_{\rm DM}^{\rm Bloch}=D \sum_i
(\bm m_i \times \bm m_{i+\hat{x}} \cdot \hat{x}
+\bm m_i \times \bm m_{i+\hat{y}} \cdot \hat{y}).
\label{eqn:DMBloch}
\end{align}
\underline{-- N\'{e}el-type skyrmions}
\begin{align}
\mathcal{H}_{\rm DM}^{\rm Neel}=D \sum_i
(\bm m_i \times \bm m_{i+\hat{x}} \cdot \hat{y}
-\bm m_i \times \bm m_{i+\hat{y}} \cdot \hat{x}).
\label{eqn:DMNeel}
\end{align}
\underline{-- Bloch-type antiskyrmions}
\begin{align}
\mathcal{H}_{\rm DM}^{\rm a\-Bloch}=D \sum_i
(\bm m_i \times \bm m_{i+\hat{x}} \cdot \hat{x}
-\bm m_i \times \bm m_{i+\hat{y}} \cdot \hat{y}).
\label{eqn:DMAntiV}
\end{align}
\underline{-- N\'{e}el-type antiskyrmions}
\begin{align}
\mathcal{H}_{\rm DM}^{\rm a\-Neel}=D \sum_i
(\bm m_i \times \bm m_{i+\hat{x}} \cdot \hat{y}
+\bm m_i \times \bm m_{i+\hat{y}} \cdot \hat{x}).
\label{eqn:DMAntiV}
\end{align}
The phase diagram of the two-dimensional model at zero temperature in a magnetic field is shown in Figs.~\ref{Fig07}(g)-(j)~\cite{Iwasaki2013a,Mochizuki2012}. A skyrmion crystal phase appears in the intermediate field-strength region which is sandwiched by the helical phase at low-field regime and the ferromagnetic phase at high-field regime. This theoretical phase diagram reproduces well the experimental phase diagrams of thin-plate samples of chiral magnets at low temperatures quantitatively. The values of the critical magnetic fields are in good agreement with the experimental results. It is worth noting that the above four DM-vector configurations give the same critical magnetic field when the value of $D$ is the same.

Spatial periodicities of the helimagnetic orders and skyrmion crystals are determined by the strength ratio $J/D$ of the ferromagnetic exchange interactions and the DM interactions. With a stronger ferromagnetic exchange interaction $J$ that favors a parallel magnetization alignment, the magnetizations rotate gradually, which requires a longer spatial length for variation of the magnetization angle from zero to $2\pi$. Consequently, the spatial period becomes longer and the size of the magnetization texture becomes larger.  Conversely, with a stronger DM interaction $D$ that favors a 90-degree rotating magnetization alignment, an opposite thing happens where the spatial period and size of the magnetization alignment become shorter and smaller.

For a uniformly rotating helimagnetic order with a pitch angle of $\theta$, the energy density associated with the ferromagnetic exchange interactions and the DM interactions can be written as a function of $\theta$. From its saddle point equation, we obtain a relation $\tan\theta$=$D/(\sqrt{2}J)$. The period of the skyrmion crystal is approximately $2/(\sqrt{3})$ times the period of the helimagnetic order. Specifically, the period of helimagnetic order is about 100 sites for $|D/J|=0.09$, and it is about 35 sites for $|D/J|=0.27$, which correspond to 50 nm and 17.5 nm, respectively, with an assumed typical lattice constant of 5 $\AA$. These values correspond to the helimagnetic periods observed in Cu$_2$OSeO$_3$ ($\lambda_m$=50 nm)~\cite{Seki2012a} and MnSi ($\lambda_m$=18 nm)~\cite{Muhlbauer2009}, respectively.

It should be mentioned that in addition to the terms discussed here, the dipolar interaction plays an important role in the crystallization of skyrmions or in the formation of skyrmion crystals. The dipolar interaction also, more or less, modifies the skyrmion size and the skyrmion magnetization configuration in competition with magnetic anisotropies in both three-dimensional bulk materials and pseudo two-dimensional systems. However, these modifications hardly affect the current-induced motion of skyrmions. Hence, we neglect the dipolar interaction in the present study for simplicity.

\subsection{8.2 Numerical Treatment of LLGS Equation}
To perform micromagnetic simulations using the LLGS equation, several steps of processing on this equation are required, that is, the nondimensionalization, linearization, and discretization of the equation. We first do the nondimensionalization by introducing a dimensionless Hamiltonian $\tilde{\mathcal{H}} \equiv \mathcal{H}/J$. Next, substituting the entire right-hand side of the equation to the time derivative $d\bm m_i/dt$ in the right-hand side, we find that the following term appears in the last term,
\begin{eqnarray}
+\alpha^2 \left[\bm m_i \times \left(\bm m_i \times \frac{d\bm m_i}{dt}\right)\right].
\end{eqnarray}
This term can be reduced to $-\alpha^2 d\bm m_i/dt$ by using the vector formula,
\begin{eqnarray}
\bm A \times (\bm B \times \bm C)=(\bm A \cdot \bm C)\bm B - (\bm A \cdot \bm B)\bm C.
\label{eq:vecform1}
\end{eqnarray}
In addition, the spatial derivatives $\bm \nabla$ that appear in the equation are converted to spatial differences using rectangular cells whose volume is $a_xa_ya_z$. After these procedures, the LLGS equation is rewritten in a form suitable for numerical simulations as follows.
\begin{widetext}
\begin{eqnarray}
\frac{d \bm m_i}{d\tau}
&=&\frac{1}{1+\alpha^2} \left\{
-\bm m_i \times \left(-\frac{\partial \tilde{\mathcal{H}}}{\partial \bm m_i} \right) +\mathcal{A}_i +\mathcal{B}_i +\mathcal{C}_i 
+\alpha \bm m_i \times \left[-\bm m_i \times \left(-\frac{\partial \tilde{\mathcal{H}}}{\partial \bm m_i} \right) 
+\mathcal{A}_i +\mathcal{B}_i +\mathcal{C}_i \right] \right\},
\label{eq:LLGM3}
\end{eqnarray}
where
\begin{eqnarray}
\tilde{\mathcal{H}}&=&\mathcal{H}/J, \quad\quad \tau=tJ\gamma/M,\\
\mathcal{A}_i
&=&\frac{\hbar p}{2eJ} \left(
 a_ya_zj_x \frac{\bm m_{i+\hat{\bm x}}-\bm m_{i-\hat{\bm x}}}{2}
+a_za_xj_y \frac{\bm m_{i+\hat{\bm y}}-\bm m_{i-\hat{\bm y}}}{2}
\right),\\
\nonumber \\
\mathcal{B}_i
&=&-\beta\frac{\hbar p}{2eJ}
\left[ \bm m_i \times \left(
 a_ya_zj_x \frac{\bm m_{i+\hat{\bm x}}-\bm m_{i-\hat{\bm x}}}{2}
+a_za_xj_y \frac{\bm m_{i+\hat{\bm y}}-\bm m_{i-\hat{\bm y}}}{2}
\right) \right],\\
\nonumber \\
\mathcal{C}_i
&=&\frac{\hbar a_xa_ya_z\theta_{\rm SH}j_{\rm e}}{2eJd}
\left[ \bm m_i \times (\bm \sigma \times \bm m_i) \right]
=\frac{\hbar\theta_{\rm SH}}{2eJ}\frac{a_xa_yj_{\rm e}}{n_z}
\left[ \bm m_i \times (\bm \sigma \times \bm m_i) \right]
\end{eqnarray}
Here $d=n_z a_z$ is the sample thickness.
\end{widetext}

For the simulations, we examine both thin-plate system (i.e., a system large enough to allow the skyrmion motion without being affected by potentials from edges) and nanotrack system (i.e., a system in which skyrmions are affected by repulsive confinement potentials from edges in the width direction). For the thin-plate system, we adopt a system of $150 \times 150$ sites (75 nm $\times$ 75 nm) with periodic boundary conditions. On the other hand, for the nanotrack system, we use a system of $600 \times 50$ sites (300 nm $\times$ 25 nm) with open boundary conditions. In parentheses, the corresponding system sizes for the assumed lattice constant of $a$=0.5 nm are given. Both clean case without impurities and dirty case with impurities are investigated. For the dirty case, the following term is added to the Hamiltonian,
\begin{align}
\mathcal{H}_{\rm imp}=A_{\rm imp} \sum_{i \in {\rm imp.}} m_{zi}^2.
\end{align}
This term describes the perpendicular magnetization anisotropy ($A_{\rm imp}<0$) at randomly selected impurity sites. In the simulations argued below, the strength of anisotropy is set to be $A_{\rm imp}=-0.2$, and the impurity concentration is set to be $0.1\%$.

\subsection{8.3 Simulations: STT-Driven Motion of FM Skyrmions}
\begin{figure*}
\includegraphics[scale=1.0]{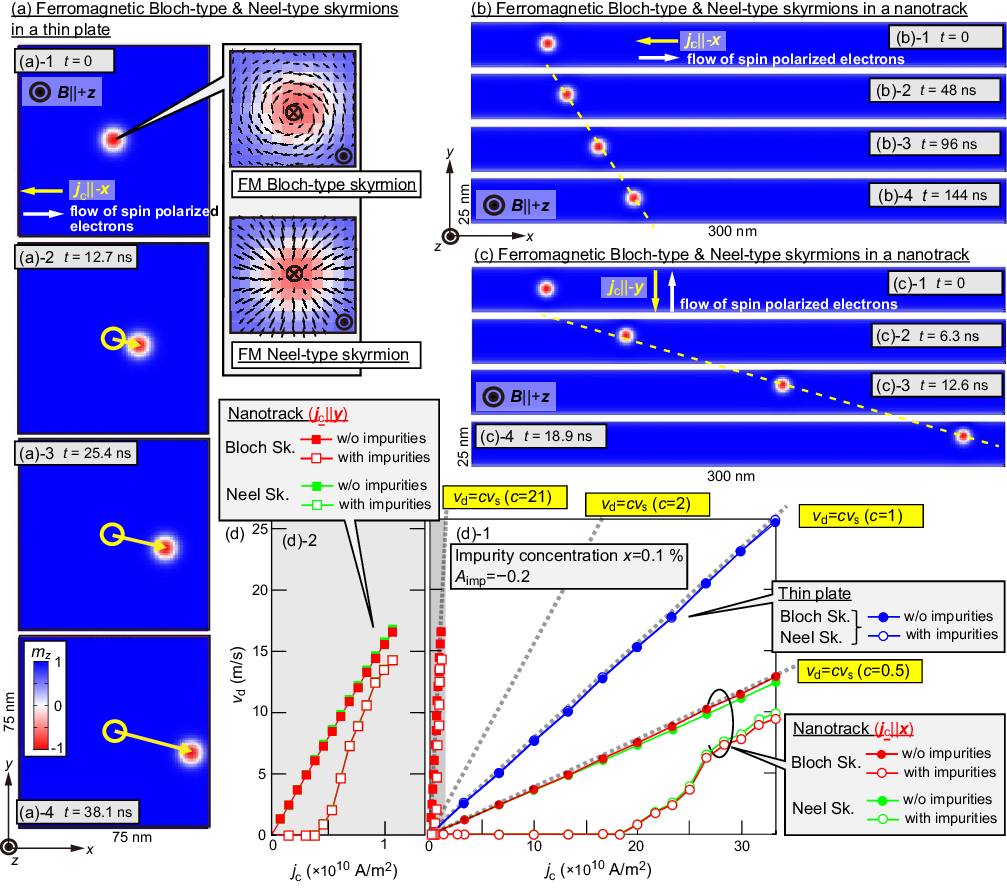}
\caption{Results of micromagnetic simulations for current-induced motion of ferromagnetic skyrmions driven by the spin-transfer torque. The Bloch-type and N\'{e}el-type skyrmions exhibit exactly the same motion when the spin-transfer torque is the driving mechanism. (a) Snapshots of the simulated motion of a skyrmion in the ferromagnetic thin-plate system when the electric current of $j_{\rm e}=1.0 \times 10^{10}$ A/m$^2$ is applied in the $-\bm x$ direction. (b),~(c) Snapshots of the simulated motion of a skyrmion in the ferromagnetic nanotrack system when the electric current of $j_{\rm e}=1.0 \times 10^{10}$ A/m$^2$ is applied in the $-\bm x$ (length) direction (b) and the $-\bm y$ (width) direction (c). The latter case shows extremely fast motion. (d) Current-density dependence of the drift velocity $v_{\rm d}$ for the Bloch-type skyrmion and the N\'{e}el-type skyrmion in the thin-plate and nanotrack systems. Both clean and dirty cases are examined. Large threshold current densities appear in the dirty nanotrack system. The plotted drift velocities for the dirty case are the averages of 16 simulation results for different patterns of random impurity distributions. Panel (d)-1 shows plots over a wide current-density range, while Panel (d)-2 shows enlarged plots for a small current-density range of $0 \leq j_{\rm e} \leq 1.2 \times 10^{10}$ A/m$^2$. The dashed lines in (d)-1 are analytical results obtained from the Thiele equation.}
\label{Fig08}
\end{figure*}
We first discuss the simulation results for the current-induced motion of ferromagnetic skyrmions, i.e., the Bloch-type and N\'{e}el-type ferromagnetic skyrmions, driven by the spin-transfer torque mechanism. These two types of skyrmions show the same motion when the spin-transfer torque is a driving mechanism. This aspect can be understood from the expressions of drift velocities derived in Sec. 5, which do not contain the helicity $\eta$ (see also Table I). For the parameters, we adopt $J=-1$ meV for the ferromagnetic exchange interactions, $D=0.18$ meV for the DM interactions, $\gamma H_z=0.0278$ meV for the external magnetic field, $\alpha=0.04$ for the Gilbert-damping coefficient, and $p=0.2$ for the spin polarization of the electric current. The parameter for the nonadiabatic torque is set to be $\beta=0.02(=0.5\alpha)$. 

Figures~\ref{Fig08}(a) show snapshots of the simulated motion of a skyrmion in the ferromagnetic thin-plate system. Here the electric current of $j_{\rm e}=1.0 \times 10^{10}$ A/m$^2$ is applied in the $-\bm x$ direction. The skyrmion moves almost in a direction antiparallel to the current, i.e., in the $+\bm x$ direction, while it is accompanied by small transverse motion in the $-\bm y$ direction due to the skyrmion Hall effect. The drift velocity $\bm v_{\rm d}=(v_{{\rm d}x},v_{{\rm d}y})$ is given by Eq.~(\ref{eq:5-3}). On the other hand, Figs.~\ref{Fig08}(b) and (c) show snapshots of the simulated motion of a skyrmion in the ferromagnetic nanotrack. Here the electric current of $j_{\rm e}=1.0 \times 10^{10}$ A/m$^2$ is applied in the $-\bm x$ direction (length direction) in Fig.~\ref{Fig08}(b) and in the $-\bm y$ direction (width direction) in Fig.~\ref{Fig08}(c). Importantly, the latter case shows extremely fast motion. The drift velocity $(v_{{\rm d}x},v_{{\rm d}y})$ for the former case is given by Eq.~(\ref{eq:5-5}), and that for the latter case is given by Eq.~(\ref{eq:5-6}).

Figure~\ref{Fig08}(d) show the calculated current-density dependence of $v_{\rm d}$ for different types of skyrmions (Bloch-type and N\'{e}el-type) and different system geometries (thin plate and nanotrack) for both clean and dirty cases. Dependence on the direction of electric current (i.e., $\bm j_{\rm e}$$\parallel$$-\bm x$ and $\bm j_{\rm e}$$\parallel$$-\bm y$) is also examined for the nanotrack system.

First we focus on the current-velocity relation for skyrmions in the thin-plate system. According to Eq.~(\ref{eq:5-3}), the velocity obeys a universal relation $v_{\rm d}=v_{\rm s}$ in the clean case, independent of the values of $\alpha$ and $\beta$. This expected relation is indeed confirmed by the plot. Note that this universal relation holds when the damping constant $\alpha$ is small and their second order quantities can be neglected. Typically, the order of $\alpha$ is 10$^{-2}$ for bulk magnetic materials and 10$^{-1}$ for synthetic magnetic systems. This means that the correction due to the damping is at most a few percent. Effects of the corrections on the velocity of current-driven skyrmions due to the damping were theoretically studied in Ref.~\cite{ZhangX2017}.

We also find that this current-velocity relation is almost not altered even in the presence of impurities as far as the current-density scale of 10$^{10}$ A/m$^2$ is concerned as in the plots of Fig.~\ref{Fig08}(d). Indeed, the threshold current density in the dirty case is so small that it is not discernible in the plot of this scale. This ultra-small threshold current density and the resulting high mobility are important properties for the current-induced skyrmion motion in thin-plate systems. Note that if we zoom up a region of small current density of the order of 10$^5$-10$^6$ A/m$^2$, the threshold current density becomes discernible in the presence of impurities, and characters of the impurities such as size, shape, and concentration should become important factors, which govern the threshold current density as well as the current-velocity relation~\cite{Stosic2017,LiangX2019}.

On the contrary, the skyrmion motion in the nanotrack system shows totally different behaviors. At first, the clear dependence on the current direction appears. When the electric current is applied in the length direction of the nanotrack (i.e., $\bm j_{\rm e}$$\parallel$$-\bm x$), the velocity obeys a relation $v_{\rm d}=(\beta/\alpha)v_{\rm s}$ according to Eq.~(\ref{eq:5-5}), which sensitively depends on both $\alpha$ and $\beta$. As we set $\beta=0.5\alpha$ in the present simulations, the relation becomes $v_{\rm d}= c_1 v_{\rm s}$ with $c_1=0.5$, which is indeed confirmed by the plot. Furthermore, a large threshold current density appears when impurities are introduced in contrast to the case of the thin-plate system. 

When the electric current is applied in the width direction (i.e., $\bm j_{\rm e}$$\parallel$$-\bm y$), the velocity obeys a relation $v_{\rm d}=-(\mathcal{G}/\alpha\mathcal{D})v_{\rm s}$ according to Eq.~(\ref{eq:5-6}). With the present parameters, the factor $c_2 \equiv \mathcal{G}/\alpha\mathcal{D}$ becomes $c_2 \approx 21$, which indicates skyrmions can move much more quickly with an enhanced speed in the nanotrack system for the peculiar current direction of $\bm j_{\rm e}$$\parallel$$\bm y$. Its extremely fast velocity can be seen in the steep slope of the plot in Fig.~\ref{Fig08}(d)-1. Note that a significant threshold current density appears again in the dirty case as seen in the enlarged plots in Fig.~\ref{Fig08}(d)-2.

\subsection{8.4 Simulations: SOT-Driven Motion of FM Skyrmions}
\begin{figure*}
\includegraphics[scale=1.0]{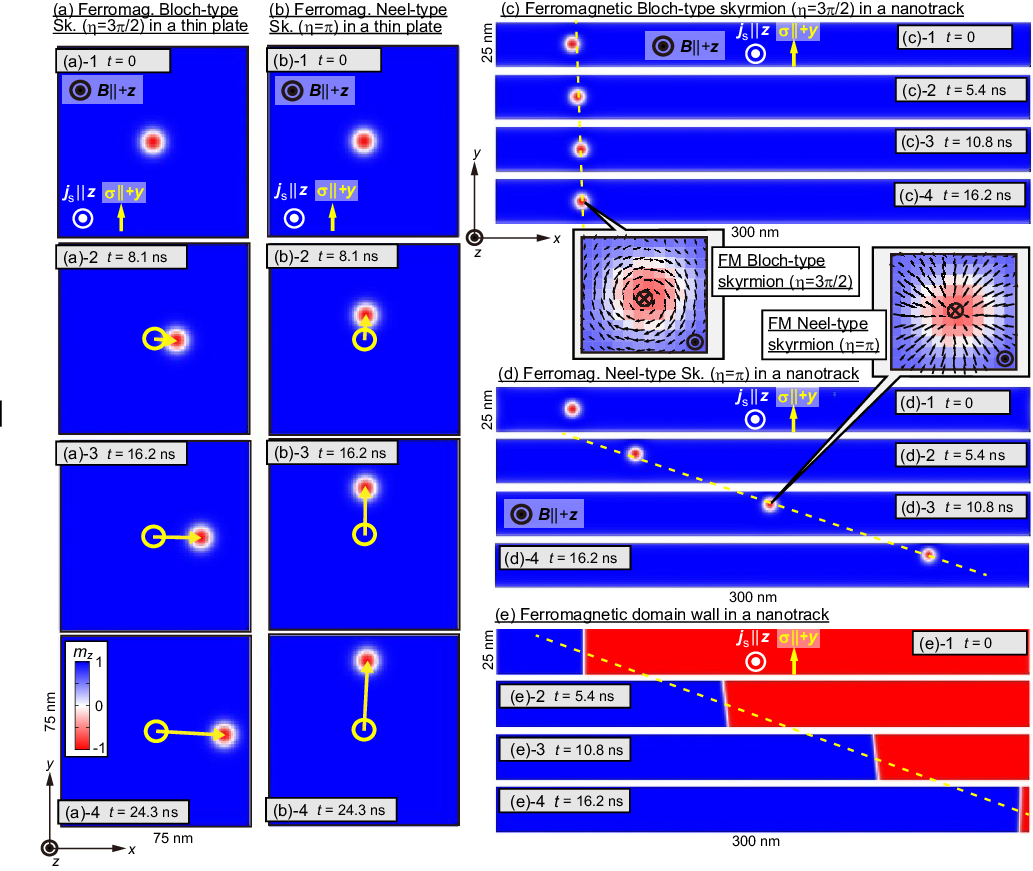}
\caption{Results of micromagnetic simulations for current-induced motion of ferromagnetic skyrmions and ferromagnetic domain wall driven by the spin-orbit torque. The system and the parameters used in the simulations are the same as those used in the simulations for Fig.~\ref{Fig08}, but the perpendicular anisotropy with $K=-0.1$ meV is introduced for the simulations of domain wall. In all the simulations, we assume a vertial spin current $\bm j_{\rm s}$$\parallel$$\bm z$ with the spin polarization $\bm \sigma$$\parallel$$+\bm y$, which is realized by injecting an electric current $\bm j_{\rm e}$$\parallel$$\bm x$ to the heavy-metal layer of the heavy-metal/ferromagnetic heterojunction. The spin-current density $j_{\rm s}(=|\theta_{\rm SH}|j_{\rm e})$ is set to be $1.0 \times 10^9$ A/m$^2$. (a),~(b) Snapshots of the simulated motion for (a) Bloch-type skyrmion and (b) N\'{e}el-type skyrmion in the ferromagnetic thin-plate system. Note that the direction of motion differs between these two skyrmions. (c)-(e) Snapshots of the simulated motion for (c) Bloch-type skyrmion, (d) N\'{e}el-type skyrmion, and (e) ferromagnetic domain wall in the ferromagnetic nanotrack system. For these simulations, a clean system without impurities is assumed. In the nanotrack system, the N\'{e}el-type skyrmions and the domain walls move quickly, whereas the Bloch-type skyrmions do not move at all when $\bm \sigma$$\parallel$$\bm y$.}
\label{Fig09}
\end{figure*}
We next discuss the simulation results for the current-induced motion of ferromagnetic skyrmions and ferromagnetic domain wall driven by the spin-orbit torque. The system and the parameters used in the simulations are the same as those used in the simulations for Fig.~\ref{Fig08}, but the perpendicular anisotropy is introduced for the simulations of domain wall to stabilize it. We set the anisotropy strength at $K$=$-0.1$ meV, while $K$=0 for the simulations of skyrmions. We assume a vertical spin current $\bm j_{\rm s}$$\parallel$$\bm z$ with the spin polarization $\bm \sigma$$\parallel$$+\bm y$. Experimentally, this spin current is realized by injecting an electric current $\bm j_{\rm e}$$\parallel$$\bm x$ to a heavy-metal layer of the heavy-metal/ferromagnetic heterojunction system. The injected electric current is converted to the vertical spin current via the spin Hall effect due to the interfacial spin-orbit coupling. Here the spin-current density $j_{\rm s}(=|\theta_{\rm SH}|j_{\rm e})$ is set to be $1.0 \times 10^9$ A/m$^2$. 

As the spin-orbit torque is the driving mechanism, the current-induced motion of skyrmion sensitively depends on the type of skyrmion (N\'{e}el-type and Bloch-type) and its helicity $\eta$ in contrast to the case of the spin-transfer torque argued above. Figures~\ref{Fig09}(a) and (b) show snapshots of the simulated motion of (a) a Bloch-type skyrmion and (b) a N\'{e}el-type skyrmion in the clean ferromagnetic thin-plate system without impurities. For the Bloch-type skyrmion, it moves nearly perpendicular to the spin-polarization direction $\bm \sigma$, that is, it moves in the $x$ direction when $\bm \sigma$$\parallel$$\bm y$ as indicated Eq.~(\ref{eq:6-9}). On the contrary, the N\'{e}el-type skyrmion moves parallel to $\bm \sigma$, that is, it moves in the $y$ direction when $\bm \sigma$$\parallel$$\bm y$ as indicated Eq.~(\ref{eq:6-10}). 

Figures.~\ref{Fig09}(c)-(e) show snapshots of the simulated motion for (c) the Bloch-type skyrmion, (d) the N\'{e}el-type skyrmion, and (e) a ferromagnetic domain wall in the clean ferromagnetic nanotrack system without impurities. In the nanotrack system, the Bloch-type skyrmion driven by the spin-orbit torque does not move at all or moves only very little as indicated by Eq.~(\ref{eq:6-14}). On the contrary, the N\'{e}el-type skyrmion as well as the ferromagnetic domain wall move very quickly along the nanotrack with the drift velocity $|v_{{\rm d}x}|=\mathcal{J}_{\rm FM}/(\alpha \mathcal{D})$ as indicated by Eq.~(\ref{eq:6-17}). Note that these aspects hold when the electric current is injected in the length direction $\bm j_{\rm e}$$\parallel$$\bm x$ and, thereby, the spin polarization of the spin current is $\bm \sigma$$\parallel$$\bm y$. On the contrary, when $\bm j_{\rm e}$$\parallel$$\bm y$ and $\bm \sigma$$\parallel$$\bm x$, these behaviors become opposite (the simulation results are not shown), that is, the Bloch-type skyrmion moves quickly along the nanotrack with $|v_{{\rm d}x}|=\mathcal{J}_{\rm FM}/(\alpha \mathcal{D})$ as indicated by Eq.~(\ref{eq:6-20})., while the N\'{e}el-type skyrmion does not move as indicated by Eq.~(\ref{eq:6-23}).

\begin{figure*}
\includegraphics[scale=1.0]{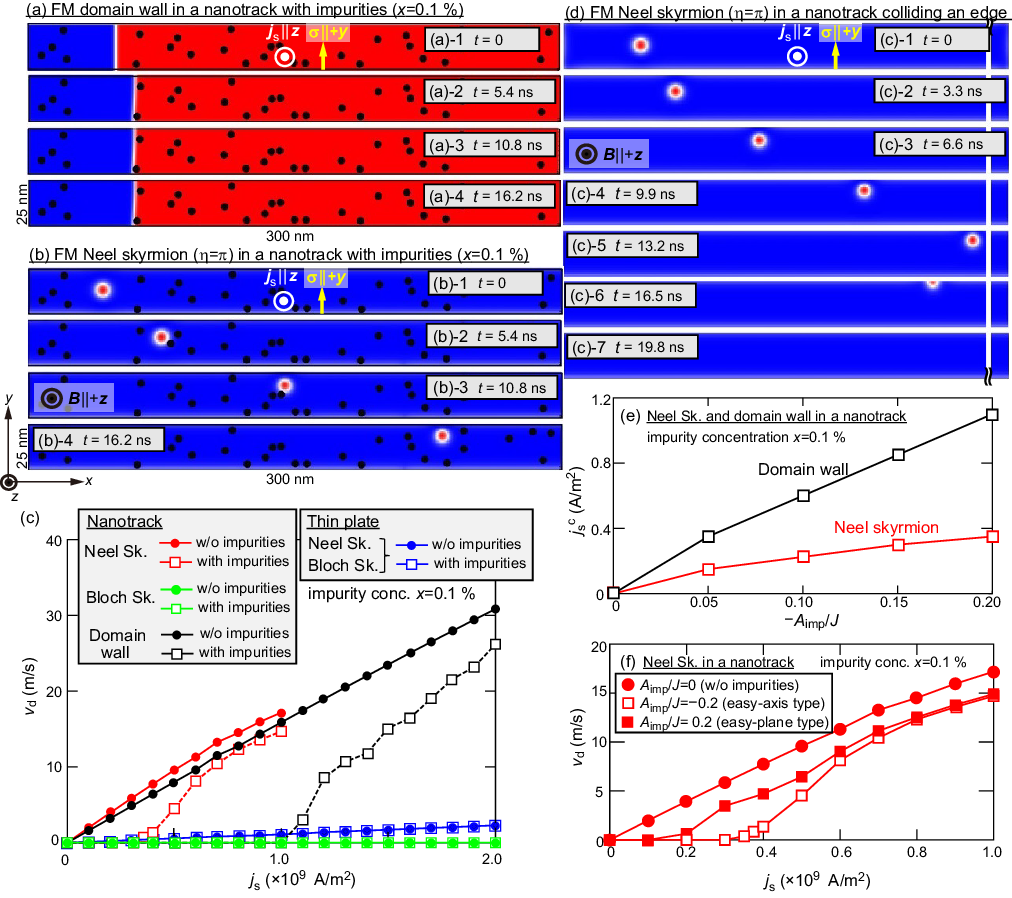}
\caption{Additional results of micromagnetic simulations for current-induced motion of ferromagnetic skyrmions and ferromagnetic domain wall driven by the spin-orbit torque. Simulations are performed for the dirty case with impurities as well. (a),~(b) Snapshots of the simulated motion of (a) a ferromagnetic domain wall and (b) a N\'{e}el-type skyrmion in the ferromagnetic nanotrack with impurities. The vertical spin current $\bm j_{\rm s}$$\parallel$$\bm z$ with the spin polarization $\bm \sigma$$\parallel$$+\bm y$ is considered again. Comparison with Figs.~\ref{Fig09}(c) and (d) indicates that the motion of domain wall is strongly affected by impurities, while the motion of N \'{e}el-type skyrmion is not. (c) Snapshot of the simulated motion of a N\'{e}el-type skyrmion with much larger spin current. The large Hall motion results in collision of the skyrmion with the upper edge of the nanotrack. The skyrmion is pushed against the edge strongly and is absorbed by the edge resulting in its annihilation. (d) Dependence of the drift velocity $v_{\rm d}$ on the spin-current density $j_{\rm s}$ for the N\'{e}el-type skyrmion, the Bloch-type skyrmion and the ferromagnetic domain wall in the thin-plate and nanotrack systems. Both clean and dirty cases are examined. A large threshold spin-current density appears for the nanotrack system in the dirty case. The plotted drift velocities in the dirty case are averages of 16 simulation results for different patterns of random impurity distributions. (e) Dependence of $j_{\rm s}$ on the strength of the impurity-site magnetic anisotropy $A_{\rm imp}$ for the N\'{e}el-type skyrmion and the ferromagnetic domain wall in the nanotrack system. (f) $j_{\rm s}$-dependence of $v_{\rm d}$ for the N\'{e}el-type skyrmion in the nanotrack system for different types of impurity-site magnetic anisotropy, i.e., $A_{\rm imp}/J=0$ (no magnetic anisotropy or clean case), $A_{\rm imp}/J=-0.2$ (perpendicular anisotropy), and $A_{\rm imp}/J=+0.2$ (easy-plane or hard-axis anisotropy).}
\label{Fig10}
\end{figure*}
Although the domain wall and the N\'{e}el-type skyrmion move at almost the same speed in the clean system as seen in Figs.~\ref{Fig09}(d) and (e), their behaviors become totally different once the system becomes dirty with impurities as seen in the comparison between Figs.~\ref{Fig10}(a) and (b). These figures show snapshots of the simulated motion for (a) the ferromagnetic domain wall and (b) the N\'{e}el-type skyrmion in the dirty ferromagnetic nanotrack system with impurities. The domain wall does not move at all being trapped by impurities at this spin-current density of $j_{\rm s}=1.0 \times 10^9$ A/m$^2$. This indicates that the motion of domain wall is significantly affected by impurities, and a large threshold spin-current density appears in the dirty system. This aspect can be seen in the plot of $v_{\rm d}$ as a function of $j_{\rm s}$ in Fig.~\ref{Fig10}(c). In contrast, the N\'{e}el-type skyrmion moves as quickly as that in the clean system, indicating that the motion of N\'{e}el-type skyrmion is not affected by impurities. In fact, the threshold spin-current density also appears for the N\'{e}el-type skyrmion when impurities are introduced, but its magnitude is much smaller than that of the domain wall as seen in Fig.~\ref{Fig10}(c).

Because of the high mobility, N\'{e}el-type skyrmions in the nanotrack system driven by the spin-orbit torque seem to be suitable for application for devices using current-driven skyrmions such as skyrmion-based racetrack memory. However, there is a crucial problem against the practical application, that is, skyrmions are annihilated at an edge of the nanotrack as shown in Fig.~\ref{Fig10}(d) when the spin-current density is relatively large. Because the vertical spin current with $\bm \sigma$$\parallel$$\bm y$ originally drives the N\'{e}el-type skyrmion in the $y$ direction (i.e., width direction), the skyrmion is pushed to the upper edge and feels a strong repulsive potential from the edge. With a moderate spin-current density, the skyrmion can move rather quickly along the length direction because it moves perpendicular to the potential gradient as indicated by the Thiele equation. However, once the spin-current density exceeds a certain threshold value, the skyrmion is absorbed by the edge resulting in its annihilation. The threshold value is approximately $j_{\rm s}=1.0 \times 10^9$ A/m$^2$ and, hence, the plots of $v_{\rm d}$ for the N\'{e}el-type skyrmion end up at this value in Fig.~\ref{Fig10}(c).

In Fig.~\ref{Fig10}(e), we plot the calculated threshold values of $j_{\rm s}$ as functions of the strength of impurity-site magnetic anisotropy $A_{\rm imp}$ for the N\'{e}el-type skyrmion and the ferromagnetic domain wall in the nanotrack system. We find that the threshold value increases as $A_{\rm imp}$ increases for both N\'{e}el-type skyrmion and domain wall, but that of the N\'{e}el-type skyrmion is significantly suppressed as compared to that of the domain wall. In Fig.~\ref{Fig10}(f), we plot the calculated $j_{\rm s}$-dependence of $v_{\rm d}$ for the N\'{e}el-type skyrmion in the nanotrack system for different types of impurity-site magnetic anisotropy, i.e., $A_{\rm imp}/J=0$ for no magnetic anisotropy or clean case, $A_{\rm imp}/J=-0.2$ for the easy-axis or perpendicular anisotropy, and $A_{\rm imp}/J=+0.2$ for the easy-plane or hard-axis anisotropy. We find that both the easy-axis and the easy-plane anisotropies give rise to nonzero threshold values of the spin-current density, while it is absent when $A_{\rm imp}/J=0$. Furthermore, the easy-axis or perpendicular anisotropy causes a larger threshold value as compared to the easy-plane anisotropy.

\subsection{8.5 Simulations: SOT-Driven Motion of AFM Skyrmions}
\begin{figure*}
\includegraphics[scale=1.0]{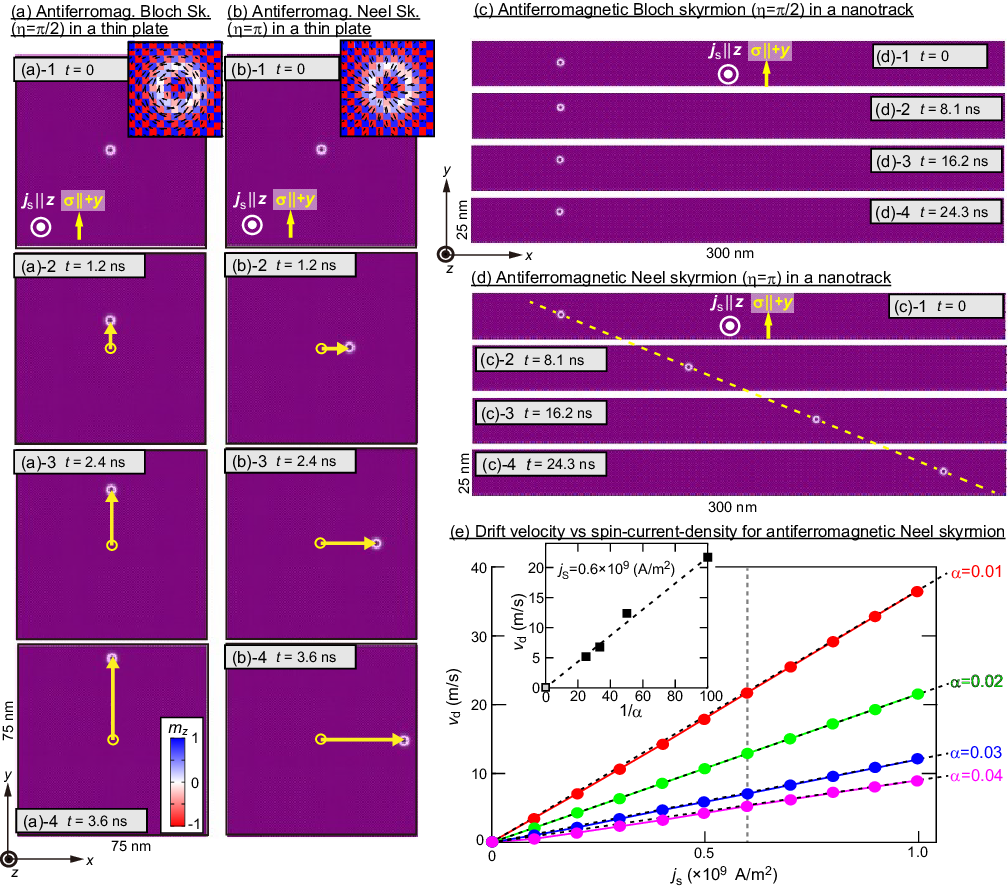}
\caption{Results of micromagnetic simulations for current-induced motion of antiferromagnetic skyrmion driven by the spin-orbit torque. The parameters used in the simulations are $J=1$ meV for the antiferromagnetic exchange interactions, $D=0.2$ meV for the DM interactions, $K=-0.1$ meV for the perpendicular anisotropy, $\gamma H_z=0.5$ meV for the external magnetic field, and $\alpha=0.04$ for the Gilbert-damping coefficient. The vertical spin current $\bm j_{\rm s}$$\parallel$$\bm z$ with spin polarization $\bm \sigma$$\parallel$$+\bm y$ is considered again, which is realized by injection of electric current $\bm j_{\rm e}$$\parallel$$\bm x$ in the heavy-metal layer of the heavy-metal/antiferromagnetic heterojunction. The spin-current density is set to be $j_{\rm s}=1.0 \times 10^9$ A/m$^2$ in the simulations for (a)-(d). (a),~(b) Snapshots of the simulated motion for (a) a Bloch-type skyrmion and (b) a N\'{e}el-type skyrmion in the antiferromagnetic thin-plate system. Note that the direction of motion is different between these two skyrmions. (c),~(d) Snapshots of the simulated motion for (c) a Bloch-type skyrmion and (d) a N\'{e}el-type skyrmion in the antiferromagnetic nanotrack system. For these simulations, a clean system without impurities is assumed. The N\'{e}el-type skyrmion moves quickly both in the thin-plate and nanotrack systems at the same speeds. On the contrary, the Bloch-type skyrmion does not move at all in the nanotrack system. (e) Dependence of the drift velocity $v_{\rm d}$ on the spin-current density $j_{\rm s}$ for the N\'{e}el-type antiferromagnetic skyrmion for various values of the Gilbert-damping coefficient $\alpha$. When $j_{\rm s}$ is fixed, $v_{\rm d}$ is inversely proportional to $\alpha$. Inset shows the plot of $v_{\rm d}$ as a function of $1/\alpha$ when $j_{\rm s}=0.6 \times 10^9$ A/m$^2$, which clearly shows the relation.}
\label{Fig11}
\end{figure*}
We finally discuss the simulation results for the current-induced motion of antiferromagnetic skyrmions in a bipartite antiferromagnet, i.e., Bloch-type and N\'{e}el-type antiferromagnetic skyrmions, driven by the spin-orbit torque. Their current-induced motion again sensitively depends on the type of skyrmion. For the parameters of the antiferromagnetic system, we adopt $J=1$ meV for the antiferromagnetic exchange interactions, $D=0.2$ meV for the DM interactions, $K=-0.1$ meV for the perpendicular anisotropy, $\gamma H_z=0.5$ meV for the external magnetic field, and $\alpha=0.04$ for the Gilbert-damping coefficient. The spin-current density is set to be $j_{\rm s}=1.0 \times 10^9$ A/m$^2$ in the simulations for Figs.~\ref{Fig11}(a)-(d), and its spin-polarization is set as $\bm \sigma$$\parallel$$+\bm y$, which is realized when the electric current is injected along the $x$ direction, i.e., $\bm j_{\rm e}$$\parallel$$\bm x$.

Figures~\ref{Fig11}(a) and (b) show snapshots of the simulated motion of (a) a Bloch-type antiferromagnetic skyrmion and (b) a N\'{e}el-type antiferromagnetic skyrmion in the clean thin-plate system without impurities. In this condition, the Bloch-type antiferromagnetic skyrmion moves in the $y$ direction, which is parallel to the spin polarization of the vertical spin current $\bm \sigma$$\parallel$$+\bm y$. On the contrary, the N\'{e}el-type antiferromagnetic skyrmion moves in the $x$ direction which is perpendicular to the spin polarization $\bm \sigma$. These behaviors are opposite to those of the ferromagnetic skyrmions driven by the spin-orbit torque. On the other hand, Figs.~\ref{Fig11}(c) and (d) show snapshots of the simulated motion of the antiferromagnetic skyrmions in the clean nanotrack system. Notably, the N\'{e}el-type antiferromagnetic skyrmion moves very quickly, whereas the Bloch-type skyrmion does not move at all. Importantly, the N\'{e}el-type antiferromagnetic skyrmion moves at the same speed both in the thin-plate and nanotrack systems as indicated by Eqs.~(\ref{eq:7-9}) and (\ref{eq:7-14}) derived from the Thiele equation.

Figure~\ref{Fig11}(e) show dependence of the drift velocity $v_{\rm d}$ on the spin-current density $j_{\rm s}$ for the N\'{e}el-type antiferromagnetic skyrmion when $\bm \sigma$$\parallel$$\bm y$ for various values of the Gilbert-damping coefficient $\alpha$. Equations~(\ref{eq:7-9}) and (\ref{eq:7-14}) in the Thiele analysis indicates that $v_{\rm d}$ is inversely proportional to $\alpha$ when $j_{\rm s}$ is fixed. Inset of Fig.~\ref{Fig11}(e) shows the plot of $v_{\rm d}$ as a function of $1/\alpha$ when $j_{\rm s}=0.6 \times 10^9$ A/m$^2$, which clearly shows the relationship.

\section{9. Summary and Discussion}
In this article, we have discussed comprehensive theoretical descriptions of general properties of current-induced motion of skyrmions driven by the spin-transfer torque and the spin-orbit torque. Using the analytical theory based on the Thiele equation, we have systematically discussed the direction and velocity of current-induced motion of skyrmions from the viewpoint of their dependencies on several factors and conditions, e.g., the type of skyrmion (Bloch type or N\'{e}el type), the system geometry (thin plate or nanotrack), the direction of applied current (length direction or width direction of the nanotrack), the type of magnet (ferromagnet or antiferromagnet). Furthermore, these analytical results have been visually demonstrated and quantitatively confirmed by micromagnetic simulations using the Landau-Lifshitz-Gilbert-Slonczewski equation. The comprehensive theoretical descriptions provided by this article are expected to make significant contributions to research on the manipulation and control of magnetic skyrmions with electric currents for future spintronics applications.

In recent years, significant progress has been made in experimental and theoretical studies on current-induced phenomena of skyrmions and related topological magnetic textures. In addition to uniform thin-plate and nanotrack systems discussed in this article, various interesting motion and behaviors exhibited by current-driven skyrmions in systems with patterned structures or spatially arranged defects have been observed or predicted~\cite{Reichhardt2015a,Reichhardt2015b,MaF2016,Fernandes2018,Menezes2019,Gonzalez-Gomez2019,Castell-Queral2019,Bhatti2019,Feilhauer2020,Fernandes2020,Arjana2020,ChenW2020a,Vizarim2021,SouzaJCB2021,Del-Valle2022,SouzaJCB2022,ZhangX2022,Rocha2024,SouzaJCB2024,Juge2021,Ohara2021,Jong2022}. Unlike ordinary classical particles, the motion of skyrmions is perpendicular to the potential force or the potential gradient, and thus nontrivial and dramatic phenomena show up when they interact with structures and defects in the systems. The theory based on the Thiele equation provided in this article is expected to be powerful and useful to investigate such phenomena as well. 

In the studies of current-induced skyrmion dynamics, there are several interesting phenomena and subjects in addition to the current-induced drift motion, that is, the creation and annihilation of skyrmions by current application~\cite{Iwasaki2013b,Sampaio2013,TchoeY2012,Yuan2016,YinG2016,Everschor-Sitte2017,Hrabec2017,YuXZ2017,YuXZ2020,Fujimoto2021,Fujimoto2022,WangW2022,Buttner2017}, the deformation, contraction, and expansion of skyrmions during the current-induced motion~\cite{Salimath2020,ChenZ2022,Msiska2022}, the domain-skyrmion transformations~\cite{ZhouY2014,JiangW2015}, and the proliferation and multiplication of skyrmions and related magnetic textures induced by currents~\cite{YuXZ2023,ZhangX2024}. The Thiele equation, in fact, cannot fully describe such phenomena because it is derived under the assumption that the shape of the magnetic texture is rigid and never changes in the process of motion. However, the equation can provide important information on their initial processes and behaviors after the sufficient duration as well and should offer a solid basis for studying these phenomena.

In addition, although the discussions in this article particularly assume skyrmions stabilized by the DM interaction in systems with broken spatial inversion symmetry, it is known that skyrmions and skyrmion-like topological magnetic textures can also be stabilized by the effect of frustrated exchange interactions among localized spins~\cite{Okubo2012}. If the skyrmions in the frustrated system couple to conduction electrons, they can also be driven by electric current~\cite{ZhangX2017,Leonov2015,Leonov2017}, and the discussion in this article can be directly applied to the descriptions of their current-induced motion. Furthermore, not only electric currents but also many other kinds of external stimuli can drive or generate skyrmions. For example, motion of skyrmions can be induced by microwave fields~\cite{WangW2015,Ikka2018,Takeuchi2018,Takeuchi2019,YuanHY2019,ChenW2020b,BoL2024b}, spin waves~\cite{Iwasaki2014b,Schutte2014,DingJ2015,LiuY2022,MaXP2023,HuangL2023,BaiX2023,LauM2024,ShenM2018}, electric fields~\cite{WangYD2022,LiZ2018,WangX2018,ShenL2018,LiuY2019,MaC2019,Yanes2019,Nakatani2016}, temperature gradient~\cite{Kovalev2009,KongL2013,Mochizuki2014,LinSZ2014,Matsuki2023,YuXZ2021,QinG2022}, Brownian thermal fluctuations~\cite{Zazvorka2019,Nozaki2019,ZhaoL2020,Ishikawa2021,Miki2021,SongC2021,ZhangX2023b}, acoustic waves~\cite{Seki2020,ChenR2023,YangY2023,YangY2024}, and light~\cite{YangW2018,Hirosawa2022}. The discussion based on the Thiele equation is also applicable to the skyrmion dynamics caused by these various physical mechanisms. In addition, it has turned out that skyrmions can be generated by various external stimuli such as magnetic fields~\cite{Koshibae2015,WooS2016,Buttner2015,Mochizuki2017}, electric fields~\cite{WangYD2022,LiZ2018,MaC2019,Mochizuki2015a,Mochizuki2015b,Okamura2016,WangL2018,HuangP2018,Kruchkov2018,WangY2020}, spin-polarized STM~\cite{Romming2013,Heinze2011,Wiesendanger2016}, light-induced local Joule heating~\cite{Ogasawara2009,Finazzi2013,Koshibae2014,Oike2016,Vallobra2018,Berruto2018}, current-induced local Joule heating~\cite{Lemesh2018}, pressures~\cite{Nii2015}, tensile strains~\cite{Seki2017,Tanaka2020}, circularly polarized microwave fields~\cite{Miyake2020}, and light with an angular momentum~\cite{Fujita2017a,Fujita2017b}. The Thiele equation is expected to provide important insights into the initial process of generation and the behaviors of skyrmions after generation. We hope that this article will contribute to the development of future research on dynamical phenomena of skyrmions and related topological magnetic textures~\cite{Gobel2021}, which have become more and more important in recent years.

\begin{widetext}
\section{Appendix: Evaluations of the quantities in the Thiele equation}
We can quantitatively evaluate the values of $\mathcal{D}$ and $\bm J^{\rm spin}$ in the Thiele equation by assuming an analytical expression of the spatial profile of skyrmion magnetizations. As discussed in Sec.2, the normalized magnetization vectors $\bm m(r, \phi)$ constituting a skyrmion is given in the two-dimensional polar representation as,
\begin{align}
\begin{pmatrix}
m_x \\
m_y \\
m_z 
\end{pmatrix}
=
\begin{pmatrix}
\sin{\Theta(r)}\cos{\Phi(\phi)} \\
\sin{\Theta(r)}\sin{\Phi(\phi)} \\
\cos{\Theta(r)}
\end{pmatrix}.
\end{align}
Here we set a center of the skyrmion as the origin of the polar coordinate $(r, \phi)$. We assume approximate spatial profiles for $\Theta(r)$ and $\Phi(\phi)$ as,
\begin{align}
&\Theta(r)=\pi\zeta_{\rm m} \left( \frac{r}{r_{\rm sk}} + \dfrac{\zeta_{\rm m} - 1}{2} \right), \\
&\Phi(\phi)=n_\Omega \phi + \eta,
\end{align}
where $r_{\rm sk}$ is the radius of the skyrmion, $n_\Omega$$(=\pm 1)$ is the vorticity, $\zeta_{\rm m}$$(=\pm 1)$ is the sign of core magnetization, and $\eta$ is the helicity (see also Fig.~\ref{Fig02}). Specifically, (anti)skyrmions with $\eta$=$\pi/2$ and $3\pi/2$ are referred to as the Bloch-type, while those with $\eta$=0 and $\pi$ are referred to as the N\'{e}el-type.

Since the following relations hold for $n_\Omega$$(=\pm 1)$,
\begin{align}
\sin(n_\Omega \phi)=n_\Omega \sin\phi,
\quad
\cos(n_\Omega \phi)=\cos\phi,
\end{align}
the spatial derivatives in the Cartesian representation with $(\xi_x, \xi_y)$ can be rewritten in the polar representation with $(r, \phi)$ as,
\begin{align}
&\frac{\partial}{\partial \xi_x}
=\cos(n_\Omega \phi) \frac{\partial}{\partial r} 
-\frac{n_\Omega \sin(n_\Omega \phi)}{r} \frac{\partial}{\partial \phi},
\\
&\frac{\partial}{\partial \xi_y}
=n_\Omega \sin(n_\Omega \phi) \frac{\partial}{\partial r} 
+\frac{\cos(n_\Omega \phi)}{r} \frac{\partial}{\partial \phi}.
\end{align}
Using these relations, we obtain the following expressions for the derivatives of $\bm m$ as,
\begin{align}
\frac{\partial \bm m}{\partial \xi_x}
&=\cos(n_\Omega \phi) \frac{\partial \Theta}{\partial r} \frac{\partial \bm m}{\partial \Theta}
-\frac{n_\Omega \sin(n_\Omega \phi)}{r} 
\frac{\partial \Phi}{\partial\phi} \frac{\partial \bm \Omega}{\partial \Phi}
\notag \\
&=\frac{\pi \zeta_{\rm m}}{r_{\rm sk}}\cos(n_\Omega \phi)
\begin{pmatrix}
 \cos{\Theta}\cos{\Phi}\\
 \cos{\Theta}\sin{\Phi}\\
-\sin{\Theta}
\end{pmatrix}
-\frac{\sin(n_\Omega \phi)}{r}
\begin{pmatrix}
-\sin{\Theta}\sin{\Phi}\\
 \sin{\Theta}\cos{\Phi}\\
 0
\end{pmatrix} 
\\
\frac{\partial \bm m}{\partial \xi_y}
&=n_\Omega \sin(n_\Omega \phi) \frac{\partial \Theta}{\partial r} \frac{\partial \bm m}{\partial \Theta}
+\frac{\cos(n_\Omega \phi)}{r} 
\frac{\partial \Phi}{\partial\phi} \frac{\partial \bm \Omega}{\partial \Phi}
\notag \\
&=n_\Omega \frac{\pi \zeta_{\rm m}}{r_{\rm sk}}\sin(n_\Omega \phi)
\begin{pmatrix}
 \cos{\Theta}\cos{\Phi}\\
 \cos{\Theta}\sin{\Phi}\\
-\sin{\Theta}
\end{pmatrix}
+n_\Omega \frac{\cos(n_\Omega \phi)}{r}
\begin{pmatrix}
-\sin{\Theta}\sin{\Phi}\\
 \sin{\Theta}\cos{\Phi}\\
 0
\end{pmatrix}.
\end{align}
Because the spatial derivatives of $\bm m$ should be zero outside of the skyrmion area, we restrict the interval of integral for the radial direction to $0 \le r \le r_{\rm sk}$ as,
\begin{align}
\displaystyle \int_{\rm UC} dxdy 
\quad \rightarrow \quad 
\int_{0}^{r_{\rm sk}} \int_{0}^{2\pi} rdrd\phi.
\end{align}

The components of the dissipation tensor $D_{\mu\nu}$ are given by,
\begin{align}
D_{\mu\nu}=\int_{\rm UC} 
\frac{\partial \bm m}{\partial \xi_\mu} \cdot
\frac{\partial \bm m}{\partial \xi_\nu} dxdy
=\int_{0}^{r_{\rm sk}} \int_{0}^{2\pi}
\frac{\partial \bm m}{\partial \xi_\mu} \cdot
\frac{\partial \bm m}{\partial \xi_\nu}
rdrd\phi.
\end{align}
After the spatial integration, we obtain,
\begin{align}
D_{xx}&=\int_{0}^{r_{\rm sk}} \int_{0}^{2\pi} \left( 
\frac{\partial \bm m}{\partial \xi_x} \cdot
\frac{\partial \bm m}{\partial \xi_x} \right)
rdrd\phi = \mathcal{D},
\\
D_{yy}&=\int_{0}^{r_{\rm sk}} \int_{0}^{2\pi} \left( 
\frac{\partial \bm m}{\partial \xi_y} \cdot
\frac{\partial \bm m}{\partial \xi_y} \right)
rdrd\phi = \mathcal{D},
\\
D_{xy}&=\int_{0}^{r_{\rm sk}} \int_{0}^{2\pi} \left( 
\frac{\partial \bm m}{\partial \xi_x} \cdot
\frac{\partial \bm m}{\partial \xi_y} \right)
rdrd\phi = 0,
\\
D_{yx}&=\int_{0}^{r_{\rm sk}} \int_{0}^{2\pi} \left( 
\frac{\partial \bm m}{\partial \xi_y} \cdot
\frac{\partial \bm m}{\partial \xi_x} \right)
rdrd\phi = 0,
\end{align}
where
\begin{align}
\mathcal{D}=\frac{\pi}{2} \left\{
\pi^2 + {\rm log}(2\pi) + C -{\rm Ci}(2\pi) 
\right\} \approx 6.15363 \pi.
\end{align}
Here $C$ is the Euler-Mascheroni constant, and ${\rm Ci}(x)$ is the cosine integral given by,
\begin{align}
{\rm Ci}(x)=-\int_x^\infty \dfrac{\cos t}{t}dt.
\end{align}

On the other hand, the quantities $J_\mu^{\rm spin}$ associated with the spin-orbit torque are given by,
\begin{align}
J_{\mu}^{\rm spin}=-\frac{v_{\rm s}^\perp}{d}\bm \sigma \cdot
\int_{\rm UC} 
\left( \frac{\partial \bm m}{\partial \xi_\mu} \times \bm m \right) rdrd\phi
\equiv -\frac{v_{\rm s}^\perp}{d} \mathcal{J}_{\mu\nu}^{\rm spin} \sigma_\nu,
\end{align}
where the quantities $\mathcal{J}_{\mu\nu}^{\rm spin}$ are defined by,
\begin{align}
\mathcal{J}_{\mu\nu}^{\rm spin}=\left[\int_{\rm UC} 
\left( \frac{\partial \bm m}{\partial \xi_\mu} \times \bm m \right)
dxdy \right]_\nu
=\left[\int_{0}^{r_{\rm sk}} \int_{0}^{2\pi}
\left( \frac{\partial \bm m}{\partial \xi_\mu} \times \bm m \right)
rdrd\phi \right]_\nu.
\end{align}
After the spatial integration, we obtain,
\begin{align}
\mathcal{J}_{xx}^{\rm spin}&=\left[\int_{0}^{r_{\rm sk}} \int_{0}^{2\pi} \left( 
\frac{\partial \bm m}{\partial \xi_x} \times
\bm m \right)
rdrd\phi \right]_x = \frac{\pi^2 r_{\rm sk}}{2}\zeta_{\rm m} \sin\eta,
\\
\mathcal{J}_{yy}^{\rm spin}&=\left[\int_{0}^{r_{\rm sk}} \int_{0}^{2\pi} \left( 
\frac{\partial \bm m}{\partial \xi_y} \times
\bm m \right)
rdrd\phi \right]_y = \frac{\pi^2 r_{\rm sk}}{2}\zeta_{\rm m} n_\Omega \sin\eta,
\\
\mathcal{J}_{xy}^{\rm spin}&=\left[\int_{0}^{r_{\rm sk}} \int_{0}^{2\pi} \left( 
\frac{\partial \bm m}{\partial \xi_x} \times
\bm m \right)
rdrd\phi \right]_y =-\frac{\pi^2 r_{\rm sk}}{2}\zeta_{\rm m} \cos\eta,
\\
\mathcal{J}_{yx}^{\rm spin}&=\left[\int_{0}^{r_{\rm sk}} \int_{0}^{2\pi} \left( 
\frac{\partial \bm m}{\partial \xi_y} \times
\bm m \right)
rdrd\phi \right]_x = \frac{\pi^2 r_{\rm sk}}{2}\zeta_{\rm m} n_\Omega \cos\eta.
\end{align}
\end{widetext}
When the vertical spin current is spin polarized in the $\bm y$ direction, i.e., 
$\bm \sigma=(0, \sigma_y, 0)$ with $\sigma_y=\pm 1$, $\bm J^{\rm spin}$ reads,
\begin{eqnarray}
\left(\begin{array}{cc} J^{\rm spin}_x \\ J^{\rm spin}_y
\end{array}
\right)=-\frac{\pi^2v_{\rm s}^\perp r_{\rm sk} \zeta_{\rm m}}{d}\sigma_y
\left(\begin{array}{cc} -\cos\eta \\ n_\Omega \sin\eta
\end{array}
\right).
\label{eq:JFM1a}
\end{eqnarray}
On the other hand, when the vertical spin current is spin polarized in the $\bm x$ direction, i.e., 
$\bm \sigma=(\sigma_x, 0, 0)$ with $\sigma_x=\pm 1$, $\bm J^{\rm spin}$ reads,
\begin{eqnarray}
\left(\begin{array}{cc} J^{\rm spin}_x \\ J^{\rm spin}_y
\end{array}
\right)=-\frac{\pi^2v_{\rm s}^\perp r_{\rm sk} \zeta_{\rm m}}{d}\sigma_x
\left(\begin{array}{cc} \sin\eta \\ n_\Omega \cos\eta
\end{array}
\right).
\label{eq:JFM1b}
\end{eqnarray}
Considering that $n_\Omega=+1$ for both Bloch-type and N\'{e}el-type skyrmions and $\zeta_{\rm m}=-1$ when $B_z>0$, we obtain the following expressions for $\bm J^{\rm spin}$,\\
\underline{-- When  $\bm \sigma \parallel \bm y$:}
\begin{eqnarray}
\left(\begin{array}{cc} J^{\rm spin}_x \\ J^{\rm spin}_y
\end{array}
\right)=\mathcal{J}_{\rm FM}\sigma_y
\left(\begin{array}{cc} -\cos\eta \\ \sin\eta
\end{array}
\right).
\label{eq:JFM2a}
\end{eqnarray}
\underline{-- When  $\bm \sigma \parallel \bm x$:}
\begin{eqnarray}
\left(\begin{array}{cc} J^{\rm spin}_x \\ J^{\rm spin}_y
\end{array}
\right)=\mathcal{J}_{\rm FM}\sigma_x
\left(\begin{array}{cc} \sin\eta \\ \cos\eta
\end{array}
\right),
\label{eq:JFM2b}
\end{eqnarray}
where $\displaystyle \mathcal{J}_{\rm FM}=\frac{\pi^2v_{\rm s}^\perp r_{\rm sk}}{d}(>0)$. Note that $\cos\eta$ vanishes for the Bloch-type skyrmions ($\eta$=$\pi/2$ and $3\pi/2$), whereas $\sin\eta$ vanishes for the N\'{e}el-type skyrmions ($\eta$=0 and $\pi$).

In the case of antiferromagnetic skyrmions, the quantities $\mathcal{D}$ and $\bm J_{\rm spin}$ in the Thiele equation [Eq.~(\ref{eq:ThieleAF2}) or Eq.~(\ref{eq:7-1}] should be those defined using the sublattice magnetization vectors $\bm m_{\rm A}$ or $\bm m_{\rm B}$, both of which should have an identical value according to the discussion in Subsection 4.3. Because the sublattice magnetizations of an antiferromagnetic skyrmion constitute a ferromagnetic skyrmion, the values of $\mathcal{D}$ and $\bm J_{\rm spin}$ in Eq.~(\ref{eq:ThieleAF2}) and Eq.~(\ref{eq:7-1}) are equivalent to those of the ferromagnetic skyrmions. One thing that we should be careful about is that the saturation magnetization $M_{\rm s}$ used in the definition of the normalized vertical spin current $v_{\rm s}^\perp$ in Eq.~(\ref{eq:vs2}) should be replaced with that of the sublattice magnetization $M_{\rm As}$ or $M_{\rm Bs}$. With thus defined $v_{\rm s}^\perp$ and $\displaystyle \mathcal{J}_{\rm AFM}=\frac{\pi^2v_{\rm s}^\perp r_{\rm sk}}{d}(>0)$, the expressions of $\bm J^{\rm spin}$ for the antiferromagnetic skyrmions are given in the same forms as Eq.~(\ref{eq:JFM2a}) and Eq.~(\ref{eq:JFM2b})  with $\mathcal{J}_{\rm FM}$ being replaced with $\mathcal{J}_{\rm AFM}$.

\section{Acknowledgments}
MM thanks Shiho Nakamura and Xiuzhen Yu for fruitful discussions. 
This work was supported by the cooperation of organization between Kioxia Corporation and Waseda University, JSPS KAKENHI (No.~20H00337 and No.~24H02231),  JST CREST (No.~JPMJCR20T1), and 
Waseda University Grant for Special Research Projects (No.~2024C-153).


\begin{thebibliography}{999}
\bibitem{Compagnoni2017}C. M. Compagnoni, A. Goda, A. S. Spinelli, P. Feeley, A. L. Lacaita, and A. Visconti, Proc. IEEE {\bf 105}, 1609-1633 (2017).

\bibitem{SunG2017}G. Sun, J. Zhao, M. Poremba, C. Xu, and Y. Xie, National Sci. Rev. {\bf 5}, 577-592 (2017).

\bibitem{AChen2016}A. Chen, Solid-State Electron. {\bf 125}, 25-38 (2016).

\bibitem{Meena2014}J. Meena, S. Sze, U. Chand, and T.-Y. Tseng, Nanosc. Res. Lett. {\bf 9}, 526 (2014).

\bibitem{Qureshi2009}M. K. Qureshi, V. Srinivasan, and J. A. Rivers, ACM SIGARCH Comput. Archit. News {\bf 37}, 24-33 (2009).
\bibitem{Barla2021}P. Barla, V. K. Joshi, and S. Bhat, J. Comput. Electron. {\bf 20}, 805-837 (2021).

\bibitem{Joshi2016}V. K. Joshi, Eng. Sci. Technol. Int. J. {\bf 19}, 1503-1513 (2016).
\bibitem{Blasing2020}R. Bl\"{a}sing, A. A. Khan, P. Ch. Filippou, C. Garg, F. Hameed, J. Castrillon, and 
S. S. P. Parkin, Proc. IEEE {\bf 108}, 1303-1321 (2020).

\bibitem{Parkin2015}S. Parkin, S. Yang, Nat. Nanotech. {\bf 10}, 195-198 (2015).

\bibitem{Parkin2008}S. S. P. Parkin, M. Hayashi, and L. Thomas, Science {\bf 320}, 190-194 (2008).
\bibitem{Ikegawa2020}S. Ikegawa, F. B. Mancoff, J. Janesky, and S. Aggarwal, IEEE Transactions on Electron Devices {\bf 67}, 1407-1419 (2020).

\bibitem{Bhatti2017}S. Bhatti, R. Sbiaa, A. Hirohata, H. Ohno, S. Fukami, S. N. Piramanayagam, Mater. Today {\bf 20}, 530-548 (2017).
\bibitem{Muhlbauer2009}S. M{\"u}hlbauer, B. Binz, F. Jonietz, C. Pfleiderer, A. Rosch,  A. Neubauer, R. Georgii, and P. B{\"o}ni, Science {\bf 323}, 915 (2009).

\bibitem{YuXZ2010}X. Z. Yu, Y. Onose, N. Kanazawa, J. H. Park, J. H. Han, Y. Matsui, N. Nagaosa, and Y. Tokura, Nature {\bf 465}, 901 (2010).

\bibitem{SekiBook2016}S. Seki and M. Mochizuki, {\it Skyrmions in Magnetic Materials} (Springer, Berlin, 2016).

\bibitem{Nagaosa2013}N. Nagaosa and Y. Tokura, Nature Nanotech. {\bf 8}, 899 (2013).

\bibitem{SeidelBook2016}{\it Topological Structures in Ferroic Materials: Domain Walls, Skyrmions and Vortices} (Springer, 2016) edited by J. Seidel, Chapter 3, ``Current-Driven Dynamics of Skyrmions" pp. 55-81, by M. Mochizuki.

\bibitem{Fert2013}A. Fert, V. Cros, and J. Sampaio, Nature Nanotech. {\bf 8}, 152 (2013).
\bibitem{Koshibae2015}W. Koshibae, Y. Kaneko, J. Iwasaki, M. Kawasaki, Y. Tokura, and N. Nagaosa, Jpn. J. Appl. Phys. {\bf 54}, 053001 (2015).

\bibitem{Finocchio2016}G. Finocchio, F. Buttner, R. Tomasello, M. Carpentieri, and M. Kl\"aui, J. Phys. D: Appl. Phys. {\bf 49}, 423001 (2016).

\bibitem{KangW2016}W. Kang, Y. Huang, X. Zhang, Y. Zhou, and W. Zhao, Proc. IEEE {\bf 104}, 2040 (2016).

\bibitem{Fert2017}A. Fert, N. Reyren, and V. Cros, Nat. Rev. Mater. {\bf 2}, 17031 (2017).

\bibitem{XichaoZ2020}X. Zhang, Y. Zhou, K. M. Song, T.-E. Park, J. Xia, M. Ezawa, X. Liu, W. Zhao, G. Zhao, and S. Woo, J. Phys. Condens. Matter {\bf 32}, 143001 (2020).
DOI: https://doi.org/10.1088/1361-648X/ab5488
\bibitem{Jonietz2010}F. Jonietz, S. M\"uhlbauer, C. Pfleiderer, A. Neubauer, W. M\"unzer, A. Bauer, T. Adams, R. Georgii, P. B\"oni, R. A. Duine, K. Everschor, M. Garst, and A. Rosch, Science {\bf 330}, 1648 (2010).

\bibitem{YuXZ2012}X. Z. Yu, N. Kanazawa, W. Z. Zhang, T. Nagai, T. Hara, K. Kimoto, Y. Matsui, Y. Onose, and Y. Tokura, Nat. Commun. {\bf 3}, 988 (2012).
\bibitem{Iwasaki2013a}J. Iwasaki, M. Mochizuki, and N. Nagaosa, Nat. Commun. {\bf 4}, 1463 (2013).

\bibitem{Iwasaki2013b}J. Iwasaki, M. Mochizuki, and N. Nagaosa, Nat. Nanotech. {\bf 8}, 742 (2013).

\bibitem{Sampaio2013}J. Sampaio, V. Cros, S. Rohart, A. Thiaville, and A. Fert, Nat. Nanotech. {\bf 8}, 839 (2013).

\bibitem{Iwasaki2014a}J. Iwasaki, W. Koshibae, and N. Nagaosa, Nano Lett. 14, 4432 (2014).



\bibitem{WooS2016}S. Woo, K. Litzius, B. Kr\"{u}ger, M.-Y. Im, L. Caretta, K. Richter, M. Mann, A. Krone, R. M. Reeve, M. Weigand, P. Agrawal, I. Lemesh, M.-A. Mawass, P. Fischer, M. Kl\"{a}ui, and G. S. D. Beach, Nat. Mater. {\bf 15}, 501-506 (2016).

\bibitem{ZhangX2017}X. Zhang, J. Xia, Y. Zhou, X. Liu, H. Zhang, and M. Ezawa, Nat. Commun. {\bf 8}, 1717 (2017).

\bibitem{Litzius2020}K. Litzius, J. Leliaert, P. Bassirian, D. Rodrigues, S. Kromin, I. Lemesh, J. Zazvorka, K.-J. Lee, J. Mulkers, N. Kerber, D. Heinze, N. Keil, R. M. Reeve, M. Weigand, B. V. Waeyenberge, G. Sch\"{u}tz, K. Everschor-Sitte, G. S. D. Beach, and M. Kl\"{a}ui, Nat. Electronics {\bf 3}, 30-36 (2020).

\bibitem{Reichhardt2022}C. Reichhardt, C.J.O. Reichhardt, and M.V. Milo\v{s}evi\'{c}, Rev. Mod. Phys. 94, 035005 (2022).

\bibitem{ZhangX2023a}X. Zhang, J. Xia, O. A. Tretiakov, M. Ezawa, G. Zhao, Y. Zhou, X. Liu, and M. Mochizuki, Phys. Rev. B {\bf 108}, 144428 (2023).

\bibitem{Pham2024}V. T. Pham, N. Sisodia, I. Di Manici, J. Urrestarazu-Larranaga, K. Bairagi, J. Pelloux-Prayer, R. Guedas, L. D. Buda-Prejbeanu, S. Auffret, A. Locatelli, T. O. Mentes, S. Pizzini, P. Kumar, A. Finco, V. Jacques, G. Gaudin, O. Boulle, Science {\bf 384}, 307-312 (2024).
\bibitem{Tomasello2014}R. Tomasello, E. Martinez, R. Zivieri, L. Torres, M. Carpentieri, and G. Finocchio, Sci. Rep. {\bf 4}, 6784 (2014).

\bibitem{ZhangX2015}X. Zhang, G. P. Zhao, H. Fangohr, J. P. Liu, W. X. Xia, J. Xia, and F. J. Morvan, Sci. Rep. {\bf 5}, 7643 (2015).

\bibitem{YuG2017}G. Yu, P. Upadhyaya, Q. Shao, H. Wu, G. Yin, X. Li, C. He, W. Jiang, X. Han, P. K. Amiri, and K. L. Wang, Nano Lett. {\bf 17}, 261 (2017).

\bibitem{Maccariello2018}D. Maccariello, W. Legrand, N. R. S. Collin, V. Cros, and A. Fert, Nat. Nanotech. {\bf 13}, 233-237 (2018).

\bibitem{ZhuD2018}D. Zhu, W. Kang, S. Li, Y. Huang, X. Zhang, Y. Zhou, and W. Zhao, IEEE Trans. Electron Devices {\bf 65}, 87 (2018).

\bibitem{HeB2023}B. He, R. Tomasello, X. Luo, R. Zhang, Z. Nie, M. Carpentieri, X. Han, G. Finocchio, and G. Yu, Nano Lett. {\bf 23}, 9482-9490 (2023).
\bibitem{ZangJ2011}J. Zang, M. Mostovoy, J. H. Han, and N. Nagaosa, Phys. Rev. Lett. {\bf 107}, 136804 (2011).

\bibitem{Everschor-Sitte2014}K. Everschor-Sitte, and M. Sitte, J. Appl. Phys. {\bf 115}, 172602 (2014).

\bibitem{JiangW2017}W. Jiang, X. Zhang, G. Yu, W. Zhang, X, Wang, M. B, Jungfleisch, J, E. Pearson, X, Cheng, O, Heinonen, K, L. Wang, Y, Zhou, A. Hoffmann, and S. G. E. te Velthuis, Nature Phys. {\bf 13}, 162 (2017).

\bibitem{Litzius2017}K. Litzius, I. Lemesh, B. Kr\"{u}ger, P. Bassirian, L. Caretta, K. Richter, F. B\"{u}ttner, K. Sato, O. A. Tretiakov, J. Forster, R. M. Reeve, M. Weigand, I. Bykova, H. Stoll, G. Schutz, G. S. D. Beach, and M. Kla\"{u}i, Nature Phys. {\bf 13}, 170 (2017).
\bibitem{XingX2020}X. Xing, J. {\AA}kerman, and Y. Zhou, Phys. Rev. B {\bf 101}, 214432 (2020).

\bibitem{SongM2020}M. Song, K.-W. Moon, S. Yang, C. Hwang, and K.-J. Kim, Appl. Phys. Express {\bf 13}, 063002 (2020).
\bibitem{Muller2017a}J. M\"{u}ller, J. Rajeswari, P. Huang, Y. Murooka, H. M. R{\o}nnow, F. Carbone, and A. Rosch, Phys. Rev. Lett. 119, 137201 (2017).

\bibitem{Knapman2021}R. Knapman, D. R. Rodrigues, J. Masell, and K. Everschor-Sitte, J. Phys. D: Appl. Phys. {\bf 54} 404003 (2021).

\bibitem{HeZ2024}Z. He, Z. Li, Z. Chen, Z. Wang, J. Shen, S. Wang, C. Song, T. Zhao, J. Cai, S.-Z. Lin, Y. Zhang, and B.Shen, Nat. Mater. {\bf } (2024).
\bibitem{Purnama2015}I. Purnama, W. L. Gan, D. W. Wong, and W. S. Lew, 
Sci. Rep. {\bf 5}, 10620 (2015).

\bibitem{Muller2017b}J. M\"{u}ller, New J. Phys. 19, 025002 (2017).

\bibitem{LaiP2017}P. Lai, G. P. Zhao, H. Tang, N. Ran, S. Q. Wu, J. Xia, X. Zhang, and Y. Zhou, 
Sci. Rep. {\bf 7}, 45330 (2017).

\bibitem{CaiN2021}N. Cai, and Y. Liu, J. Phys. Appl. Phys. {\bf 54}, 125001 (2021).

\bibitem{Toscano2020}D. Toscano, J. P. A. Mendonca, A. L. S. Miranda, C. I. L. de Araujo, F. Sato, P. Z. Coura, and S. A. Leonel, J. Magn. Magn. Mater. {\bf 504}, 166655 (2020).

\bibitem{Kern2022}L.-M. Kern, B. Pfau, V. Deinhart, M. Schneider, C. Klose, K. Gerlinger, S. Wittrock, D. Engel, I. Will, C. M. G\"{u}nther, R. Liefferink, J. H. Mentink, S. Wintz, M. Weigand, M.-J. Huang, R. Battistelli, D. Metternich, F. B\"{u}ttner, K. H\"{o}flich, and S. Eisebitt, Nano Lett. {\bf 22}, 4028 (2022).
\bibitem{ZhangX2021}X. Zhang, J. Xia, K. Shirai, H. Fujiwara, O. A. Tretiakov, M. Ezawa, Y. Zhou, and X. Liu, Commun. Phys. {\bf 4}, 255 (2021).

\bibitem{ZhaoL2024}L. Zhao, C. Hua, C. Song, W. Yu, W. Jiang, Sci. Bull., In Press (2024).
\bibitem{Barker2016}J. Barker and O. A. Tretiakov, Phys. Rev. Lett. {\bf 116}, 147203 (2016).

\bibitem{ZhangX2016a}X. Zhang, Y. Zhou, and M. Ezawa, Sci. Rep. {\bf 6}, 24795 (2016).

\bibitem{Gobel2017}B. G\"{o}obel, A. Mook, J. Henk, and I. Mertig, Phys. Rev. B {\bf 96}, 060406(R) (2017).

\bibitem{Akosa2018}C. A. Akosa, O. A. Tretiakov, G. Tatara, and A. Manchon, Phys. Rev. Lett. {\bf 121}, 097204 (2018).
\bibitem{ZhangX2016b}X. Zhang, Y. Zhou, and M. Ezawa, Nature Commun. {\bf 7}, 10293 (2016).

\bibitem{XiaJ2019}J. Xia, X. Zhang, M. Ezawa, Z. Hou, W. Wang, X. Liu, and Y. Zhou, Phys. Rev. Appl. {\bf 11}, 044046 (2019).

\bibitem{Dohi2019}T. Dohi, S. DuttaGupta, S. Fukami, and H. Ohno, Nat. Commun. {\bf 10}, 5153 (2019).

\bibitem{MaM2022}M. Ma, K. Huang, Y. Li, S. Li, Q. Feng, C. C. I. Ang, T. Jin, Y. Lu, Q. Lu, W.S. Lew, F. Ma, and X. R. Wang, Appl. Phys. Rev. {\bf 9}, 021404 (2022).

\bibitem{ChenR2022}R. Chen, Q. Cui, L. Han, X. Xue, J. Liang, H. Bai, Y. Zhou, F. Pan, H. Yang, and C. Song, Adv. Funct. Mater. {\bf 32}, 2111906 (2022).

\bibitem{Panigrahy2022}S. Panigrahy, S. Mallick, J. Sampaio, and S. Rohart, Phys Rev. B {\bf 106}, 144405 (2022).
\bibitem{WooS2018}S. Woo, K. M. Song, X. Zhang, Y.Zhou, M. Ezawa, X. Liu, S. Finizio, J. Raabe, N. J. Lee, S.-I. Kim, S.-Y. Park, Y. Kim, J.-Y. Kim, D. Lee, O. Lee, J. W. Choi, B.-C. Min, H. C. Koo, and J. Chang, Nat. Commun. {\bf 9}, 959 (2018).

\bibitem{Hirata2019}Y. Hirata, D.-H. Kim, S. K. Kim, D.-K. Lee, S.-H. Oh, D.-Y. Kim, T. Nishimura, T. Okuno, Y. Futakawa, H. Yoshikawa, A. Tsukamoto, Y. Tserkovnyak, Y. Shiota, T. Moriyama, S.-B. Choe, K.-J. Lee, and T. Ono, Nat. Nanotech. {\bf 14}, 232-236 (2019).

\bibitem{LiuY2023}Y. Liu, T. T. Liu, Z. P. Hou, D. Y. Chen, Z. Fan, M. Zeng, X.B. Lu, X. S. Gao, M. H. Qin, and J.-M. Liu, Appl. Phys. Lett. {\bf 122}, 172405 (2023).

\bibitem{XuT2023}T. Xu, Y. Zhang, Z. Wang, H. Bai, C. Song, J. Liu, Y. Zhou, S.-G. Je, A. T. N' Diaye, M.-Y. Im, R. Yu, Z. Chen, and W. Jiang, ACS Nano {\bf 17}, 7920-7928 (2023).

\bibitem{BoL2024a}L. Bo, X. Zhang, M. Mochizuki, and X. Zhang, Phys. Rev. Res. {\bf 6}, 023199 (2024).

\bibitem{Silva2024}R. C. Silva, R. L. Silva, J. C. Moreira, W. A. Moura-Melo, and A. R. Pereira, J. Appl. Phys. {\bf 135}, 183902 (2024).
\bibitem{Gobel2019}B. G\"{o}bel, A. Mook, J. Henk, and I. Mertig, Phys. Rev. B {\bf 99}, 020405(R) (2019).

\bibitem{Akosa2019}C. A. Akosa, H. Li, G. Tatara, and O. A. Tretiakov, Phys. Rev. Appl. {\bf 12}, 054032 (2019).
\bibitem{Thiele1973}A. A. Thiele, Phys. Rev. Lett. {\bf 30}, 230 (1973).

\bibitem{Everschor2012}K. Everschor, M. Garst, B. Binz, F. Jonietz, S. M\"uhlbauer, C. Pfleiderer, and A. Rosch, Phys. Rev. B {\bf 86}, 054432 (2012).

\bibitem{Schulz2012}T. Schulz, R. Ritz, A. Bauer, M. Halder, M. Wagner, C. Franz, C. Pfleiderer, K. Everschor, M. Garst, and A. Rosch, Nat. Phys. {\bf 8}, 301-304 (2012).
\bibitem{Dzyaloshinskii1957}I. E. Dzyaloshinsky, Sov. Phys. JETP {\bf 5}, 1259-1262 (1957).

\bibitem{Moriya1960a}T. Moriya, Phys. Rev. {\bf 120}, 91-98 (1960).

\bibitem{Moriya1960b}T. Moriya, Phys. Rev. Lett. {\bf 4}, 228 (1960).

\bibitem{Fert1980}A. Fert, and P. A. Levy, Phys. Rev. Lett. {\bf 44}, 1538-1541 (1980).
\bibitem{Bogdanov1989}A. N. Bogdanov, and D. A. Yablonskii, Sov. Phys. JETP {\bf 68}, 101-103 (1989).

\bibitem{Bogdanov1994}A. Bogdanov, and A. Hubert, J. Magn. Magn. Mater. {\bf 138}, 255-269 (1994).

\bibitem{Munzer2010}W. M\"{u}nzer, A. Neubauer, T. Adams, S. M\"{u}hlbauer, C. Franz, F. Jonietz, R. Georgii, P. B\"{o}ni, B. Pedersen, M. Schmidt, A. Rosch, and C. Pfleiderer, Phys. Rev. B {\bf 81}, 041203(R) (2010).

\bibitem{YuXZ2011}X. Z. Yu, N. Kanazawa, Y. Onose, K. Kimoto, W. Z. Zhang, S. Ishiwata, Y. Matsui, and Y. Tokura, Nature Mater. {\bf 10}, 106 (2011).

\bibitem{Seki2012a}S. Seki, X. Z. Yu, S. Ishiwata, and Y. Tokura, Science {\bf 336}, 198 (2012).

\bibitem{Seki2012b}S. Seki, S. Ishiwata, and Y. Tokura, Phys. Rev. B {\bf 86}, 060403(R) (2012).

\bibitem{Tokunaga2015}Y. Tokunaga, X. Z. Yu, J. S. White, H. M. Ronnow, D. Morikawa, Y. Taguchi, and Y. Tokura, Nat. Commun. {\bf 6}, 7638 (2015).

\bibitem{Karube2017}K. Karube, J. S. White, D. Morikawa, M. Bartkowiak, A. Kikkawa, Y. Tokunaga, T. Arima, H. M. R{\o}nnow, Y. Tokura, and Y. Taguchi, Phys. Rev. Mater. {\bf 1}, 074405 (2017).
\bibitem{Kezsmarki2015}I. K\'{e}zsm\'{a}rki, S. Bord\'{a}cs, P. Milde, E. Neuber, L. M. Eng, J. S. White, H. M. R{\o}nnow, C. D. Dewhurst, M. Mochizuki, K. Yanai, H. Nakamura, D. Ehlers, V. Tsurkan, and A. Loidl, Nat. Mater. {\bf 14}, 1116 (2015).

\bibitem{Fujima2017}Y. Fujima, N. Abe, Y. Tokunaga, and T. Arima, Phys. Rev. B {\bf 95}, 180410(R) (2017).

\bibitem{Kurumaji2017}T. Kurumaji, T. Nakajima, V. Ukleev, A. Feoktystov, T.-h. Arima, K. Kakurai, and Y. Tokura, Phys. Rev. Lett. {\bf 119}, 237201 (2017).

\bibitem{Kurumaji2021}T. Kurumaji, T. Nakajima, A. Feoktystov, E. Babcock, Z. Salhi, V. Ukleev, T.-h. Arima, K. Kakurai, and Y. Tokura, J. Phys. Soc. Jpn. {\bf 90}, 024705 (2021).
\bibitem{ChenG2015}G. Chen, A. Mascaraque, A. T. N'Diaye, and A. K. Schmid, Appl. Phys. Lett. {\bf 106}, 242404 (2015).

\bibitem{Moreau-Luchaire2016}C. Moreau-Luchaire, C. Moutafis, N. Reyren, J. Sampaio, C. A. F. Vaz, N. Van Horne, K. Bouzehouane, K. Garcia, C. Deranlot, P. Warnicke, P. Wohlh\"uter, J.-M. George, M. Weigand, J. Raabe, V. Cros, and A. Fert, Nat. Nanotech. {\bf 11}, 444 (2016).

\bibitem{Soumyanarayanan2017}A. Soumyanarayanan, M. Raju, A. L. G. Oyarce, A. K. C. Tan, M.-Y. Im, A. P. Petrovi\'c, P. Ho, K. H. Khoo, M. Tran, C. K. Gan, F. Ernult, and C. Panagopoulos, Nat. Mater. {\bf 16}, 898-904 (2017).
\bibitem{Romming2013}N. Romming, C. Hanneken, M. Menzel, J. E. Bickel, B. Wolter, K. von Bergmann, A. Kubetzka, and R. Wiesendange, Science {\bf 341}, 636 (2013).

\bibitem{Heinze2011}S. Heinze, K. von Bergmann, M. Menzel, J. Brede, A. Kubetzka, R. Wiesendanger, G. Bihlmayer, and Stefan Bl\"{u}gel, Nat. Phys. {\bf 7}, 713-718 (2011).

\bibitem{Wiesendanger2016}R. Wiesendanger, Nat. Rev. Mater. {\bf 1}, 16044 (2016).
\bibitem{Nayak2017}A. K Nayak, V. Kumar, T. Ma, P. Werner, E. Pippel, R. Sahoo, F. Damay, U. K R\"{o}{\ss}ler, C. Felser, and S. S. P. Parkin, Nature 548, 561-566 (2017).

\bibitem{Karube2021}K. Karube, L. C. Peng, J. Masell, X.Z. Yu, F. Kagawa, Y. Tokura, and Y. Taguchi, Nat. Mater. {\bf 20}, 335 (2021).

\bibitem{Karube2022}K. Karube, L. C. Peng, J. Masell, M. Hemmida, H.-A. Krug von Nidda, I. K\'ezsm\'arki, X. Z. Yu, Y. Tokura, and Y. Taguchi, Adv. Mater. {\bf 34}, 2108770 (2022).
\bibitem{Brataas2012}A. Brataas, A. D. Kent, and H. Ohno, Nat Mater. {\bf 11}, 372-381 (2012).

\bibitem{Slonczewski1996}J. C. Slonczewski, J. Magn. Magn. Mater. {\bf 159}, L1-L7 (1996).

\bibitem{ZhangLi2004}S. Zhang and Z. Li, Phys. Rev. Lett. {\bf 93}, 127204 (2004).

\bibitem{Tatara2004}G. Tatara, and H. Kohno, Phys. Rev. Lett. {\bf 92}, 086601 (2004).

\bibitem{Manchon2019}A. Manchon, J. \v{Z}elezn\'{y}, I. M. Miron, T. Jungwirth, J. Sinova, A. Thiaville, K. Garello, and P. Gambardella, Rev. Mod. Phys. {\bf 91}, 035004 (2019).

\bibitem{ShaoQ2021}Q. Shao, P. Li, L. Liu, H. Yang, S. Fukami, A. Razavi, H. Wu, K. Wang, F. Freimuth, Y. Mokrousov, M. D. Stiles, S. Emori, A. Hoffmann, J. {\AA}kerman, K. Roy, J.-P. Wang, S.-H. Yang, K. Garello, and W. Zhang, IEEE Trans. Magn. {\bf 57}, 1-39 (2021).

\bibitem{KimKW2024}K.-W. Kim, B.-G. Park, and K.-J. Lee, npj Spintronics {\bf 2}, 8 (2024).


\bibitem{Makhfudz2012}I. Makhfudz, B. Kr\"{u}ger, and O. Tchernyshyov, Phys. Rev. Lett. {\bf 109}, 217201 (2012).

\bibitem{Tveten2013}E. G. Tveten, A. Qaiumzadeh, O. A. Tretiakov, and A. Brataas, Phys. Rev. Lett. {\bf 110}, 127208 (2013).

\bibitem{Schutte2014b}C. Sch\"{u}tte, J. Iwasaki, A. Rosch, and N. Nagaosa, Phys. Rev. B {\bf 90}, 174434 (2014).

\bibitem{ZhangX2017b}X. Zhang, J. Xia, G. P. Zhao, X. Liu, and Y. Zhou, IEEE Transactions on Magnetics {\bf 53}, 1500206 (2017).

\bibitem{LuoJ2023}J. Luo, J.-H. Guo, Y.-H. Hou, J.-L. Wang, Y.-B. Xu, Y. Zhou, P. W. T. Pong, and G.-P. Zhao, Chin. Phys. Lett. {\bf 40}, 097501 (2023).

\bibitem{ShenL2019a}L. Shen, X. Li, Y. Zhao, J. Xia, G. Zhao, and Y. Zhou, Phys. Rev. Appl. {\bf 12}, 064033 (2019).

\bibitem{ShenL2019b}L. Shen, J. Xia, G. Zhao, X. Zhang, M. Ezawa, O. A. Tretiakov, X. Liu, and Y. Zhou, Appl. Phys. Lett. {\bf 114}, 042402 (2019).

\bibitem{Bak1980}P. Bak, and M. H. Jensen, J. Phys. C {\bf 13}, L881 (1980).

\bibitem{YiSD2009}S. D. Yi, S. Onoda, N. Nagaosa, and J. H. Han, Phys. Rev. B {\bf 80}, 054416 (2009).

\bibitem{Buhrandt2013}S. Buhrandt and L. Fritz, Phys. Rev. B {\bf 88}, 195137 (2013).

\bibitem{Mochizuki2012}M. Mochizuki, Phys. Rev. Lett. {\bf 108}, 017601 (2012).


\bibitem{Stosic2017}D. Stosic, T. B. Ludermir, and M. V. Milo\v{s}evi\'{c}, Phys. Rev. B {\bf 96}, 214403 (2017).

\bibitem{LiangX2019}X. Liang, G. Zhao, L. Shen, J. Xia, L. Zhao, X. Zhang, and Y. Zhou, Phys. Rev. B {\bf 100}, 144439 (2019).

\bibitem{Reichhardt2015a}C. Reichhardt, D. Ray, and C. J. O. Reichhardt, Phys. Rev. Lett. {\bf 114}, 217202 (2015).

\bibitem{Reichhardt2015b}C. Reichhardt, D. Ray, and C. J. O. Reichhardt, New J. Phys. {\bf 17}, 073034 (2015).

\bibitem{MaF2016}F. Ma, C. Reichhardt, W. Gan, C. J. O. Reichhardt, and W. S. Lew, Phys. Rev. B {\bf 94}, 144405 (2016).

\bibitem{Fernandes2018}I. L. Fernandes, J. Bouaziz, S. Blugel, and S. Lounis, Nat. Commun. {\bf 9}, 4395 (2018).

\bibitem{Menezes2019}R. M. Menezes, J. Mulkers, C. C. de Souza Silva, and M. V. Milosevic, Phys. Rev. B {\bf 99}, 104409 (2019).

 \bibitem{Gonzalez-Gomez2019}L. Gonzalez-Gomez, J. Castell-Queralt, N. Del-Valle, A. Sanchez, and C. Navau, Phys. Rev. B {\bf 100}, 054440 (2019).

\bibitem{Castell-Queral2019}J. Castell-Queralt, L. Gonzalez-Gomez, N. Del-Valle, A. Sanchez, and C. Navau, Nanoscale {\bf 11}, 12589 (2019).

\bibitem{Bhatti2019}S. Bhatti and S. N. Piramanayagam, Phys. Status Solidi RRL {\bf 13}, 1900090 (2019).

\bibitem{Feilhauer2020}J. Feilhauer, S. Saha, J. Tobik, M. Zelent, L. J. Heyderman, and M. Mruczkiewicz, Phys. Rev. B {\bf 102}, 184425 (2020).

\bibitem{Fernandes2020}I. L. Fernandes, J. Chico, and S. Lounis, J. Phys.: Condens. Matter {\bf 32}, 425802 (2020).

\bibitem{Arjana2020}I. G. Arjana, I. L. Fernandes, J. Chico, and S. Lounis, Sci. Rep. {\bf 10}, 14655 (2020).

\bibitem{ChenW2020a}W. Chen, L. Liu, and Y. Zheng, Phys. Rev. Appl. {\bf 14}, 064014 (2020).

\bibitem{Vizarim2021}N. P. Vizarim, J. C. Bellizotti Souza, C. Reichhardt, C. J. O. Reichhardt, and P. A. Venegas, J. Phys.: Condens. Matter {\bf 33}, 305801 (2021).

\bibitem{SouzaJCB2021}J. C. Bellizotti Souza, N. P. Vizarim, C. J. O. Reichhardt, C. Reichhardt, and P. A. Venagas, Phys. Rev. B {\bf 104}, 054434 (2021).

\bibitem{Del-Valle2022}N. Del-Valle, J. Castell-Queralt, L. Gonzalez-Gomez, and C. Navau, APL Mater. {\bf 10}, 010702 (2022).

\bibitem{SouzaJCB2022}J. C. Bellizotti Souza, N. P. Vizarim, C. J. O. Reichhardt, C. Reichhardt, and P. A. Venegas, New J. Phys. {\bf 24}, 103030 (2022).

\bibitem{ZhangX2022}X. Zhang, J. Xia, and X. Liu, Phys. Rev. B {\bf 106}, 094418 (2022).

\bibitem{Rocha2024}F. S. Rocha, J. C. Bellizotti Souza, N. P. Vizarim, C. J. O. Reichhardt, and C. Reichhardt, J. Phys. Condens Matter {\bf 36}, 115801 (2024).

\bibitem{SouzaJCB2024}J. C. Bellizotti Souza, N. P. Vizarim, C. J. O. Reichhardt, C. Reichhardt, and P. A. Venegas, Phys. Rev. B {\bf 109}, 054407 (2024).
\bibitem{Juge2021}R. Juge, K. Bairagi, K. G. Rana, J. Vogel, M. Sall, D. Mailly, V. T. Pham, Q. Zhang, N. Sisodia, M. Foerster, L. Aballe, M. Belmeguenai, Y. Roussign\'{e}, S. Auffret, L. D. Buda-Prejbeanu, G. Gaudin, D. Ravelosona, and O. Boulle, Nano Lett. {\bf 21}, 2989 (2021).

\bibitem{Ohara2021}K. Ohara, X. Zhang, Y. Chen, Z. Wei, Y. Ma, J. Xia, Y. Zhou, and X. Liu, Nano Lett. {\bf 21}, 4320 (2021).

\bibitem{Jong2022}M. C. H. de Jong, M. J. Meijer, J. Lucassen, J. van Liempt, H. J. M. Swagten, B. Koopmans, and R. Lavrijsen, Phys. Rev. B {\bf 105}, 064429 (2022).
\bibitem{TchoeY2012}Y. Tchoe, and J. H. Han, Phys. Rev. B {\bf 85}, 174416 (2012).



\bibitem{Yuan2016}H. Y. Yuan, and X. R. Wang, Sci. Rep. {\bf 6}, 22638 (2016).

\bibitem{YinG2016}G. Yin, Y. Li, L. Kong, R. K. Lake, C. L. Chien, and J. Zang, Phys. Rev. B {\bf 93}, 174403 (2016).

\bibitem{Everschor-Sitte2017}K. Everschor-Sitte, M. Sitte, T. Valet, A. Abanov, and J. Sinova, New J. Phys. {\bf 19}, 092001 (2017).

\bibitem{Hrabec2017}A. Hrabec, J. Sampaio, M. Belmeguenai, I. Gross, R. Weil, S.M. Ch\'{e}rif, A. Stashkevich, V. Jacques, A. Thiaville, and S. Rohart, Nat. Commun. {\bf 8}, 15765 (2017).

\bibitem{YuXZ2017}X. Yu, D. Morikawa, Y. Tokunaga, M. Kubota, T. Kurumaji,H. Oike, M. Nakamura, F. Kagawa, Y. Taguchi, T.-h. Arima, M. Kawasaki, and Y. Tokura, Adv. Mater. {\bf 29}, 1606178 (2017).

\bibitem{YuXZ2020}X. Z. Yu, D. Morikawa, K. Nakajima, K. Shibata, N. Kanazawa, T. ArimaN. Nagaosa, and Y. Tokura, Sci. Adv. {\bf 6}, eaaz9744 (2020).

\bibitem{Fujimoto2021}J. Fujimoto, W. Koshibae, M. Matsuo, and S. Maekawa, Phys. Rev. B {\bf 103}, L220402 (2021).

\bibitem{Fujimoto2022}J. Fujimoto, H. Funaki, W. Koshibae, M. Matsuo, and S. Maekawa, IEEE Trans. Mag. {\bf 58}, 1500407 (2022).

\bibitem{WangW2022}W. Wang, D. Song, W. Wei, P. Nan, S. Zhang, B. Ge, M. Tian, J. Zang, and H. Du, Nat. Commun. {\bf 13}, 1593 (2022).
\bibitem{Buttner2017}F. B\"{u}ttner, I. Lemesh, M. Schneider, B. Pfau, C. M. G\"{u}nther, P. Hessing, J. Geilhufe, L. Caretta, D. Engel, B. Kr\"{u}ger, J. Viefhaus, S. Eisebitt, and G. S. D. Beach, Nature Nanotech. {\bf 12}, 1040-1044 (2017).
\bibitem{Salimath2020}A. Salimath, Fengjun Zhuo, R. Tomasello, G. Finocchio, and A. Manchon, Phys. Rev. B {\bf 101}, 024429 (2020).

\bibitem{ChenZ2022}Z. Chen, X. Zhang, Y. Zhou, and Q. Shao, Phys. Rev. Appl. {\bf 17}, L011002 (2022).

\bibitem{Msiska2022}R. Msiska, D. R. Rodrigues, J. Leliaert, and K. Everschor-Sitte, Phys. Rev. Appl. {\bf 17}, 064015 (2022).
\bibitem{ZhouY2014}Y. Zhou, and M. Ezawa, Nat. Commun. {\bf 5}, 4652 (2014).

\bibitem{JiangW2015}W. Jiang, P. Upadhyaya, W. Zhang, G. Yu, M.B.Jungfleisch, F. Y. Fradin, J. E. Pearson, Y. Tserkovnyak, K. L. Wang, O. Heinonen, S. G. E. te Velthuis, and A. Hoffmann, Science {\bf 349}, 283 (2015).
\bibitem{YuXZ2023}X. Z. Yu, N. Kanazawa, X. Zhang, Y. Takahashi, K. V. Iakoubovskii, K. Nakajima, T. Tanigaki, M. Mochizuki, and Y. Tokura, Adv. Mat. {\bf 36}, 2306441 (2023).

\bibitem{ZhangX2024}X. Zhang, Y. Zhou, X. Z. Yu, and M. Mochizuki, Aggregate (2024).
\bibitem{Okubo2012}T. Okubo, S. Chung, and H. Kawamura, Phys. Rev. Lett. {\bf 108}, 017206 (2012).

\bibitem{Leonov2015}A. O. Leonov and M. Mostovoy, Nat. Commun. {\bf 6}, 8275 (2015).

\bibitem{Leonov2017}A. O. Leonov and M. Mostovoy, Nat. Commun. {\bf 8}, 14394 (2017).

\bibitem{WangW2015}W. Wang, M. Beg, B. Zhang, W. Kuch, and H. Fangohr, Phys. Rev. B {\bf 92}, 020403(R) (2015).

\bibitem{Ikka2018}M. Ikka, A. Takeuchi, and M. Mochizuki, Phys. Rev. B {\bf 98}, 184428 (2018).

\bibitem{Takeuchi2018}A. Takeuchi, and M. Mochizuki, Appl. Phys. Lett. {\bf 113}, 072404 (2018).

\bibitem{Takeuchi2019}A. Takeuchi, S. Mizushima, and M. Mochizuki, Sci. Rep. {\bf 9}, 9528 (2019).

\bibitem{YuanHY2019}H. Y. Yuan, X. S. Wang, M.-H. Yung, and X. R. Wang, Phys. Rev. B {\bf 99}, 014428 (2019).

\bibitem{ChenW2020b}W. Chen, L. Liu, and Y. Zheng, Phys. Rev. Appl. {\bf 14}, 064014 (2020).

\bibitem{BoL2024b}L. Bo, R. Zhao, X. Zhang, M. Mochizuki, X. Zhang, J. Appl. Phys. {\bf 135}, 063905 (2024).
\bibitem{Iwasaki2014b}J. Iwasaki, A. J. Beekman, and N. Nagaosa, Phys. Rev. B {\bf 89}, 064412 (2014).

\bibitem{Schutte2014}C. Sch\"{u}tte and M. Garst, Phys. Rev. B {\bf 90}, 094423 (2014).

\bibitem{DingJ2015}J. Ding, X. Yang, and T, Zhu, IEEE Trans. Mag. {\bf 51}, 1500504 (2015).

\bibitem{LiuY2022}Y. Liu, T. T. Liu, Z. Jin, Z. P. Hou, D. Y. Chen, Z. Fan, M. Zeng, X. B. Lu, X. S. Gao, M. H. Qin, and J.-M. Liu, Phys. Rev. B {\bf 106}, 064424 (2022).

\bibitem{MaXP2023}X.-P. Ma, X. Ai, X.-X. Yang, M.-X. Cai, J.-H. Shim, H.-G. Piao, J. Mag. Mag. Mat. {\bf 581}, 170665 (2023).

\bibitem{HuangL2023}L. Huang, G. Burnell, and C. H. Marrows, Phys. Rev. B {\bf 107}, 224418 (2023).

\bibitem{BaiX2023}X. Bai, J. Wang, J. Yang, H. Liu, S. Zhang, and Q. Liu,  J. Mag. Mag. Mat. {\bf 586}, 171231 (2023).

\bibitem{LauM2024}M. Lau, W. H\"{a}usler, and M. Thorwart, Phys. Rev. B {\bf 109}, 014435 (2024).

\bibitem{ShenM2018}M. Shen, Y. Zhang, J. Ou-Yang, X. Yang, and L. You, Appl. Phys. Lett. 112, 062403 (2018).

\bibitem{WangYD2022}Y.-D. Wang, Z.-J. Wei, H.-R. Tu, C.-H. Zhang, and Z.-P. Hou, Rare Met. {\bf 41}, 4000-4014 (2022).

\bibitem{LiZ2018}Z. Li, Y. Zhang, Y. Huang, C. Wang, X. Zhang, Y. Liu, Y. Zhou, W. Kang, S. C. Koli, and N Lei, J Magn. Magn. Mater. {\bf 455}, 19-24 (2018).

\bibitem{WangX2018}X. Wang, W. L. Gan, J. C. Martinez, F. N. Tan, M. B. A. Jalilc, and  W. S. Lew, Nanoscale {\bf 10}, 733 (2018).

\bibitem{ShenL2018}L. Shen, J. Xia, G. Zhao, X. Zhang, M. Ezawa, O. A. Tretiakov, X. Liu, and Y. Zhou, Phys. Rev. B {\bf 98}, 134448 (2018).

\bibitem{LiuY2019}Y. Liu, N. Lei, C. Wang, X. Zhang, W. Kang, D. Zhu, Y. Zhou, X. Liu, Y. Zhang, and W. Zhao,
Phys. Rev. Applied {\bf 11}, 014004 (2019).

\bibitem{MaC2019}C. Ma, X. Zhang, J. Xia, M. Ezawa, W. Jiang, T. Ono, S. N. Piramanayagam, A. Morisako, Y. Zhou, and X. Liu, Nano Lett. {\bf 19}, 353 (2019).

\bibitem{Yanes2019}R. Yanes, F. Garcia-Sanchez, R. F. Luis, E. Martinez, V. Raposo, L. Torres, and L. Lopez-Diaz, Appl. Phys. Lett. {\bf 115}, 132401 (2019).

\bibitem{Nakatani2016}Y. Nakatani, M. Hayashi, S. Kanai, S. Fukami, and H. Ohno, Appl. Phys. Lett. {\bf 108}, 152403 (2016).
\bibitem{Kovalev2009}A. A. Kovalev and Y. Tserkovnyak, Phys. Rev. B {\bf 80}, 100408(R) (2009).

\bibitem{KongL2013}L. Kong and J. Zang, Phys. Rev. Lett. {\bf 111}, 067203 (2013).

\bibitem{Mochizuki2014}M. Mochizuki, X. Z. Yu, S. Seki, N. Kanazawa, W. Koshibae, J. Zang, M. Mostovoy, Y. Tokura, and N. Nagaosa, Nat. Mat. {\bf 13}, 241 (2014).

\bibitem{LinSZ2014}S. Z. Lin, C. D. Batista, C. Reichhardt, and A. Saxena, Phys. Rev. Lett. {\bf 112}, 187203 (2014).


\bibitem{Matsuki2023}J. Matsuki and M. Mochizuki, Phys. Rev. B {\bf 107}, L100408 (2023).

\bibitem{YuXZ2021}X. Z. Yu, F. Kagawa, S. Seki, M. Kubota, J. Masell, F. S. Yasin, K. Nakajima, M. Nakamura, M. Kawasaki, N. Nagaosa, and Y. Tokura, Nat. Commun. {\bf 12}, 5079 (2021).

\bibitem{QinG2022}G. Qin, X. Zhang, R. Zhang, K. Pei, C. Yang, C. Xu, Y. Zhou, Y. Wu, H. Du, and R. Che, Phys. Rev. B {\bf 106}, 024415 (2022).
\bibitem{Zazvorka2019}J. Zazvorka, F. Jakobs, D. Heinze, N. Keil, S. Kromin, S. Jaiswal, G. Jakob, P. Virnau, D. Pinna, K. Everschor-Sitte, L. Rosza, A. Donges, U. Nowak, and M. Kl\"{a}ui, Nat. Nanotech. {\bf 14}, 658 (2019).

\bibitem{Nozaki2019}T. Nozaki, Y. Jibiki, M. Goto, E. Tamura, T. Nozaki, H. Kubota, A. Fukushima, S. Yuasa, and Y. Suzuki, Appl. Phys. Lett. {\bf 114}, 012402 (2019).

\bibitem{ZhaoL2020}L. Zhao, Z. Wang, X. Zhang, X. Liang, J. Xia, K. Wu, H. Zhou, Dong, G. Yu, K. L. Wang, X. Liu, Y. Zhou, and W. Jiang, Phys. Rev. Lett. {\bf 125}, 027206 (2020).

\bibitem{Ishikawa2021}R. Ishikawa, M. Goto, H. Nomura, and Y. Suzuki, Appl. Phys. Lett. {\bf 119}, 072402 (2021).

\bibitem{Miki2021}S. Miki, Y. Jibiki, E. Tamura, M. Goto, M. Oogane, J. Cho, R. Ishikawa, H. Nomura, and Y. Suzuki, J. Phys. Soc. Jpn. {\bf 90}, 083601 (2021).

\bibitem{SongC2021}C. Song, N. Kerber, J. Rothorl, Y. Ge, K. Raab, B. Seng, M. A. Brems, F. Dittrich, R. M. Reeve, J. Wang, Q. Liu, P. Virnau, and M. Kl\"{a}ui, Adv. Funct. Mater. {\bf 31}, 2010739 (2021).

\bibitem{ZhangX2023b}X. Zhang, J. Xia, O. A. Tretiakov, M. Ezawa, G. Zhao, Y. Zhou, X. Liu, and M. Mochizuki, Nano Lett. 23, 11793 (2023).
\bibitem{Seki2020}S. Seki, M. Garst, J. Waizner, R. Takagi, N. D. Khanh, Y. Okamura, K. Kondou, F. Kagawa, Y. Otani, and Y. Tokura, Nat. Commun. {\bf 11}, 256 (2020).

\bibitem{ChenR2023}R. Chen, C. Chen, L. Han, P. Liu, R. Su, W. Zhu, Y. Zhou, F. Pan, and C. Song, Nat. Commun. {\bf 14}, 4427 (2023).

\bibitem{YangY2023}Y. Yang, Y. Ji, C. Zhang, and T. Nan, J. Phys. D: Appl. Phys. {\bf 56}, 084002 (2023).

\bibitem{YangY2024}Y. Yang, L. Zhao, D. Yi, T. Xu, Y. Chai, C. Zhang, D. Jiang, Y. Ji, D. Hou, W. Jiang, J. Tang, P. Yu, H. Wu, and T. Nan, Nat. Commun. {\bf 15}, 1018 (2024).
\bibitem{YangW2018}W. Yang, H. Yang, Y. Cao, and P. Yan, Opt. Exp. {\bf 26}, 8778-8790 (2018).

\bibitem{Hirosawa2022}T. Hirosawa, J. Klinovaja, D. Loss, and S. A. D\'{i}az, Phys. Rev. Lett. {\bf 128}, 037201 (2022).


\bibitem{Buttner2015}F. B\"{u}ttner, C. Moutafis, M. Schneider, B. Kr\"{u}ger, C. M. G\"{u}nther, J. Geilhufe, C. v. Korff Schmising, J. Mohanty, B. Pfau, S. Schaffert, A. Bisig, M. Foerster, T. Schulz, C. A. F. Vaz, J. H. Franken, H. J. M. Swagten, M. Kl\"{a}ui, and S. Eisebitt, Nat. Phys. {\bf 11}, 225 (2015).



\bibitem{Mochizuki2017}M. Mochizuki, Appl. Phys. Lett. {\bf 111}, 092403 (2017).
\bibitem{Mochizuki2015a}M. Mochizuki and Y. Watanabe, Appl. Phys. Lett. {\bf 107}, 082409 (2015).

\bibitem{Mochizuki2015b}M. Mochizuki, Adv. Electron. Mater. {\bf 1}, 1500180 (2015).

\bibitem{Okamura2016}Y. Okamura, F. Kagawa, S. Seki, and Y. Tokura, Nat. Commun. {\bf 7}, 12669 (2016).



\bibitem{WangL2018}L. Wang, Q. Feng, Y. Kim, R. Kim, K. H. Lee, S. D. Pollard, Y. J. Shin, H. Zhou, W. Peng, D. Lee, W. Meng, H. Yang, J. H. Han, M. Kim, Q. Lu, and T. W. Noh, Nat. Mater. {\bf 17}, 1087 (2018).

\bibitem{HuangP2018}P. Huang, M. Cantoni, A. Kruchkov, J. Rajeswari, A. Magrez, F. Carbone, and H. M. R{\o}nnow, Nano Lett. {\bf 18}, 5167 (2018).

\bibitem{Kruchkov2018}A. J. Kruchkov, J. S. White, M. Bartkowiak, I. \u{Z}ivkovi'c, A. Magrez, and H. M. R{\o}nnow, Sci. Rep. {\bf 8}, 10466 (2018).


\bibitem{WangY2020}Y. Wang, L. Wang, J. Xia, Z. Lai, G. Tian, X. Zhang, Z. Hou, X. Gao, W. Mi, C. Feng, M. Zeng, G. Zhou, G. Yu, G. Wu, Y. Zhou, W. Wang, X.-X. Zhang, and J. Liu, Nat. Commun. {\bf 11}, 3577 (2020).
\bibitem{Ogasawara2009}T. Ogasawara, N. Iwata, Y. Murakami, H. Okamoto, and Y. Tokura, Appl. Phys. Lett. {\bf 94}, 162507 (2009).

\bibitem{Finazzi2013}M. Finazzi, M. Savoini, A. R. Khorsand, A. Tsukamoto, A. Itoh, L. Duo, A. Kirilyuk, Th. Rasing, and M. Ezawa, Phys. Rev. Lett. {\bf 110}, 177205 (2013).

\bibitem{Koshibae2014}W. Koshibae, and N. Nagaosa, Nat. Commun. {\bf 5}, 5148 (2014).

\bibitem{Oike2016}H. Oike, A. Kikkawa, N. Kanazawa, Y. Taguchi, M. Kawasaki, Y. Tokura, and F. Kagawa, Nat. Phys. {\bf 12}, 62-66 (2016).

\bibitem{Vallobra2018}S. G. Je, P. Vallobra, T. Srivastava, J. C. Rojas-S\'{a}nchez, T. H. Pham, M. Hehn, G. Malinowski, C. Baraduc, S. Auffret, G. Gaudin, S. Mangin, H. B\'{e}a, and O. Boulle, Nano Lett. {\bf 18}, 7362 (2018).

\bibitem{Berruto2018}G. Berruto, I. Madan, Y. Murooka, G.-M. Vanacore, E. Pomarico, J. Rajeswari, R. Lamb, P. Huang, A.-J. Kruchkov, Y. Togawa, T. LaGrange, D. McGrouther, H.-M. R{\o}nnow, and F. Carbone, Phys. Rev. Lett. {\bf 120}, 117201 (2018).
\bibitem{Lemesh2018}I. Lemesh, K. Litzius, M. B\"{o}ttcher, P. Bassirian, N. Kerber, D. Heinze, J. Z\'{a}zvorka, F. B\"{u}ttner, L. Caretta, M. Mann, M. Weigand, S. Finizio, J. Raabe, M.-Y. Im, H. Stoll, G. Sch\"{u}tz, B. Dup\'{e}, M. Kl\"{a}ui, and G. S. D. Beach, Adv. Mater. {\bf 30}, 1870372 (2018).
\bibitem{Nii2015}Y. Nii, T. Nakajima, A. Kikkawa, Y. Yamasaki, K. Ohishi, J. Suzuki, Y. Taguchi, T. Arima, Y. Tokura, and Y. Iwasa, Nat. Commun. {\bf 6}, 8539 (2015).
\bibitem{Seki2017}S. Seki, Y. Okamura, K. Shibata, R. Takagi, N. D. Khanh, F. Kagawa, T. Arima, and Y. Tokura, Phys. Rev. B {\bf 96}, 220404(R) (2017).

\bibitem{Tanaka2020}K. Tanaka, R. Sugawara, and M. Mochizuki, Phys. Rev. Mater. {\bf 4}, 034404 (2020).
\bibitem{Miyake2020}M. Miyake and M. Mochizuki, Phys. Rev. B. {\bf 101}, 094419 (2020).
\bibitem{Fujita2017a}H. Fujita, and M. Sato, Phys. Rev. B {\bf 95}, 054421 (2017).

\bibitem{Fujita2017b}H. Fujita, and M. Sato, Phys. Rev. B {\bf 96}, 060407(R) (2017).
\bibitem{Gobel2021}B. G\"{o}bel, I. Mertig, and O. A. Tretiakov, Phys. Rep. {\bf 895}, 1-28 (2021).
\end{thebibliography}
\end{document}